\documentclass[11pt]{article}
\usepackage{jcapmod}

\usepackage{mathtools}
\usepackage{booktabs}
\usepackage[english]{babel}
\usepackage{amsmath,amssymb,amsbsy,amstext, amsthm, amsfonts}
\usepackage{hyperref}
\usepackage{graphicx}
\usepackage[small]{caption}
\usepackage{siunitx}
\usepackage{framed}
\usepackage{wrapfig}
\usepackage{multirow}
\usepackage[svgnames,dvipsnames,x11names]{xcolor}
\usepackage{selinput}
\usepackage{bm}
\usepackage{float}
\usepackage{geometry}
\usepackage{yfonts}
\usepackage{caption}
\usepackage{subcaption}
\usepackage{subfloat}
\usepackage{sidecap}
\usepackage{longtable}
\usepackage{physics}
\usepackage[compat=1.0.0]{tikz-feynman}
\usepackage{tikz}
\usepackage{bbm}
\usepackage{threeparttable}
\usepackage{lipsum}

\usetikzlibrary{shapes.geometric}

\newcommand{\hS}{\hat{S}}
\newcommand{\nnm}{\nonumber}
\newcommand{\mr}{\text}
\newcommand{\mc}{\mathcal}
\usepackage{colortbl}
\definecolor{summersky}{cmyk}{0.71,0.33,0,0.5}
\definecolor{flamingo}{cmyk}{0,0.51,0.71,0.5}
\definecolor{rp}{cmyk}{0.2, 1, 0.6, 0}
\definecolor{pacificblue}{cmyk}{0.95,0.3,0, 0.5}
\definecolor{gray60}{cmyk}{0.4,0.4,0,0.8}

\usepackage[framemethod=default]{mdframed}
\newmdenv[skipabove=7pt,
skipbelow=7pt,
rightline=false,
leftline=false,
topline=false,
bottomline=false,
backgroundcolor=pacificblue!10,
linecolor=gray,
innerleftmargin=5pt,
innerrightmargin=5pt,
innertopmargin=2pt,
innerbottommargin=10pt,
leftmargin=0cm,
rightmargin=0cm,
linewidth=4pt]{eBox}
\newmdenv[skipabove=7pt,
skipbelow=7pt,
rightline=false,
leftline=false,
topline=false,
bottomline=false,
backgroundcolor=gray!10,
linecolor=gray,
innerleftmargin=5pt,
innerrightmargin=5pt,
innertopmargin=2pt,
innerbottommargin=10pt,
leftmargin=0cm,
rightmargin=0cm,
linewidth=4pt]{eBox2}

\definecolor{blue3}{RGB}{31, 119, 180}
\definecolor{red3}{RGB}{    214, 39, 40}
\definecolor{orange3}{RGB}{255, 127, 14}
\definecolor{green3}{RGB}{44, 160, 44}
\definecolor{repBlue}{RGB}{31, 119, 180}
\definecolor{repRed}{RGB}{  214, 39, 40}
\definecolor{repGreen}{RGB}{44, 160, 44}
\definecolor{MyRed}{RGB}{208,57,75}
\definecolor{MyBlue}{RGB}{68,123,178}
\definecolor{MyYellow}{RGB}{238,140,59}

\renewcommand{\(}{\left(}
\renewcommand{\)}{\right)}
\renewcommand{\[}{\left[}
\renewcommand{\]}{\right]}

\def\be{\begin{equation}}
\def\ee{\end{equation}}
\newcommand{\bea}{\begin{eqnarray}}
\newcommand{\eea}{\end{eqnarray}}

\newcommand{\blue}[1]{\textcolor{blue3}{#1}}
\newcommand{\green}[1]{\textcolor{green3}{#1}}
\newcommand{\red}[1]{\textcolor{red3}{#1}}
\newcommand{\purple}[1]{\textcolor{Purple}{#1}}
\newcommand{\orange}[1]{\textcolor{orange3}{#1}}

\newcommand{\rom}[1]{\uppercase\expandafter{\romannumeral #1\relax}}

\makeatletter

\newcommand{\Rmnum}[1]{\expandafter\@slowromancap\romannumeral #1@}
\makeatother

\usepackage{colortbl}
\definecolor{lightgreen}{cmyk}{0.2, 0, 0.2, 0.2}
\definecolor{lightgray}{cmyk}{0.1,0.2,0,0.1}
\definecolor{lightgray2}{cmyk}{0.1,0.1,0,0.1}

\makeatletter
\newlength{\apb@width}
\newcommand{\autoparbox}[2][c]{\settowidth{\apb@width}{#2}\parbox[#1]{\apb@width}{#2}}

\makeatother


\def\beq{\begin{equation}}
\def\eeq{\end{equation}}

\allowdisplaybreaks[1]
\title{
Dynamical Tidal Response of Kerr Black Holes from Scattering Amplitudes
}

\author[1a,b]{M. V. S. Saketh\note{msaketh@umd.edu},}
\author[2c]{Zihan Zhou\note{zihanz@princeton.edu},}
\author[3d,e]{Mikhail M. Ivanov\note{ivanov99@mit.edu}}

\affiliation[a]{Department of Physics, University of Maryland, College Park, MD 20742, USA}
\affiliation[b]{Max Planck Institute for Gravitational Physics (Albert Einstein Institute), Am M{\" u}hlenberg 1, Potsdam 14476, Germany}
\affiliation[c]{Department of Physics, Princeton University, Princeton, NJ 08540, USA}
\affiliation[d]{Center for Theoretical Physics, Massachusetts Institute of Technology, Cambridge, MA 02139, USA}
\affiliation[e]{School of Natural Sciences, Institute for Advanced Study, 1 Einstein Drive, Princeton, NJ 08540, USA}

\date{}
\abstract{
We match scattering amplitudes 
in point particle effective field theory (EFT) and general relativity 
to extract low frequency dynamical tidal responses 
of rotating (Kerr) black holes to all orders in spin.  
In the conservative sector, we study local worldline couplings 
that correspond to the 
time-derivative expansion of the 
black hole tidal response function. 
These are dynamical (frequency-dependent) 
generalizations of the static Love numbers. 
We identify and extract couplings of three types 
of subleading local worldline 
operators: the curvature time derivative terms, 
the spin - curvature time derivative couplings, and 
quadrupole - octupole mixing operators that arise due to the 
violation of spherical symmetry.
The first two subleading couplings are non-zero and exhibit a classical 
renormalization group running; we explicitly 
present their scheme-independent beta functions. 
The conservative mixing terms, however, vanish 
as a consequence of vanishing static Love numbers.
In the non-conservative sector, 
we match the dissipation numbers at 
next-to-leading and next-to-next-to leading orders 
in frequency.
In passing, we identify terms in the general relativity 
absorption probabilities
that originate from tails
and short-scale logarithmic 
corrections to the lowest order
dissipation contributions. 
}

\begin{document}

\maketitle

\clearpage

\section{Introduction}
\label{sec:int}

The growing list of gravitational wave (GW) events observed by the LIGO-Virgo-KAGRA collaboration has sparked a new era in the study of strongly gravitating compact objects \cite{LIGOScientific:2014pky,VIRGO:2014yos,LIGOScientific:2016aoc,LIGOScientific:2018mvr,LIGOScientific:2020ibl,LIGOScientific:2021usb,LIGOScientific:2021djp,KAGRA:2020agh}. 
Parameter inference from this data requires accurate 
waveform templates, which must include, in particular, tidal effects~\cite{Flanagan:2007ix,Buonanno:2022pgc}.
These effects are most prominent for neutron stars, 
whose static tidal deformations allow one 
to test their equation of state~\cite{Flanagan:2007ix,Vines:2011ud,Bini:2012gu}.
These practical applications motivated 
further
theoretical studies of tidal deformations of compact bodies, 
especially black holes (BHs). 
In Newtonian theory, leading order tidal deformations of celestial bodies 
are captured by static Love numbers~\cite{love1909yielding,poisson2014gravity}, which depend only on the 
internal structure of the body. 
The post-Newtonian generalization of Love numbers in the 
case of static metric was done in~\cite{Fang:2005qq,Damour:2009vw,Binnington:2009bb}. Further generalizations to the case of spin and 
dynamical perturbations were made in~\cite{Poisson:2014gka,Landry:2015zfa,Landry:2015snx,Poisson:2020mdi,Poisson:2020vap,LeTiec:2020spy,LeTiec:2020bos,Chia:2020yla,Poisson:2021yau} where certain 
conceptual issues were encountered, such as difficulties 
in the source-response split of the tidal moments, 
the presence of frame dragging, 
concerns about the coordinate dependence of the results, 
and the presence of logarithmic non-localities. 

A conceptually clean definition of static Love numbers and 
more general tidal deformations can be given within the 
point particle worldline effective field theory (EFT), 
also known as non-relativistic perturbative general relativity (GR)~\cite{Goldberger:2004jt,Goldberger:2005cd,Goldberger:2009qd,Rothstein:2014sra,Porto:2016pyg,Levi:2018nxp,Goldberger:2020fot,Goldberger:2022ebt,Goldberger:2022rqf}. 
In this approach, in the first approximation, 
a compact object is described as a point 
particle on a worldline. 
The true finite-size effects related to the internal 
structure are captured by local worldline couplings that 
start at quadratic order in curvature, e.g. 
for a spherically-symmetric background we have 
\be 
\label{eq:LNwdef}
S_{\rm finite-size}=c_E MR^4 \int d\tau~E_{\mu\nu}E^{\mu\nu} + 
c_B MR^4 \int d\tau~B_{\mu\nu}B^{\mu\nu}\,,
\ee
where $R$ and $M$ are the radius and the mass of a compact body, 
$\tau$ is proper time, while $E_{\mu\nu}$ and $B_{\mu\nu}$
are electric and magnetic parts of the Weyl tensor.
Performing linear static response calculations~\cite{Kol:2011vg,Hui:2020xxx,Charalambous:2021mea}
one can see that the Wilson coefficients $c_{E,B}$ in~\eqref{eq:LNwdef}
reproduce the classic definition 
of static Love numbers in the Newtonian limit,
whereby defining the Love numbers in full GR.
This definition of conservative tidal responses 
as worldline couplings is free of difficulties
related to the non-linear structure of GR, 
the presence of spin\footnote{Note that the spin effects were included in the EFT in~\cite{Porto:2005ac,Porto:2007px,Steinhoff:2007mb,Porto:2008jj,Porto:2008tb,Steinhoff:2009tk,Porto:2010tr,Porto:2010zg,Wang:2011bt,Hartung:2011ea,Hartung:2011te,Porto:2012as,Marsat:2014xea,Steinhoff:2014kwa,Levi:2014gsa,Levi:2015msa,Steinhoff:2015ksa,Vines:2016qwa,Levi:2016ofk,Vines:2016unv,Siemonsen:2017yux,Levi:2019kgk,Siemonsen:2019dsu,Liu:2021zxr, Levi:2020kvb, Cho:2022syn, Levi:2020lfn, Levi:2020uwu,Kim:2022pou,Levi:2022dqm,Kim:2022bwv, Levi:2022rrq, Saketh:2022wap}.},
and gauge invariance issues. 
In particular, in order to include spin, one has to 
promote the Love numbers to spin-dependent tensors in order to account 
for the breaking of the spherical symmetry by the underlying 
gravitational background~\cite{LeTiec:2020spy,LeTiec:2020bos,Goldberger:2020fot,Charalambous:2021mea}. 
Importantly, the non-analytic (logarithmic) corrections to the 
static Love numbers can readily be interpreted 
within the definition~\eqref{eq:LNwdef} as a familiar renormalization
group running of Wilson coefficients.\footnote{The Love numbers of BHs in four dimensional GR do not run as a result of 
the fine-tuning hidden symmetry~\cite{Kol:2011vg,Porto:2016zng,Charalambous:2022rre,Ivanov:2022hlo}. They run, however, in higher dimensions~\cite{Kol:2011vg}. }
 Within the EFT framework, it has been shown that the 
 static Love numbers of Scwarzschild and Kerr black holes 
 vanish identically in four dimensional general relativity~\cite{Kol:2011vg,Hui:2020xxx,Chia:2020yla,Charalambous:2021mea,Ivanov:2022hlo,Ivanov:2022qqt}\footnote{Love numbers  have also been studied in the context of BHs in higher dimensions~\cite{Kol:2011vg,Hui:2021vcv,Ivanov:2022qqt,longpaper,Charalambous:2023jgq,Rodriguez:2023xjd}, supergravity solutions~\cite{Cvetic:2021vxa}, and black string systems~\cite{Rodriguez:2023xjd}.   Non-linear static tidal Love numbers were studied in \cite{DeLuca:2023mio}. }, 
 which implies a fine-tuning from the EFT perspective~\cite{Porto:2016zng}\footnote{Proposals for a hidden symmetry resolution of this paradox have been made in \cite{Charalambous:2021kcz,Hui:2021vcv,Hui:2022vbh,Charalambous:2022rre}.}. 

The time-dependent tides can also be consistently
defined in the EFT.
One can also introduce time-dependent (dynamical) Love numbers
as EFT couplings in front of operators like~\cite{Porto:2016pyg,Goldberger:2022rqf}
\be 
\label{eq:NLOspher}
c_{\dot E} M R^6 \int d\tau~u^\sigma \nabla_\sigma E_{\mu\nu} u^\rho \nabla_\rho E^{\mu\nu} = 
c_{\dot E} M R^6 \int d\tau~\dot E^{ij} ~\dot E_{ij}\,,
\ee
where in the last equality we have switched to the local comoving frame.
In analogy to static Love numbers, the Wilson coefficient
$c_{\dot E}$ captures the leading order frequency-dependent 
conservative response to external perturbations. 
If the compact body carries spin, $c_{\dot E}$ should be promoted 
to a tensor. In addition, the breaking of spherical symmetry 
generates new local worldline couplings, e.g.~\cite{Bini:2012gu,Chakrabarti:2013lua,Goldberger:2020fot,Charalambous:2021mea}
\be 
\label{eq:NLOspin}
M R^5 \int d\tau~ \lambda_{ij,kl} E^{ij} \dot E_{kl} \,,\quad 
MR^5 \int d\tau~ (\nu)_{ij,kl,m} B^{ij} \nabla^{\langle k} E^{lm\rangle}\,
\ee
where 
$\lambda_{ij,kl}$ and $\nu_{ij,kl,m}$
are spin-dependent tensors.\footnote{The symbol $\langle...\rangle$ in Eq.~\eqref{eq:NLOspin} denotes symmetrization over relevant indices 
and consequent subtraction of traces.} 
The first term in Eq.~\eqref{eq:NLOspin}
describes the leading order interaction between the spin and the
curvature time derivative, while the last one captures 
the spin-induced quadrupole-octupole mixing.

As far as the non-conservative effects are concerned, the worldline EFT 
elegantly takes them into account by means of introducing 
internal worldline degrees of freedom that couple to ``bulk''
gravitational fields.
The leading order effects of these couplings, such as the horizon absorption, 
superradiance, tidal torquing, heating, etc. have been extensively 
studied in the literature~\cite{Starobinsky:1973aij,Starobinskil:1974nkd,Maldacena:1996ix,Goldberger:2005cd,LeTiec:2020spy,LeTiec:2020bos,Chia:2020yla,Charalambous:2021mea,Goldberger:2020fot, Saketh:2022xjb}. 
In terms of the BH tidal response function, these effects 
may be parametrized by the so-called dissipation numbers, introduced 
by analogy with the Love numbers, but capturing the non-conservative 
part of the response. In this paper we will focus on their 
generalizations beyond the leading order.

The conservative effects associated with operators~\eqref{eq:NLOspher}
and~\eqref{eq:NLOspin} are formally 
suppressed in the PN regime compared to the leading order (LO) static 
deformations, and therefore have not yet been extensively studied in
the literature.\footnote{It is worth mentioning that Ref.~\cite{Chakrabarti:2013lua} was the first one to show that the operators~\eqref{eq:NLOspher} are not zero for Schwarzschild black holes. Ref.~\cite{Charalambous:2021mea}
showed, for the first time, that the first operator in~\eqref{eq:NLOspin} (the dynamical Love number) is not zero 
for Kerr black holes. See also \cite{Landry:2015snx,Steinhoff:2016rfi,Gupta:2020lnv,Steinhoff:2021dsn,Gupta:2023oyy,Mandal:2023lgy,Poisson:2020vap}.}
The non-conservative tidal effects also remain largely unexplored beyond 
the LO. 
There are strong reasons to consider these 
next-to-leading order (NLO)
and next-to-next-to leading order (NNLO)
tidal effects both in the conservative and dissipative sectors. In general, calculations of subleading effects put the LO results
on firm ground. 
This is especially relevant for BHs, for which 
LO conservative tidal effects are absent for all multipoles,
and hence the genuine finite-size tidal effects  
appear for the first time at NLO (for Kerr) and NNLO (for Schwarzschild) 
in the low frequency expansion. 
Since the vanishing of static (LO)
Love numbers has recently attracted some attention in the context of 
hidden symmetries, it is worth noticing that the non-vanishing 
of NLO tidal couplings may provide important information 
about breaking mechanisms for these hidden symmetries. The higher order 
dynamical tidal effects could also shed light on the dispersive representation of BH tidal response function and its relation with the vanishing static Love numbers for 4D BHs~\cite{Rothstein:2014sra,Porto:2016pyg,Goldberger:2022rqf,Kim:2020dif}.~\footnote{
	 In addition to this, there is significant motivation to study dynamical responses even in the context of Newtonian fluid bodies. For 
	 example, it can provide us with a microscopic understanding of Love numbers in terms of the fluid's fundamental $f$-mode \cite{poisson2014gravity,Andersson:2019yve}. In astrophysics, dynamical tides govern the evolution of an oscillating star which then backreact to the orbital motion \cite{rathore2003variational, Chakrabarti:2013lua}, even with resonances \cite{Flanagan:2006sb,Ma:2020oni,Poisson:2020eki}. 
}

With these motivations in mind, in this work, we systematically study 
the frequency-dependent corrections to the 
tidal response function of rotating (Kerr) black holes
perturbatively in $GM\omega$ (with $\omega$ being the frequency of the external source\footnote{In a particular case of an inspiraling binary, this would be the orbital frequency of the satellite body.}, and $G$ is 
the Newton's constant), analogous to the post-Minkowskian (PM) expansion, up to next-to-next-to leading order (NNLO). 
We accomplish this by matching various EFT observables 
produced by tidal operators to known 
analytic GR results. 
Specifically, we match the phase shifts due to 
scattering of gravitational waves off Kerr 
black holes in the EFT and full general relativity. 
This on-shell matching method allows us to avoid gauge and coordinate 
dependence issues
as the amplitudes are manifestly gauge invariant objects. 
For analytical expressions of the scattering phase shifts in GR, 
we use the black hole perturbation theory (BHPT) methods of Mano, Suzuki, and Takasugi (MST) for the analytic solution of the Teukolsky master equation~\cite{Mano:1996gn,Mano:1996mf,Mano:1996vt,Tagoshi:1997jy,Sasaki:2003xr}\footnote{Note that similar in spirit matching calculations were recently performed by \cite{Saketh:2022xjb} in worldline theory and \cite{Bautista:2021wfy,Bautista:2022wjf,Aoude:2023fdm} in an on-shell-amplitudes approach.}.
One important result of our calculations is that the NLO and NNLO tidal
worldline couplings of Kerr BHs do not vanish and also exhibit a renormalization group 
running behavior, which is expected on the basis of 
Wilsonian naturalness and power counting~\cite{Kol:2011vg,Ivanov:2022hlo,Charalambous:2022rre}. Thus, 
the fine tuning of the worldline EFT for Kerr BHs 
takes place for static couplings only.
As far as the quadrupole-octupole 
mixing local couplings are concerned, 
we will show that they vanish identically as a consequence of the 
vanishing of static Love numbers.

There are important technical and conceptual 
challenges what we have resolved in our study. 
First of all, we take into account the BH spin formally to all 
orders in its value. 
To that end, we use the most general expansion of the BH response function
that is consistent with axial symmetry and parity, 
and decompose the unknown 
response tensors over a finite basis of master tensors. 
We then separate the conservative and dissipative parts by studying the behavior of the response under time reversal symmetry that includes spin flip~\cite{Goldberger:2020fot,Charalambous:2021mea,Saketh:2022xjb}. 
The second important problem is the isolation of tidal contribution in the 
BHPT scattering expressions. This difficulty is overcome with the 
application of the near-far 
factorization arguments presented in~\cite{Ivanov:2022qqt}, which we extend here 
to NLO and NNLO. The third important aspect is the tail effect~\cite{Blanchet:1993ec,Poisson:1994yf,Goldberger:2009qd}
due to the wave scattering off the Newtonian potential before or after scattering off the tidal moments.
If unaccounted for, the tail effect would obscure 
the matching of the dissipative numbers 
beyond LO. Using the EFT, we explicitly calculate tail
corrections to BH absorption and identify them in the BHPT result. 
Finally, we systematically study the mentioned above 
quadrupole-octupole mixing effect on the conservative and dissipative 
parts of the tidal response.

This paper is structured as follows.
In Sec.~\ref{sec:EFT}, we first explicitly construct the most general building blocks for the investigation of the tidal response of compact rotating objects. We then define all tidal response coefficients and discuss the wave scattering framework that will be used in this work. 
In Sec.~\ref{sec:BHPT}, we provide a brief overview of the MST method. We also briefly present the near-far factorization arguments that allow for the extraction of the phase shift from dynamical tidal effects in BHPT. 
In Sec.~\ref{sec:matching}, we match the EFT amplitude calculations with the BHPT phase shifts. This includes explicit matching of the static tidal Love numbers and the LO+NLO dissipation numbers in Sec.~\ref{subsec:match love and dissipation} and Sec.~\ref{subsec:match NLO dissipation}, respectively. In Sec.~\ref{subsec:tail} we focus on the tail effects on the tidal response. Sec.~\ref{subsec:RGabs} and Sec.~\ref{subsec: RG elastic} focus on the RG running of dynamical Love numbers and dissipation numbers. 
We present the explicit scheme-independent beta functions for 
relevant local couplings there. 
In Sec.~\ref{sec:mix}, we match the quadrupole-octupole mixing terms in the tidal response for rotating black holes. 
Sec.~\ref{sec:constants} presents a discussion of the finite, scheme-dependent 
parts of the conservative couplings. 
Finally, in Sec.~\ref{sec:con}, we summarize the main results and discuss the future directions. Supplementary material is collected in two appendices. App.~\ref{app:tetrads}
contains details of our tetrad choice and its relationship 
to that of Ref.~\cite{Goldberger:2020fot}.
In Appendix~\ref{app:love rep} we set up 
conventions for the plane waves, 
spherical states, and the transformations between them.

\section{Tidal Effects in Worldline Effective Field Thoery}
\label{sec:EFT}

When the relevant length scales are much larger than the size of a BH (more generally, any compact body), its dynamics can be described by a worldline EFT, which can be thought of as an expansion around the point particle approximation. 
In this approach, one typically writes down a universal point particle action 
and then perturbatively adds non-minimal coupling terms, consisting of higher powers or higher derivatives of the relevant fields to include finite size effects, e.g. spin-induced multipole moments and tidal effects. 
In this work, we will restrict ourselves to the tidal effects induced by linear external perturbations, which may subsequently be compared with results obtained via BHPT. These correspond to the quadratic curvature terms in the 
effective worldline action. 
In the worldline EFT finite size effects are taken into account by approximating the matter distribution inside the compact object via multipolar expansion. 
In the following, we will briefly go over this approach, starting with the simple case of a Schwarzschild BHs. 

\subsection{Spherically Symmetric Compact Objects}
\label{sec:ScwBH}
In the case of spherical compact objects, the starting point for the action is simply that of a structure-less point particle, moving along a path that maximizes proper time given by \cite{Goldberger:2004jt,Goldberger:2005cd,Goldberger:2009qd,Rothstein:2014sra,Porto:2016pyg,Levi:2018nxp,Goldberger:2020fot,Goldberger:2022ebt,Goldberger:2022rqf}
\begin{alignat}{3}
S_{\text{p.p}}= -M \int  d\tau = -M \int  d\sigma \sqrt{-\frac{dz^{\mu}}{d\sigma}\frac{dz^{\nu}}{d\sigma}g_{\mu\nu}},
\label{eq:basic}
\end{alignat}
where $M$ is the mass of the body, and $\tau$ is proper time. $z^{\mu}(\sigma)$ is the worldline with $\sigma$ as a suitable parameter.  p.p stands for point particle. Going forward, we will not explicitly write the dependence of the worldline coordinates $z^{\mu}$ on $\sigma$, and use an overdot to denote derivatives w.r.t proper-time, so that the 4-velocity $u^{\mu}=\dot{z}^{\mu}$.

As mentioned earlier, finite size effects may be included perturbatively via effective operators that consist of appropriate 
derivatives and powers of of the metric tensor $g_{\mu\nu}$. 
Due to general covariance, we cannot add terms with arbitrary dependence on derivatives of the metric, but only those consisting of the curvature tensors and their covariant derivatives. Furthermore, there is no need to add terms made up of Ricci tensors as they can be eliminated via a field redefinition. Thus, at leading order in finite size effects, we can write \cite{Bailey:1975fe,Damour:1998jk}
\begin{alignat}{3}
S= S_{\text{p.p}}+ \int d\tau~ J_Q^{\mu\nu\rho\sigma}C_{\mu\nu\rho\sigma},
\end{alignat}
where $C_{\mu\nu\rho\sigma}$ is the Weyl tensor and $J_Q^{\mu\nu\rho\sigma}$ is an arbitrary tensor representing a generic quadrupole moment. The intuition for this effective action comes from studying finite size effects in Newtonian gravity, where the leading order term arises from the quadrupolar deformation of a body \cite{poisson2014gravity}. Extending the intuition further, we may write $J_Q^{\mu\nu\rho\sigma}$ in different forms, built of different tensors to include various finite size effects. For example, in the spinning case, we can build the tensor $J_Q^{\mu\nu\rho\sigma}$ out of the spin tensor $S_{\mu\nu}$, and the 4-velocity $u^{\mu}$, to include the effect of spin-induced quadrupolar deformation. More generally, tidal deformation due to external gravitational fields can also be included by writing $J_Q^{\mu\nu\rho\sigma}$ in terms of the curvature tensors in the equations of motion. To make the parity transformation property manifest, it is convenient to first split up the finite size effective action into electric and magnetic parts as \cite{bel1958definition,Marsat:2014xea}
\begin{alignat}{3}
S = S_{\text{p.p}}+\int d\tau \Big(Q^{\mu\nu}_EE_{\mu\nu} + Q^{\mu\nu}_BB_{\mu\nu}\Big),
\label{eq:genl2scw}
\end{alignat}
where $E^{\mu\nu}=C^{\mu\rho\nu\sigma}u_{\rho}u_{\sigma}$ and $B^{\mu\nu} =(1/2)\epsilon_{\gamma\langle\mu}{}^{\alpha\beta}C_{\alpha\beta|\nu\rangle\delta}u^{\gamma}u^{\delta}$ are the electric and magnetic parts of the Weyl tensor, also referred to as the quadrupolar electric and magnetic tidal fields. By taking spatial derivatives on the quadrupolar electric and magnetic tidal fields, one can straightforwardly generalize the definition to higher order multipole moments \cite{Marsat:2014xea}. 

The influence of finite size effects on the dynamics can then be easily obtained by varying the above action and substituting suitable ansatz for the quadrupole moments. In this work, we will focus on the case where the ansatz for the multipole moments is linear in the tidal fields, thus neglecting spin induced multipole moments or any non-linear effects. We can more generally extend the above action to include other multipole moments as 
\begin{alignat}{3}
\label{eq:general eft action}
S= S_{\text{p.p}} + \sum_{\ell=2}^{\infty}\int d\tau(Q_{E}^{\mu_L} E_{\mu_L} + Q_{B}^{\mu_L}B_{\mu_L}),
\end{alignat}
where we are using the notation $\mu_L=\mu_1\mu_2...\mu_\ell$ with multipole index $\ell$. Noting that under parity transformation 
\begin{equation}
    E_{\mu_L} \rightarrow (-1)^\ell E_{\mu_L} ~, \quad B_{\mu_L} \rightarrow (-1)^{\ell + 1}B_{\mu_L} ~,
\end{equation}
thus the linear perturbations of different parities will not mix in this context. According to the linear response theory, we can write a general ansatz as
\begin{alignat}{3}
Q_E^{\mu_L} = -M(GM)^4\sum_{n=0}^{\infty}(-1)^n(GM)^n\lambda^E_{\omega^n}\frac{d^nE^{\mu_L}}{d\tau^n},~(E\leftrightarrow B)
\label{eq:scwgans}
\end{alignat}
which is an expansion about the adiabatic limit of static tides. By explicitly looking at the time-reversal symmetry, one immediately see that time-reversal even terms capture the conservative effects, e.g. tidal Love numbers, while time-reversal odd terms contribute to dissipation, e.g. horizon absorption. Transforming into the frequency domain, the above parametrization makes manifest that $G M \omega \ll 1$ is dimensionless power counting parameter. Since the multipole moments and the tidal fields are orthogonal to the 4-velocity, It is convenient to express them in an orthogonal tetrad $e_{I=0,1,2,3}^{\mu}$, and identify $e_{0}^{\mu}=u^{\mu}$. In this way, we can replace the covariant indices $\mu_L$ with purely spatial indices $i_L$.

\subsubsection*{From Response Function to Love Numbers and Dissipation Numbers}

The time-reversal even part of the response function Eq.~\eqref{eq:scwgans} can also be written as the higher dimensional operator in the worldline EFT. For example, in the case of static tides, one can get the effective action
\begin{alignat}{3}
S = S_{\text{p.p}} - \frac{1}{2} \int d\tau M(GM)^4\lambda^E E_{i_L}E^{i_L} + (E\leftrightarrow B)\,,
\end{alignat}
so the conservative tidal response is fully
captured by local contact terms on the worldline EFT. 
$\lambda^{E/B}$ can then be identified with the Schwarzschild static tidal Love numbers that are known to be zero~\cite{Fang:2005qq,Damour:2009vw,Binnington:2009bb,Kol:2011vg,Landry:2015zfa,Hui:2020xxx}. 
Furthermore, the next-to-next-to-leading order (NNLO) term in Eq.~\eqref{eq:scwgans}, i.e. $(G M \omega)^2$ order term, is also invariant under time reversal symmetry, and therefore can be absorbed into the action via the local contact terms of the form
\begin{alignat}{3}
S = S_{\text{p.p}} +  \frac{1}{2} \int d\tau M(GM)^6\lambda^E_{\omega^2}  \dot{E}_{i_L}\dot{E}^{i_L} + (E\leftrightarrow B).
\end{alignat}
$\lambda^{E/B}_{\omega^2}$ will be referred to as dynamical tidal Love numbers for Schwarzschild BH.

The time-reversal odd terms, e.g. the next-to-leading order (NLO) term in Eq.~\eqref{eq:scwgans}
\begin{alignat}{3}
Q_E^{i_L}\Big|_{\text{NLO}}= M(GM)^5 \lambda_{\omega}^E \dot{E}^{i_L}.~ (E\leftrightarrow B).
\end{alignat}
cannot be absorbed into the local contact term on the worldline. We can re-write it in a non-local way as 
(see Ref.~\cite{Ivanov:2022hlo} for details)
\begin{alignat}{3}
S=S_{\text{p.p}} - i \frac{1}{2} M(GM)^5\lambda^E_{\omega} \int \frac{d\omega}{2\pi} |\omega|E_{i_L}(\omega){E}^{*i_L}(\omega),
\end{alignat}
which reflects that the imaginary part of the effective action gives rise to non-conservative effects such as absorption. 
The coefficient $\lambda_{\omega}^E$ will be called a LO tidal dissipation number for Schwarzschild BH in what follows.

\subsection{Rotating Compact Objects}

The worldline EFT construction is conceptually similar in the spinning case, except for two critical distinctions. First, the EFT cannot be built solely upon a structure-less point particle. Instead, it is necessary to incorporate spin-dependent interactions, which lead to the pole-dipole Mathisson-Papapetrou-Dixon (MPD) equations of motion \cite{mathisson1937neue,Papapetrou:1951pa,Corinaldesi:1951pb,Dixon:1970zz,Dixon:1970zza}. These equations form the universal segment of the action for a body with spin. 
Second, the presence of spin not only modifies tidal effects but also introduces another kind of finite size effects: spin-induced multipole moments. As noted in previous studies \cite{Levi:2015msa,Siemonsen:2017yux,Siemonsen:2019dsu}, these two effects can be explicitly distinguished, as spin-induced multipole moments are linear in curvature, while tidal finite-size effects are quadratic in curvature. In this paper, our focus will be strictly on the tidal effects.

\subsubsection*{Building Blocks and ${\rm SO}(3)$ Representation}

In the spinning case, instead of simple proportionality relations as $Q_E^{ij}\sim \lambda_\omega^E E^{ij}$, we expect tensor relations of the form $Q_E^{ij}\sim(\lambda_\omega^E)^{ij}{}_{kl}E^{kl}$, where the unknown tensors in the tidal response (say $\lambda_\omega^E$) can only have a non-trivial structure owing to the spin of the particle.

To parametrize the kind of tensors entering the tidal response in the spinning case, let us briefly revisit the algebraic structure of spins. In the rest frame of the central compact object, we find it beneficial to use the spin tensor $S_{ij}$, the Pauli-Lubanski spin vector $s^i=(1/2)\epsilon^{ijk}S_{jk}$, and the identity tensor $\delta_{ij}$ as our building blocks. The spin magnitude is defined as $J \equiv \chi G M^2 \equiv \sqrt{(1/2)S^{ij}S_{ij}}$, where $\chi$ is the dimensionless spin parameter. For sub-extremal Kerr BHs, we must require $\chi \leq 1$. In principle, one can now build up the effective action by combining these tensorial building blocks into scalar quantities. For convenience, we further introduce the unit spin tensor $\hat{S}_{ij} \equiv S_{ij}/J$, $\hat{S}^{ij}\hat{S}_{ij} = 2$ and the unit spin vector $\hat{s}^i \equiv s^i / J$, $\hat{s}^i \hat{s}_i = 1$.

Naively, one could make infinitely many combinations of
building blocks $\{\hat{S}_{ij}, \hat{s}_i, \delta_{ij}\}$, to get tensors of the form $(\lambda_{\omega}^E)^{ij}{}_{kl}$. However, it turns out that only a finite number of them are linearly independent because of the identities
\begin{equation}
    \begin{aligned}
        \hat{S}^{ij}\hat{s}_j & = 0 ~, \\
        \hS^{i}{}_{k}\hS^{j}{}_{l} & = -\delta^{i}{}_{l}\delta^{j}{}_{k}+\delta^{ij}\delta_{kl}-\delta_{kl}\hat{s}^{i}\hat{s}^{j}+\delta^{i}{}_{l}\hat{s}^{j}\hat{s}_{k}+\delta^{j}{}_{k}\hat{s}^{i}\hat{s}_{l}-\delta^{ij}\hat{s}_{k}\hat{s}_{l} ~.
    \end{aligned}
\end{equation}
Since the spin breaks the SO(3) symmetry into a smaller SO(2) subgroup consisting of rotations about the spin-axis, it is convenient to organize our basis set of tensors based on the number of spin vectors and tensors used to construct them. We start from the Kronecker delta tensors with no spins and therefore respecting SO(3) invariance. Then we gradually break the SO(3) symmetry by adding additional spin vectors and tensors such that they still have axial invariance. 
Finally, we symmetrize and remove the traces as required for the response tensors.
Specifically, in the quadrupolar case with $\ell=2$, we are able to define the following five linearly independent basis tensors:
\begin{alignat}{3}
\Big\{\delta^{\langle i}_{\langle k}\delta^{j\rangle}_{l\rangle},~\hS^{\langle i}{}_{\langle k}\delta^{j\rangle}_{l\rangle},~\hat{s}^{\langle i}\hat{s}_{\langle k}\delta^{j\rangle}_{l\rangle}, \hS^{\langle i}{}_{\langle k}\hat{s}^{j\rangle}\hat{s}_{l\rangle},~\hat{s}^{\langle i}\hat{s}_{\langle k}\hat{s}^{j\rangle}\hat{s}_{\l\rangle}\Big\},
\label{eq:fivebasis}
\end{alignat}
where $\langle i,j\rangle$ denotes the symmetrization and trace-removal of contained indices. Intuitively, the above base tensors can be distinguished by how many spin tensors (vectors) they have. Furthermore, according to the ${\rm SO}(3)$ representation theory, we explicitly show in App.~\ref{app:love rep} that the above tensors can be re-rewritten into the operator form
\begin{alignat}{3}
\label{eq:building block}
\delta^{\langle i}_{\langle k}\delta^{j\rangle}_{l\rangle} &\equiv \bold{I}, ~ \hS^{\langle i}{}_{\langle k}\delta^{j\rangle}_{l\rangle}\equiv \frac{i}{2}\bold{J}_z, ~ \hat{s}^{\langle i}\hat{s}_{\langle k}\delta^{j\rangle}_{l\rangle}\equiv-\frac{1}{6}(\bold{J}_z^2-4~\bold{I}), \nonumber
\\ \hS^{\langle i}{}_{\langle k}\hat{s}^{j\rangle}\hat{s}_{l\rangle} &\equiv -\frac{i}{6}(\bold{J}_z^3-4 \bold{J}_z),~\hat{s}^{\langle i}\hat{s}_{\langle k}\hat{s}^{j\rangle}\hat{s}_{\l\rangle}\equiv \frac{1}{6}(\bold{J}_z^2-4~\bold{I})(\bold{J}_z^2-\bold{I}).
\end{alignat}
where $\bold{J}_z$ and $\bold{I}$ are respectively the ($z$-component) angular momentum operator and identity operator in rank-2 STF tensor space. The above equation not only makes the spin structure manifest, but also makes it easy to transform between tensor basis and spherical basis. This also makes it convenient later to switch scattering amplitudes from plane-wave basis to spherical basis.

\subsubsection*{Tensorial Response Function}

In this paper, our focus is on the dynamical tidal response of Kerr BHs. 
As a first step, we write down 
the following leading interactions in the 
quadrupolar and octupolar sectors,
\begin{equation}
\label{eq: Kerr tidal action}
    S = S_{(0)} + \int d\tau \[ Q_{ij}^E E^{ij} + Q_{ij}^B B^{ij} + Q_{ijk}^E E^{ijk} + Q_{ijk}^B B^{ijk} \] ~,
\end{equation}
where we use $S_{(0)}$ for the spinning point particle effective action. 
Note again that we write the above action for a choice of tetrads 
that corresponds to a locally-flat comoving inertial frame.
This basis is different from the one used in Ref.~\cite{Goldberger:2020fot}, and we discuss the 
technical differences w.r.t. this references in App.~\ref{app:tetrads}.
As a next step, we parametrize the  electric and magnetic tidally induced multipole moments in the most general way as \cite{Saketh:2022xjb}
    \begin{equation}
    \label{eq:KerrAns}
       \begin{aligned}
    Q_{ij}^E & = - M\Bigg[ (G M)^4 (\lambda^E)_{ijkl} E^{kl} - (G M)^5 (\lambda^E_{\omega})_{ijkl} \frac{D}{D \tau} E^{kl} + (G M)^6 (\lambda_{\omega^2}^E)_{ijkl} \frac{D^2}{D\tau^2} E^{kl} \\
    & + (G M)^5 (\nu^E)_{ij \langle kl} \hat{s}_{m\rangle} B^{klm} + \cdots  \Bigg]~, \\
    Q_{ij}^B & = - M\Bigg[(G M)^4 (\lambda^B)_{ijkl} B^{kl} - (G M)^5 (\lambda_\omega^B)_{ijkl} \frac{D}{D \tau} B^{kl} + (G M)^6 (\lambda_{\omega^2}^E)_{ijkl} \frac{D^2}{D\tau^2} B^{kl}  \\
    & + (G M)^5 (\nu^B)_{ij \langle kl} \hat{s}_{m \rangle} E^{klm} + \cdots \Bigg] ~, \\
    Q^{B}_{ijk} & = - M\[ (G M)^5\hat{s}_{\langle k} (\xi^B)_{ij \rangle lm }E^{lm}  + \cdots \] ~,\\
    Q^E_{ijk} & = -M \[ (G M)^5\hat{s}_{\langle k} (\xi^E)_{ij\rangle lm }B^{lm} + \cdots \] ~,
    \end{aligned}
    \end{equation}
where we have promoted all the coefficients in Eq.~\eqref{eq:scwgans} into tensors. Expanding these tensors over the master tensors from Eq.~\eqref{eq:building block}, we get
\begin{equation}
        \begin{aligned}
            (\lambda^{E/B})^{ij}{}_{kl} & = \Lambda_{\hat{s}^0}^{E/B} \delta_{\langle k}^{\langle i } \delta_{l \rangle}^{j \rangle}+ H_{\hat{s}^1}^{E/B} \hat{S}^{\langle i}{ }_{\langle k} \delta_{l \rangle}^{j \rangle}+ \Lambda_{\hat{s}^2}^{E/B}  \hat{s}^{\langle i} \hat{s}_{\langle k} \delta_{l \rangle}^{j \rangle} + H_{\hat{s}^3}^{E/B}  \hat{s}^{\langle i} \hat{s}_{\langle k} \hat{S}^{j\rangle}{ }_{l \rangle} + \Lambda_{\hat{s}^4}^{E/B} \hat{s}^{\langle i} \hat{s}_{\langle k} \hat{s}^{j \rangle} \hat{s}_{l \rangle} ~,
        \end{aligned}
        \label{eq : LO Kerr ans}
\end{equation}
\begin{equation}
        \begin{aligned}
            (\lambda_\omega^{E/B})^{ij}{}_{kl} & = H_{\hat{s}^0,\omega}^{E/B} \delta_{\langle k}^{\langle i } \delta_{l \rangle}^{j \rangle}+ \Lambda_{\hat{s}^1, \omega}^{E/B} \hat{S}^{\langle i}{ }_{\langle k} \delta_{l \rangle}^{j \rangle}+ H_{\hat{s}^2,\omega}^{E/B} \hat{s}^{\langle i} \hat{s}_{\langle k} \delta_{l \rangle}^{j \rangle} + \Lambda_{\hat{s}^3,\omega}^{E/B}  \hat{s}^{\langle i} \hat{s}_{\langle k} \hat{S}^{j\rangle}{ }_{l \rangle}+ H_{\hat{s}^4, \omega}^{E/B} \hat{s}^{\langle i} \hat{s}_{\langle k} \hat{s}^{j \rangle} \hat{s}_{l \rangle}  ~,
        \end{aligned}
        \label{eq : NLO Kerr ans}
\end{equation}
\begin{equation}
    \begin{aligned}
        (\lambda_{\omega^2}^{E/B})^{ij}{}_{kl} & = \Lambda_{\hat{s}^0,\omega^2}^{E/B} \delta_{\langle k}^{\langle i } \delta_{l \rangle}^{j \rangle}+ H_{\hat{s}^1, \omega^2}^{E/B} \hat{S}^{\langle i}{ }_{\langle k} \delta_{l \rangle}^{j \rangle}+ \Lambda_{\hat{s}^2, \omega^2}^{E/B} \hat{s}^{\langle i} \hat{s}_{\langle k} \delta_{l \rangle}^{j \rangle} + H_{\hat{s}^3, \omega^2}^{E/B} \hat{s}^{\langle i} \hat{s}_{\langle k} \hat{S}^{j\rangle}{ }_{l \rangle}+ \Lambda_{\hat{s}^4, \omega^2}^{E/B} \hat{s}^{\langle i} \hat{s}_{\langle k} \hat{s}^{j \rangle} \hat{s}_{l \rangle} ~.
    \end{aligned}
\end{equation}
In this expression, all the $\Lambda^{E/B}$ terms correspond to the conservative tidal deformations, whereas all the $H^{E/B}$ terms account for the tidal dissipation. This follows from their behavior under time reversal transformations. Additionally, the above ansatz for the tidal response obeys parity invariance, 
which explains the absence of some other possible combinations one can add to the response :
\begin{itemize}
    \item \textbf{Time-reversal transformation property}: In our parametrization of the response function, operators $D/D\tau$, $\hat{S}_{ij}$ and $\hat{s}^i$ are odd under time-reversal transformations. Thus, all terms with pre-coefficients $H^{E/B}$ break the time-reversal symmetry and thus correspond to tidal dissipation \footnote{In astrophysical context, this is sometimes known as tidal heating.}. Similar to the discussion in the Schwarzschild case, these coefficients cannot be absorbed into local contact terms on the worldline. However, terms proportional to $\Lambda^{E/B}$ are time-reversal even and hence contribute to the conservative tidal deformations. Furthermore, for the external perturbations, the electric tidal field $E_{i_L}$ is even under time-reversal transformation while the magnetic tidal field $B_{i_L}$ is odd. Note that it is the behavior under time-reversal of the combination $Q^{E}_{i_L}E^{i_L}$ ($Q^{B}_{i_L}B^{i_L}$) that dictates whether a given term in the response is conservative or dissipative.

    \item \textbf{Parity transformation property}: The linear tidal response of Kerr BHs is parity invariant \footnote{In the case of Kerr BHs, both the background metric and the boundary conditions at the event horizon, which we will consider during our scattering studies, respect parity invariance. Therefore, we anticipate that tidal response will also respect parity invariance. However, from the perspective of EFT, there is no fundamental principle mandating parity invariance, see discussions in \cite{Modrekiladze:2022ioh} about the finite size effects from parity violating constituents.}. Therefore, to maintain parity invariance, the induced electric quadrupole moment $Q_{ij}^E$ should contain contributions from $B^{klm}$, but not from $B^{ij}$, and vice versa. As a result, we incorporate the free tensors $(\nu^{E/B})_{ijkl}$ and $(\xi^{E/B})_{ijkl}$ to encapsulate these mixing effects. We can further decompose these tensors into our spin building blocks, enabling us to consistently track time-reversal properties
\begin{equation}
\begin{aligned}
(\nu^{E/B})^{ij}{}_{kl} &= \tilde{\Lambda}^{E/B}_{\hat{s}^0}\delta^{\langle i}_{\langle k}\delta^{j\rangle}_{l\rangle} + \tilde{H}_{\hat{s}^1 }^{E/B}\hat{S}^{\langle i}{}_{\langle k}\delta^{j\rangle}_{l\rangle} + \tilde{\Lambda}_{\hat{s}^2}^{E/B} \hat{s}^{\langle i}\hat{s}_{\langle k}\delta^{j\rangle}_{l\rangle} \\
& \quad + \tilde{H}_{\hat{s}^3}^{E/B} \hat{S}^{\langle i}{}_{\langle k}\hat{s}^{j\rangle}\hat{s}_{l\rangle} + \tilde{\Lambda}_{s^4}^E \hat{s}^{\langle i}\hat{s}_{\langle k}\hat{s}^{j\rangle}\hat{s}_{l\rangle} ~,
\end{aligned}
\end{equation}
\begin{equation}
    \begin{aligned}
        (\xi^{E/B})^{ij}{}_{kl} & = {\Lambda'}^{E/B}_{\hat{s}^0}\delta^{\langle i}_{\langle k}\delta^{j\rangle}_{l\rangle} + {H'}_{\hat{s}^1 }^{E/B}\hat{S}^{\langle i}{}_{\langle k}\delta^{j\rangle}_{l\rangle} + {\Lambda'}_{\hat{s}^2}^{E/B} \hat{s}^{\langle i}\hat{s}_{\langle k}\delta^{j\rangle}_{l\rangle} \\
& \quad + {H'}_{\hat{s}^3}^{E/B} \hat{S}^{\langle i}{}_{\langle k}\hat{s}^{j\rangle}\hat{s}_{l\rangle} + {\Lambda'}_{s^4}^E \hat{s}^{\langle i}\hat{s}_{\langle k}\hat{s}^{j\rangle}\hat{s}_{l\rangle} ~.
    \end{aligned}
\end{equation}
Similar to the previous analysis, $\tilde{\Lambda}^{E/B},{\Lambda'}^{E/B}$s capture the conservative tidal response while all $\tilde{H}^{E/B}, {H'}^{E/B}$s describe the dissipative tidal response.
\end{itemize}

\subsection{Dynamical Tidal Love Numbers, Dissipation Numbers and Mixing Coefficients}
\label{subsec: tidal coefficients}

Having established the symmetry properties of the tensorial tidal response function, let us now clarify some of the terminologies used.

\subsubsection*{Static Tidal Love Numbers and Dynamical Tidal Love Numbers}

Following previous literature, e.g.~\cite{Charalambous:2021mea}, we refer to the following coefficients as {\it static tidal Love numbers} (TLNs)
\begin{eBox2}
\begin{equation}
\label{eq: def static Love}
    \textbf{Static tidal Love numbers}: \quad \Lambda_{\hat{s}^0}^{E/B} ~, \quad \Lambda_{\hat{s}^2}^{E/B} ~, \quad \Lambda_{\hat{s}^4}^{E/B} ~.
\end{equation}
\end{eBox2}
As shown in \cite{Chia:2020yla,Charalambous:2021mea,Ivanov:2022qqt}, all static Love numbers of Kerr BHs vanish to all orders in the value of the BH spin. In this paper, we generalize the conservative tidal response to the dynamical region where we have introduced 5 more parameters. We will refer to them as the {\it dynamical tidal Love numbers} (DTLNs),
\begin{eBox2}
    \begin{equation}
    \label{eq: def dynamical Love}
        \textbf{Dynamical tidal Love numbers}: \quad \Lambda_{\hat{s}^1,\omega}^{E/B} ~, \quad \Lambda_{\hat{s}^3,\omega}^{E/B} ~, \quad \Lambda_{\hat{s}^0,\omega^2}^{E/B} ~, \quad \Lambda_{\hat{s}^2,\omega^2}^{E/B} ~, \quad \Lambda_{\hat{s}^4,\omega^2}^{E/B} ~.
    \end{equation}
\end{eBox2}
We show in Sec.~\ref{sec:matching} that all the dynamical tidal Love numbers exhibit 
the RG running behavior, similar to Love numbers in higher dimensions, e.g.~\cite{Kol:2011vg,Hui:2020xxx}.
Furthermore, similar to the Schwarzschild case, all the static tidal Love numbers and DTLNs can be absorbed into the worldline effective action via local contact terms as 
\begin{equation}
\begin{aligned}
    S = S_{(0)} &- \frac{1}{2} \int d\tau M(GM)^4 \Big[\Lambda^E_{\hat{s}^0}E^{ij}E_{ij}  + \Lambda^E_{\hat{s}^2} E^{ij}E_{kj}\hat{s}_{i}\hat{s}^k +  \Lambda_{s^4}^E E^{ij}E_{kl}\hat{s}_{i}\hat{s}_{j}\hat{s}^{k}\hat{s}^{l}\Big] \\
& +  \frac{1}{2} \int d\tau M(GM)^5 \Big[\Lambda_{\hat{s}^1,\omega}^{E}\hat{S}_{ i k}\dot{E}^{k}{}_{j}E^{ij}+\Lambda_{\hat{s}^3,\omega}^{E}\hat{S}_{ i k}\hat{s}_{j
}\hat{s}_{l}\dot{E}^{kl}E^{ij} \Big] \\
& + \frac{1}{2} \int d\tau M(GM)^6 \Big[\Lambda^E_{\hat{s}^0,\omega^2}\dot{E}^{ij}\dot{E}_{ij}  + \Lambda^E_{\hat{s}^2,\omega^2} \dot{E}^{ij}\dot{E}_{kj}\hat{s}_{i}\hat{s}^k +  \Lambda_{s^4,\omega^2}^E \dot{E}^{ij}\dot{E}_{kl}s_{i}s_{j}s^{k}s^{l}\Big] \\
& + (E\leftrightarrow B) ~.
\end{aligned}
\label{eq : countercons}
\end{equation}
Note that this expression is missing the quadrupole-octupole 
mixing terms. They will be discussed shortly. 

\subsubsection*{Tidal Dissipation Numbers}

Similar to the Schwarzschild case, we call the following coefficients as leading order Kerr {\it tidal dissipation numbers} (TDNs)
\begin{eBox2}
    \begin{equation}
    \label{eq: def LO diss}
        \textbf{LO tidal dissipation numbers}: \quad H_{\hat{s}^1}^{E/B} ~, \quad H_{\hat{s}^3}^{E/B} ~.
    \end{equation}
\end{eBox2}
These two numbers have been computed before with various methods \cite{Starobinsky:1973aij,Starobinskil:1974nkd,Page:1976df,Page:1976ki,Maldacena:1997ih,Bredberg:2009pv,Chia:2020yla,Goldberger:2020fot,Saketh:2022xjb}. 
In this work, we have extended the tidal response to $(G M \omega)^2$ order, so we need to introduce 5 more parameters to capture the tidal dissipations. Here, we call the following three {\it next-to-leading order (NLO) tidal dissipation numbers} 
\begin{eBox2}
    \begin{equation}
    \label{eq: def NLO diss}
        \textbf{NLO tidal dissipation numbers}: \quad H_{\hat{s}^0,\omega}^{E/B} ~, \quad H_{\hat{s}^2,\omega}^{E/B} ~, \quad H_{\hat{s}^4,\omega}^{E/B} ~.
    \end{equation}
\end{eBox2}
These numbers were first matched in \cite{Saketh:2022xjb} that helped resolve 
the mismatch of horizon flux between Refs.~\cite{Chatziioannou:2016kem} and \cite{Tagoshi:1997jy} in the extremal-mass-ratio limit. 
In Sec.~\ref{sec:matching}, we will show that these three NLO tidal dissipation numbers exhibit divergent behavior near the extremal limit, which gives large spin corrections that modify the superradiance condition. Furthermore, we also need the following {\it next-to-next-to-leading order (NNLO) tidal dissipation numbers}
\begin{eBox2}
    \begin{equation}
    \label{eq: def NNLO diss}
        \textbf{NNLO tidal dissipation numbers}: \quad H_{\hat{s}^1,\omega^2}^{E/B} ~, \quad H_{\hat{s}^3,\omega^2}^{E/B} ~.
    \end{equation}
\end{eBox2}
In Sec.~\ref{sec:matching}, we will show that all the NNLO tidal dissipation numbers also exhibit the RG running behavior that comes from the ultraviolet (UV) divergence in the EFT 2-loop corrections with two mass insertions. 
Note that the origin of the RG running here is different from that of the dynamical Love numbers.

\subsubsection*{Mixing Coefficients}

Consistent with the parity transformation property discussed previously, we introduced the free tensors $(\nu^{E/B})_{ijkl}$ and $(\xi^{E/B})_{ijkl}$ to capture the mixing effects. Here, we denote that the following 6 coefficients as {\it conservative tidal mixing numbers} (CTMNs)
\begin{eBox2}
    \begin{equation}
        \textbf{conservative tidal mixing numbers}: \tilde{\Lambda}_{\hat{s}^0}^{E/B} ~, \quad \tilde{\Lambda}_{\hat{s}^2}^{E/B} , \quad \tilde{\Lambda}_{\hat{s}^4}^{E/B} , \quad {\Lambda'}_{\hat{s}^0}^{E/B} , \quad {\Lambda'}_{\hat{s}^2}^{E/B} , \quad {\Lambda'}_{\hat{s}^4}^{E/B}\,.
    \end{equation}
\end{eBox2}
These effect of the corresponding operators can be reproduced with the 
following effective action, 
\be
\begin{split}
S_{\text{mixing}} = &-\frac{1}{2}\Bigg[M (GM)^5\int d\tau E_{ij}\hat{s}^{m}B_{klm}(N^{B})^{ij,kl}  
-M (GM)^5\int d\tau B_{ij}\hat{s}^{m}E_{klm}(N^{B})^{ij,kl}\Bigg]\,,
\label{eqd : mixing}
\end{split}
\ee 
with 
\be
\begin{split}
& (N^{E})^{ij,kl}   =
(\nu^{E})^{(ij,kl)} + (\xi^{B})^{(kl,ij)}\,,\\
& (N^{B})^{(ij,kl)}   =
(\nu^{B})^{(ij,kl)} + (\xi^{E})^{(kl,ij)}\,,\\
\end{split}
\ee
where time-reversal invariance dictates that only the symmetric (under $ij\leftrightarrow kl$) parts of the tensors $(\nu^E)^{ij,kl}$, $(\xi^E)^{ij,kl}$ should contribute.

The other 4 coefficients will be referred to as {\it dissipative tidal mixing numbers} (DTMNs)
\begin{eBox2}
    \begin{equation}
        \textbf{dissipative tidal mixing numbers}: \tilde{H}_{\hat{s}^1}^{E/B} , \quad \tilde{H}_{\hat{s}^3}^{E/B} , \quad {H'}_{\hat{s}^1}^{E/B} , \quad {H'}_{\hat{s}^3}^{E/B}.
    \end{equation}
\end{eBox2}
The first matching of these dissipative tidal mixing numbers was performed in \cite{Saketh:2022xjb}. 
They originate from the perturbative expansion of spheroidal harmonics in terms of spherical harmonics. Here in Sec.~\ref{sec:mix}, we present a more straightforward way to match these coefficients in both conservative and dissipative sectors in terms of amplitudes.

\subsection{Wave Scattering off Compact Objects}

We have completed the discussion of the various free coefficients in the dynamical tidal response for rotating (axi-symmetric, parity-preserving) compact objects. 
In the remaining part of the paper, we explicitly fix these coefficients by matching the scattering amplitudes obtained in the EFT with the scattering phases obtained from BHPT. Before we go into the detailed discussion of the matching procedure, let us first set up the formalism to study the wave scatterings off compact objects. In the worldline EFT language, this corresponds to the following scattering process 
\begin{equation}
    \text{worldline + particle A   $\rightarrow$ worldline + particle~A,}
\end{equation}
where we use the double solid line to denote the worldline and the yellow wavy lines for gravitons
\begin{equation}
            \vcenter{\hbox{\begin{tikzpicture}[scale=0.7]
        \begin{feynman}
            \vertex (i) at (0,0);
            \vertex (e) at (0,3);
            \vertex(f1) at (1.5,2.9);
            \vertex (fs) at (1.5,0.1); 
            \vertex[blob, scale=1.3] (w1) at (0.5, 1.5) {};

            \diagram*{
                (i) -- [double, double, thick] (w1),
                (e) -- [double, double, thick] (w1),
                (f1) -- [boson, MyYellow, ultra thick] (w1),
                (fs) -- [boson, MyYellow, ultra thick] (w1),
            };
        \end{feynman}
    \end{tikzpicture}}} 
\end{equation}
We choose the worldline to be static,
which is 
motivated by the assumption 
that the BH mass $M$ is much larger than the energy of the incoming massless particle $\omega$. This also naturally leads to the power counter parameter $G M \omega \ll 1$ which we have been using in the parametrization of dynamical tidal response and will be later used as an expansion parameter in the BHPT solution.
The scattering matrix now is defined as 
\begin{equation}
    {}_{\rm out}\langle \boldsymbol{k}', h'| \boldsymbol{k}, h \rangle_{\rm in} \equiv  \langle \boldsymbol{k}',h'| S | \boldsymbol{k}, h \rangle ~,
\end{equation}
where $h$ is the helicity of the particle, and the normalization of 
the bulk particle state is $\langle \boldsymbol{k}, h| \boldsymbol{k}', h' \rangle = 2 |\boldsymbol{k}| \times (2 \pi)^{3} \delta^{3}(\boldsymbol{k}-\boldsymbol{k}') \delta_{hh'}$ in four dimensional spacetime. 
The on-shell constraint for massless fields is $\omega^2 = \boldsymbol{k}^2$. 
In what follows we will formally decompose $S$-matrix as $S = 1 + i T$.
In the worldline theory, the time translation invariance dictates that scattering processes are constrained by energy conservation. 
Therefore, it is convenient to define the dimensionful scattering amplitude 
$\mathcal{M}$ as 
\begin{equation}
    \langle \boldsymbol{k}', h'| i T| \boldsymbol{k}, h \rangle \equiv (2\pi) \delta(\omega' - \omega) \times i \mathcal{M}(\omega,\boldsymbol{k} \rightarrow \boldsymbol{k'},h \rightarrow h') ~.
\end{equation}
The dimensionality of $\mathcal{M}(\omega, \boldsymbol{k} \rightarrow \boldsymbol{k}',h\rightarrow h')$ is [energy]$^{-1}$. In this paper, we find it more convenient to calculate the scattering amplitude directly in the spherical basis. We define the spherical wave state $| \omega, \ell,m,h \rangle$ which describes the spherical wave with frequency $\omega$, angular quantum number $\ell$, magnetic quantum number $m$ and helicity $h$. The normalization of such state is $\langle \omega, \ell,m,h| \omega', \ell',m',h' \rangle = (2\pi) \delta(\omega - \omega') \delta_{\ell \ell'} \delta_{mm'} \delta_{hh'}$. Now, we can compute the scattering amplitude in this basis
\begin{equation}
    \langle \omega', \ell', m',h'| i T | \omega, \ell,m,h \rangle  = (2\pi) \delta(\omega' - \omega) \times i \mathcal{A}(\omega, \ell, m, h \rightarrow \omega,\ell',m', h') ~,
\end{equation}
where the scattering matrix $\mathcal{A}$ is dimensionless. The relation between $\mathcal{A}$ and $\mathcal{M}$ is given by
\begin{equation}
    \begin{aligned}
        & \quad \mathcal{A}(\omega, \ell ,m,h \rightarrow \omega, \ell', m', h') \\
        & = \int \frac{d^3 \boldsymbol{k}_1 d^3 \boldsymbol{k}_2}{(2\pi)^6\times 4|\boldsymbol{k}_1| |\boldsymbol{k}_2|} \sum_{h_1,h_2} \langle \omega, \ell' ,m',h'|\boldsymbol{k}_2,h_2 \rangle \mathcal{M}(\omega, \boldsymbol{k}_1 \rightarrow \boldsymbol{k}_2, h_1 \rightarrow h_2) \langle \boldsymbol{k}_1 ,h_1| \omega, \ell,m,h \rangle ~,
    \end{aligned}
\end{equation}
where the transformation matrix \cite{Goldberger:2020fot} is given by
\begin{equation}
    \langle \omega, \ell, m,h | \boldsymbol{k},h \rangle = (2\pi)^2 \sqrt{\frac{2\ell + 1}{2\pi\omega}}\delta(\omega - |\boldsymbol{k}|)\delta_{hh'} D_{mh}^\ell(\phi,\theta,0) ~, 
\end{equation}
with $D_{mh}^\ell(\phi,\theta,0)$ the Wigner-D matrix. Here $(\theta,\phi)$ denotes the orientation of $\boldsymbol{k}$. In a diagonal basis, one can easily relate the amplitude with the scattering phase shift as 
\begin{equation}
    i\mathcal{A} = 1 - \eta_{\ell m} \exp(2 i \delta_{\ell m}) ~,
\end{equation}
where $\delta_{\ell m}$ is the elastic scattering phase shift and $1 -\eta_{\ell m}^2$ is the absorption probability. For gravitational perturbations, one can further write the polarization tensor into the spherical basis as
\begin{equation}
    \epsilon_{ij}^{h = \pm 2}(\boldsymbol{k}) = \sum_{m=-2}^2 \langle i,j|\ell = 2,m \rangle D_{mh=\pm 2}^{\ell =2}(\phi,\theta,0) ~.
\end{equation}
More detailed discussions are included in App.~\ref{app : single particle}. 

Now, we are equipped with all the tools for studying the dynamical response on Kerr BHs. In the following three sections, we shall first present the wave scattering phase shifts of Kerr BHs from the analytic solution to the Teukolsky equation in Sec~\ref{sec:BHPT}. Then, we will explicitly match the various tidal response coefficients mentioned before in Sec.~\ref{sec:matching} and Sec.~\ref{sec:mix}.

\section{Kerr Perturbations and Near-Far Factorization}
\label{sec:BHPT}

As discussed in Sec.~\ref{sec:EFT}, various Love numbers and dissipation numbers are the unknown parameters in the ansatz for the tidal response, relating tidal fields to induced multipole moments. Love numbers may be further viewed as the Wilson coefficients in front of the quadratic-in-curvature terms in the worldline EFT. Determining these coefficients requires the computation of physical observables such as scattering amplitudes, which we then match with the corresponding results from the ultraviolet (UV) theory. In this context, as far as linear tidal effects are concerned, the UV theory is essentially linear BHPT. 
An important observation in this context is the so-called 
near-far factorization in black hole scattering amplitudes~\cite{Ivanov:2022qqt}
that can be used to extract Love numbers, dissipation numbers, and their RG running 
without the interference from the post-Minkowskian correction from 
non-linearities of GR.
In Sec.~\ref{subsec:MST}, we briefly review the Mano, Suzuki, and Takasugi (MST) method and discuss its factorization properties. 
This approach is particularly useful since 
it yields an explicit Taylor expansion in $GM\omega$, 
just as the perturbative EFT calculation. 
Subsequently, in Sec.~\ref{subsec:BHPT phase shift}, we present results for Kerr BHs for the quadrupolar sector $\ell = 2$.

\subsection{Scattering of Test Fields by Black Holes and the Near-Far Factorization}
\label{subsec:MST}

The study of wave scattering off Kerr Black Holes (BHs) is a longstanding field of research with foundational works by R. A.Matzner and others \cite{matzner1968scattering, Chrzanowski:1976jb, Matzner:1977dn, Handler:1980un, Futterman:1988ni}. Recently, this area of research has been revived 
thanks to novel insights from the S-Matrix theory, see e.g.~\cite{Bautista:2021wfy, Bautista:2022wjf}.
In this section, we will outline some key components of this theory relevant 
for the matching of dynamical Love numbers and dissipation numbers. 
A comprehensive discussion will be presented elsewhere~\cite{longpaper}.

We begin our analysis by considering a test field of a positive integer spin $s$ propagating on a Kerr BH background, 
and choose the $\hat{z}$ direction to align 
with the BH spin. If the incoming wave of the test spin-$s$ field has an incident angle of $(\gamma, \phi_0)$ with respect to the $\hat{z}$ direction, we can express the scattering differential cross section as follows
\begin{equation}
    \frac{d\sigma}{d\Omega} = |f_s(\theta, \phi)|^2 + |g_s(\theta,\phi)|^2 ~,
\end{equation}
with the helicity conserving amplitude \cite{Glampedakis:2001cx, Dolan:2008kf,Stratton:2020cps, Bautista:2022wjf, longpaper}
\begin{equation}\label{eq: general incident angle master formula}
    f_s(\theta,\phi) = \frac{\pi}{i\omega} \sum_{\ell = s}^\infty \sum_{m = -\ell}^\ell {}_{-s} S_\ell^m(\cos\gamma, a\omega) {}_{-s} S_\ell^m(\cos\theta, a\omega) e^{i m (\phi - \phi_0)} ({}_{s} \eta_{\ell m} e^{2 i {}_{s}\delta_{\ell m}} - 1) \times 2 ~.
\end{equation}
Here, ${}_{-s} S_\ell^m(\cos\theta, a\omega)$ is the spin-weighted spheroidal harmonics with orbital quantum number $\ell$ and magnetic quantum number $m$. The tortoise coordinate of Kerr BH is defined as
\begin{equation}
     \frac{dr_*}{dr} \equiv \frac{(r^2 + a^2)}{\Delta}~, \quad \Delta \equiv (r - r_+)(r - r_-)
\end{equation}
where $r_+$ and $r_-$ are outer and inner horizon respectively.
By analyzing the Teukolsky master variable \cite{teukolsky1974perturbations,Mano:1996vt,Sasaki:2003xr}
\begin{equation}
\psi^{[-s]} = e^{-i\omega t} \sum_{\ell m} e^{im\phi} {}_{-s} S_\ell^m(\theta;a\omega)\, {}_{-s} R_{\ell m}(r)\, ~,
\end{equation}
with the following asymptotic behavior in the tortoise coordinate
\begin{equation}\label{eq: radial asymptotic}
    {}_{-s} R_{\ell m} (r) \rightarrow B_{-s\ell m}^{\rm (inc)} r^{-1} e^{-i \omega r_*} + B_{-s \ell m}^{\rm (refl)} r^{-1+2s} e^{i\omega r_*}~, \quad r_* \rightarrow +\infty ~,
\end{equation}
we can get the scattering phase shift
\begin{equation}\label{eq: phase shift master formula}
    {}_{s} \eta_{\ell m} e^{2 i {}_{s} \delta_{\ell m}} = (-1)^{\ell + 1} \frac{{\rm Re}({}_{s}C_\ell^m(a\omega))}{(2\omega)^{2s}} \times \frac{ B_{-s \ell m}^{(\rm refl)} }{ B_{-s \ell m}^{(\rm inc)} } ~,
\end{equation}
where ${}_sC_\ell^m(a\omega)$ is the Teukolsky-Starobinsky constant, ${}_s \delta_{\ell m}$ the elastic scattering phase shift and $1 - ({}_s \eta_{\ell m})^2$ the absorption probability. The helicity-reversing amplitude $g_{s}(\theta,\phi)$ vanishes for spin-$0,1$ field \cite{longpaper,Bautista:2021wfy}, but it is non-vanishing for spin-2 field due to the imaginary part of spin-2 Teukolsky-Starobinsky constant \cite{Glampedakis:2001cx, Dolan:2008kf,Stratton:2020cps, Bautista:2022wjf, longpaper}
\begin{equation}\label{eq: spin-2 helicity-reversing partial wave amplitudes}
    g_2(\theta) = \frac{\pi}{i \omega} \sum_{\ell = 2}^\infty (-1)^\ell {}_{-2}S_\ell^m(\cos\gamma, a\omega) {}_{-2}S_\ell^m(-\cos\theta, a\omega) \times \( (-1)^{\ell + 1+m} \frac{12 i M \omega}{16 \omega^4} \frac{B_{-2 \ell m}^{\rm (refl)}}{B_{-2 \ell m}^{\rm (inc)}} \) \times 2~.
\end{equation}

As the EFT is a low-frequency and long-wavelength expansion, its results have to be matched to the relevant low-frequency observables in BHPT. 
A solution to the Teukolsky equation in this context has been systematically developed through the matching of asymptotic expansions using the MST method \cite{Mano:1996gn,Mano:1996mf,Mano:1996vt,Sasaki:2003xr}. The key idea in this method is that one can construct the near zone solution based on the double-sided infinite series of hypergeometric function which converges within $r_+ \leq r < \infty$ and the far zone solution based on the double-sided infinite series of Coulomb wave function which converges within $r_+ < r \leq \infty$. The two solutions can be matched in the overlapping region. In order to ensure the convergence of solutions and the successful execution of the matching procedure, an auxiliary non-integer parameter, $\nu$, is introduced. This parameter, often referred to as the ``renormalized angular momentum," is a fundamental necessity in the matched asymptotic expansion method, as mentioned in previous mathematical studies \cite{el2008solutions}. Ultimately, this process yields the wave amplitude ratio as follows
\be
\label{eq:MST}
\frac{B^{\text{(refl)}}_{-s \ell m}}{B^{\text{(inc)}}_{-s \ell m}}= \omega^{2s}
\blue{\underbrace{\frac{
    1+ie^{i\pi\nu}\frac{K_{-\nu-1;-s}}{K_{\nu;-s}}}
    {1-ie^{-i\pi\nu}\frac{\sin(\pi(\nu+s+i\epsilon))}{\sin(\pi(\nu-s-i\epsilon))}\frac{K_{-\nu-1;-s}}{K_{\nu;-s}}}}_{
    \text{Near zone}} }
    \red{
    \underbrace{\times \frac{A_{-;-s}^\nu}{A_{+;-s}^\nu}
    e^{i\epsilon(2\log \epsilon -({1- \kappa}))}}_{
    \text{Far zone}}} ~,
\ee
which manifestly factorizes into two parts, which is a property introduced in~\cite{Ivanov:2022qqt} as the near-far factorization. 
Here, $\epsilon = 2 G M \omega, \kappa = \sqrt{1-\chi^2}$.
The explicit expression for the functions entering this ratio can be found in \cite{Sasaki:2003xr}. By closely examining the low-frequency behavior, we find that the near zone part scales as, schematically, 
\begin{equation}
    \blue{\text{Near Zone}} \sim (GM \omega)^{2\nu + 1} (1 + GM \omega + (G M \omega)^2 + \cdots )~,
\end{equation}
which features a non-integer power of $G$ for general $\nu$, which we will shortly relate to the analytically continued angular momentum $\ell$, 
while the far zone parts always feature only the integer power of $G$ except for the tail effect that appears as the logarithmic function $\epsilon \log \epsilon$ in the phase shift \cite{Blanchet:1993ec, Poisson:1994yf},
\begin{equation}
    \red{\text{Far Zone}} \sim  (G M \omega) \log (G M \omega) + (G M \omega) + (G M \omega)^2 + (G M \omega)^3 + \cdots 
\end{equation}
With this formula, one can naturally separate the relativistic post-Minkowskian corrections, i.e. loop diagrams in the EFT and the BH finite size effects simply by the power counting $G$ without ambiguity. Furthermore, we find it useful to parametrize the scattering phase shift as 
\begin{equation}
    {}_s \eta_{\ell m} e^{2 i {}_s \delta_{\ell m}} = {}_s \eta_{\ell m} e^{2 i {}_s \delta_{\ell m}^{\rm NZ}} \times e^{2 i {}_s \delta_{\ell m}^{\rm FZ}} \,.
\end{equation}
where NZ stands for \blue{near zone} and FZ stands for \red{far zone}. Based on the 
near-far factorization, the near zone elastic phase shift can be written as
\begin{equation}\label{eq: near zone phase shift}
    {}_s \delta_{\ell m}^{\rm NZ} = \frac{1}{2} {\rm Arg} \[  \frac{
    1+ie^{i\pi\nu}\frac{K_{-\nu-1;-s}}{K_{\nu;-s}}}
    {1-ie^{-i\pi\nu}\frac{\sin(\pi(\nu+s+i\epsilon))}{\sin(\pi(\nu-s-i\epsilon))}\frac{K_{-\nu-1;-s}}{K_{\nu;-s}}} \]\,,
\end{equation}
with \red{far zone} phase elastic shift
\begin{equation}\label{eq: far zone phase shift}
    \begin{aligned}
    {}_s \delta_{\ell m}^{\rm FZ} & = \frac{1}{2} {\rm Arg} \[\frac{A_{-;-s}^\nu}{A_{+;-s}^\nu}
    e^{i\epsilon(2\ln \epsilon -({1- \tilde \kappa}))} \] + \frac{\ell + 1}{2}\pi ~,
    \end{aligned}
\end{equation}
and the absorption probability
\begin{equation}\label{eq: absorption probability}
    1 - {}_s \eta_{\ell m}^2 = 1 - \Bigg| \frac{
    1+ie^{i\pi\nu}\frac{K_{-\nu-1;-s}}{K_{\nu;-s}}}
    {1-ie^{-i\pi\nu}\frac{\sin(\pi(\nu+s+i\epsilon))}{\sin(\pi(\nu-s-i\epsilon))}\frac{K_{-\nu-1;-s}}{K_{\nu;-s}}} \Bigg|^2 ~,
\end{equation}
where we have used the non-trivial identity \cite{Mano:1996gn}
\begin{equation}
     \frac{|A_{-;-s}^\nu/A_{+;-s}^\nu|}{ 2^{2s} |{}_s C_\ell^m(a\omega)|^{{-1}}} = 1  \quad  \text{when} \quad \nu \in \mathbb{R}
\end{equation}
This identity has been numerically tested in the Schwarzschild gravitational scatterings in \cite{longpaper}. For the Kerr BH, this identity has been perturbatively tested order by order in $(G M \omega)$ in \cite{Saketh:2022xjb}. Using the low-frequency expansion of
the ``renormalized angular momentum'' 
$\nu = \ell + \mathcal{O}((G M \omega)^2)$ around a generic angular momentum $\ell$, 
Ref.~\cite{Ivanov:2022qqt}, 
presented an on-shell proof of the vanishing of tidal Love numbers of Kerr BHs. 
Furthermore, that work also pointed out that the near zone elastic and inelastic absorption probabilities receive logarithmic corrections due to the low-frequency expansion of $\nu$, i.e., schematically we have 
\begin{equation}
    (G M \omega)^{2\nu+ 1}\Big|_{\nu=\ell +\mathcal{O}((G M \omega)^2)+... } = (G M \omega)^{2\ell + 1} (1 + (G M\omega)^2 \log(G M \omega) + \cdots ) ~.
\end{equation}
Power counting in $G$, 
we notice that the near-far factorization presented in Eq.~\eqref{eq:MST} holds for the analytically continued $\ell \in \mathbb{C}$. 
As discussed in \cite{longpaper} and \cite{Ivanov:2022qqt},
the near-far factorization will break down if one starts from the physical $\ell \in \mathbb{N}$ because the \blue{near zone} and \red{far zone} terms now both scale as integer powers of $G$ in the low frequency expansion. 
But this will not affect the study the logarithmic pieces because they 
always come from the near zone piece.
The apparent factorization breakdown
will affect the constant, scheme-dependent parts of the local 
worldline couplings, which we discuss separately in Sec.~\ref{sec:constants}.
Note that the factorization breakdown
does not affect the matching of dissipation numbers. From Eq.~\eqref{eq: absorption probability}, we see that the low-frequency absorption 
terms originate entirely from the near zone piece.

\subsection{BHPT Phase Shifts}
\label{subsec:BHPT phase shift}
With the factorization mentioned above, we can successfully get the phase shift from tidal effects (\blue{near zone}) for gravitational perturbations $\ell = 2,m=2$ as ($r_s = 2 G M$)
\begin{equation}\label{eq: 22 abs}
    \begin{aligned}
        {}_2 \eta_{2 2} & = 1 + \frac{2}{225}(\chi + 3\chi^3) (r_s \omega)^5 \\
        & \quad + \frac{1}{2025}\Bigg\{\orange{18 \pi  \left(3 \chi^3+\chi\right)}  -36 \left(3 \chi^3+\chi \right) \Im\left(H\(\frac{2 i \chi}{\sqrt{1-\chi^2}}-3\)\right)+ \\
        & +\left[6 \left(9 \sqrt{1-\chi^2}+1\right) \chi^2+117 \sqrt{1-\chi^2}-97\right] \chi^2+9 \left(\sqrt{1-\chi^2}-1\right)\Bigg\} (r_s \omega)^6 \\
        & + \Bigg\{ \purple{- \frac{214}{23625} (\chi + 3\chi^3) \log(2 \sqrt{1 - \chi^2} r_s \omega)} + \orange{\frac{\pi}{2025}\Bigg[-36 \left(3 \chi^3+\chi \right) \Im\left(H\(\frac{2 i \chi}{\sqrt{1-\chi^2}}-3\)\right)} \\
        & \orange{+\left(6 \left(9 \sqrt{1-\chi^2}+1\right) \chi^2+117 \sqrt{1-\chi^2}-97\right) \chi^2+9 \left(\sqrt{1-\chi^2}-1\right)\Bigg]} + \mathcal{A}_{22}(\chi)\Bigg\} (r_s \omega)^7~,
    \end{aligned}
\end{equation}
and 
\begin{equation}
    \begin{aligned}\label{eq: 22 elastic}
        {}_2 \delta_{22} & = \[ \green{\frac{2 \chi \left(3 \chi^2+1\right) }{225} \log \left(2\sqrt{1-\chi^2} r_s \omega \right)} + \mathcal{B}_{22}(\chi)\] (r_s \omega)^6 \\ 
        & \quad +  \Bigg\{ \frac{1}{2025} \[ \green{\Big( -9 - 97 \chi^2 + 6 \chi^4 \Big) \log(2\sqrt{1 - \chi^2} r_s \omega) } +  \orange{18\pi(\chi + 3 \chi^3) \log (2\sqrt{1 - \chi^2} r_s \omega)} \] \\
        & \quad + \orange{\mathcal{B}_{22}(\chi) \pi} + \mathcal{C}_{22}(\chi)\Bigg\} (r_s \omega)^7 ~.
    \end{aligned}
\end{equation}
In the two expressions provided, $1- ({}_2 \eta_{22})^2$ calculates the absorption probability for the $\ell = 2, m = 2$ mode, while ${}_2 \delta_{22}$ represents the elastic phase shift. We illustrate in Sec.~\ref{sec:matching} that the \orange{orange} terms in Eq.~\eqref{eq: 22 abs} are related to the tail effect brought about by the gravitational wave scattering off the long-range Newtonian potential. Translating this into field theory terms, this effect is equivalent to a 1-loop IR divergence of the leading order tidal scattering. The \purple{purple} logarithmic terms originate from the 2-loop UV divergences in the EFT loop diagrams during the computation of BH absorption, leading to the RG running of dissipation numbers. As for the \green{green} segment, the logarithmic pieces also correspond to UV divergences in the EFT loop diagrams, but they contribute to the elastic scattering pieces. For the Schwarzschild case, simple power counting in $G$ reveals that the first logarithm will appear at $G^{7}$, i.e., at the 6-loop order. Kerr cases will be much more complicated and will be discussed in the next section.

Finally, there are also constant pieces $\mathcal{A}_{22}(\chi)$, $\mathcal{B}_{22}(\chi)$, and $\mathcal{C}_{22}(\chi)$ alongside the logarithmic terms in the above equations. These constant pieces can be 
extracted analytically, but their expressions are quite cumbersome and 
does not provide any insights into the structure of BH scattering amplitudes. 
Therefore, we only present in this work their numerical 
estimates, given in Sec.~\ref{sec:constants}, along with fitting 
formulas that may be used for all practical applications.

We also provide here the phase shift for the 
gravitational perturbations $\ell = 2,m=1$ case
\begin{equation}
    \begin{aligned}
    \label{n21}
        {}_2\eta_{21} & = 1 + \frac{1}{900} (4\chi - 3 \chi^3) (r_s \omega)^5 + \frac{1}{8100} \Bigg\{  \orange{9 \pi  \left(4-3 \chi ^2\right) \chi} + 18 \left(3 \chi ^2-4\right) \chi  \Im\left(H\(\frac{i \chi }{\sqrt{1-\chi
   ^2}}-3\)\right) \\
   & \quad +\left[\left(39-54 \sqrt{1-\chi ^2}\right) \chi ^2+63 \sqrt{1-\chi ^2}-43\right] \chi^2 +36 \left(\sqrt{1-\chi^2}-1\right) \Bigg\} (r_s \omega)^6 \\
   & \quad + \Bigg\{ \purple{\frac{107 (-4\chi + 3 \chi^3)}{94500} \log\( 2 \sqrt{1-\chi^2} r_s \omega \)} + \orange{ \frac{\pi}{8100} \Bigg[18 \left(3 \chi ^2-4\right) \chi  \Im\left(H\(\frac{i \chi }{\sqrt{1-\chi
   ^2}}-3\)\right)}  \\
   & \quad \orange{+\left(\left(39-54 \sqrt{1-\chi ^2}\right) \chi ^2+63 \sqrt{1-\chi ^2}-43\right) \chi^2 +36 \left(\sqrt{1-\chi^2}-1\right) \Bigg]}  + \mathcal{A}_{21}(\chi)\Bigg\} (r_s \omega)^7 ~,
    \end{aligned}
\end{equation}
and
\begin{equation}
    \begin{aligned}
    \label{eq: 21 elastic}
        {}_2 \delta_{21} & = \[ \green{\frac{\chi  \left(4 - 3 \chi ^2\right) \log \left( 2 \sqrt{1 - \chi^2} r_s \omega \right)}{900}} + \mathcal{B}_{21}(\chi) \] (r_s \omega)^6 \\
   & \quad + \Bigg[ \green{\frac{39 \chi ^4-43 \chi ^2 - 36}{8100} \log(2 \sqrt{1 - \chi^2} r_s \omega) }  +\orange{ \frac{\pi}{900} \left(4-3 \chi ^2\right) \chi \log(2\sqrt{1 - \chi^2} r_s \omega)}\\
   & \quad + \orange{\mathcal{B}_{21}(\chi)\pi} + \mathcal{C}_{21}(\chi) \Bigg] (r_s \omega)^7 ~.
    \end{aligned}
\end{equation}
For $\ell=2, m=0$ case, we have
\begin{equation}
    {}_2 \eta_{20} = 1  -\frac{1}{225} \left( 1 - \chi^2 \right)^2 \left(\sqrt{1-\chi^2}+1\right) (r_s \omega)^6 - \orange{\frac{1}{225} \pi (1 - \chi^2)^2 (\sqrt{1 - \chi^2} + 1) (r_s \omega)^7} ~,
\end{equation}
and
\begin{equation}
\label{eq: 20 elastic}
    {}_2 \delta_{20} = \[\green{-\frac{(1-\chi^2)^2 \log (2 \sqrt{1 - \chi^2} r_s \omega )}{225}}  + \mathcal{C}_{20}(\chi)\] (r_s \omega)^7 ~.
\end{equation}

Now, with the above BHPT result, we are prepared for the matching to the EFT amplitude calculations to get the tidal response coefficients that we introduced in Sec.~\ref{subsec: tidal coefficients}. 
One important aspect to remember is that BHPT phase shift values are not in the spherical harmonic basis, but rather in the spheroidal harmonic basis. 
This distinction becomes important when we consider non-zero spin and frequencies. In Sec.~\ref{sec:mix}, we shall explore the impact of the mixing of spherical and spheroidal harmonics on conservative and dissipative tidal mixing numbers, crucial to accurately determinate the horizon flux at the 4PN order \cite{Saketh:2022xjb}. We also show in Sec.~\ref{sec:mix} that our findings in Sec.~\ref{sec:matching} are not affected by these mixing effects, therefore we can rely on spherical harmonics for these calculations.

\section{Matching EFT with BHPT}
\label{sec:matching}

The advantage of the near-far factorization is that separates
scattering
contributions from the near zone (tidal effect) and the far zone (background metric). 
We now use the worldline EFT outlined previously in Sec.~\ref{sec:EFT} to better understand each term in the near zone scattering amplitudes of the full theory. 
In Sec.~\ref{subsec:match love and dissipation}, we briefly outline the procedure of calculating scattering amplitudes in the worldline EFT and use it to match the static tidal Love numbers and LO dissipation numbers, which appear in the ansatz in Eq.~(\ref{eq:KerrAns}) and further explicitly defined in Eqs.~\eqref{eq: def static Love} and \eqref{eq: def LO diss}. 
In Sec.~\ref{subsec:match NLO dissipation}, we explicitly match the NLO dissipation numbers mentioned in Eq.~\eqref{eq: def NLO diss} and discuss their contribution in the case of near-extremal BHs and their effect on superradiance. 
Then, in Sec.~\ref{subsec:tail}, we compute the one-loop correction to the leading order tidal scattering in the worldline EFT, corresponding to the tail effect arising from the scattering of gravitational waves off the background metric before or after scattering off the quadrupole moment. 
This yields an un-physical IR divergence in the EFT which, 
obviously, does not appear in the full theory or in any observables. 
This IR singularity, however, also generates 
a finite observable contribution associated with Sommerfield enhancement which modifies the scattering amplitude at the next order in $GM\omega$ \cite{Goldberger:2005cd}. 
We also find that these contributions correspond to the \orange{orange} terms in BHPT, which disentangles them from genuine NLO dissipation numbers. 
In Sec.~\ref{subsec:RGabs}, we push similar calculation in the EFT to 2-loops for dissipation which has a physical UV divergence and leads to the RG running of the NNLO dissipation numbers corresponding to the logarithm in the \purple{purple} terms in BHPT. 
Finally in Sec.~\ref{subsec: RG elastic}, we give an interpretation to the logarithms of the \green{green} terms in BHPT. 
We argue that these logarithms arise from the RG flow of dynamical Love numbers introduced in Eq.~\eqref{eq: def dynamical Love}. 
A power counting analysis shows that the EFT diagram which yields the corresponding UV divergence arises from non-linearities in GR via higher order loop corrections (6-loop order in the Scwarzschild case) to the scattering off the background metric. 

\subsection{Static Love and LO Dissipation Numbers}
\label{subsec:match love and dissipation}

As explained in Sec.~\ref{sec:EFT}, the leading order tidal effects are included in the tidal tensor $(\lambda^{E/B})^{ij}{}_{kl}$, in which the conservative effects are encoded in static tidal Love numbers $\Lambda_{\hat{s}^0}, \Lambda_{\hat{s}^2}$ and $\Lambda_{\hat{s}^4}$ while dissipative effects are included in the LO dissipation numbers $H_{\hat{s}^1}$ and $H_{\hat{s}^3}$. Equipped with the BHPT scattering phase shift given in Sec.~\ref{subsec:BHPT phase shift}, we can get the explicit expressions for these two sets of coefficients through comparison of the scattering phases between the EFT and BHPT. Such a matching has been previously done in 
\cite{Charalambous:2021mea} in the off-shell way, and 
in \cite{Goldberger:2020fot, Ivanov:2022qqt, Saketh:2022xjb} using on-shell observables. Some of these known results will be 
reproduced below for completeness. 

We consider now the scattering of gravitational waves off a Kerr black hole in the worldline EFT and focus on the scattering phase due to the induced quadrupole moment $Q_{E/B}^{ij}(\tau)$. 
It couples to the graviton through in the interaction 
term $\int d\tau Q_E^{ij}E_{ij}$,~$(E\leftrightarrow B)$ in the action. Classically, this process needs to be understood in a causal way: 1) external incoming waves with four momenta $(\omega, \boldsymbol{k}_{\rm in})$ generate the induced quadrupole in accordance with the ansatz in Eq.~(\ref{eq:KerrAns}); 2) the quadrupole moment will radiate waves with four momenta $(\omega, \boldsymbol{k}_{\rm out})$. In the field theory language, this process can be nicely described by using the retarded Green function 
\begin{equation}
     \vcenter{\hbox{\begin{tikzpicture}[scale=0.7]
        \begin{feynman}
            \node[circle, draw=Orange, fill = Orange, scale=0.5, label=left:$Q_{E/B}$] (w1) at (0.0, 0.0);
            \node[circle, draw=Orange, fill = Orange, scale=0.5, label=right:$Q_{E/B}$] (w2) at (2.0, 0.0);

            \diagram*{
                (w1) -- [dashed, Green, ultra thick] (w2),
            };
        \end{feynman}
    \end{tikzpicture}}}
    \equiv i \langle [ Q^{E/B}_{ij}(\tau) Q^{E/B}_{kl}(0)] \rangle \theta(\tau) ~,
\end{equation}
which makes causality manifest. In the frequency domain, the retarded Green function is given by the Fourier transform of Eq.~(\ref{eq:KerrAns}). Then, from the effective action in Eq.~\eqref{eq: Kerr tidal action}, the LO scattering amplitude reads
\begin{equation}
  i \mathcal{M}(\boldsymbol{k}_{\rm in}\rightarrow \boldsymbol{k}_{\rm out},h \rightarrow h) = 
  \vcenter{\hbox{\begin{tikzpicture}[scale=0.7]
        \begin{feynman}
            \vertex (i) at (0,0);
            \vertex (e) at (0,3);
            \node[circle, draw=Orange, fill = Orange, scale=0.5] (w1) at (0, 1.0);
            \node[circle, draw=Orange, fill = Orange, scale=0.5] (w2) at (0, 2.0);
            \vertex (f1) at (1.5,2.8) {};
            \vertex (fs) at (1.5,0.2) {};

            \diagram*{
                (i) -- [double, double, thick] (w1),
                (w1) -- [dashed, Green, ultra thick,edge label = $\lambda^{E/B}$] (w2),
                (w2) -- [double, double, thick] (e),
                (f1) -- [boson, MyYellow, ultra thick] (w2),
                (fs) -- [boson, MyYellow, ultra thick] (w1)
            };
        \end{feynman}
    \end{tikzpicture}}}
    = - i \frac{\omega^4}{4 M_{\rm pl}^2}({\lambda^E})_{ij,kl}\epsilon_h^{kl}(\boldsymbol{k}_{\rm in}){\epsilon}_{h}^{*ij}(\boldsymbol{k}_{\rm out}) M (GM)^4 + {\rm magnetic} ~,
\end{equation}
where the yellow wavy lines represent gravitons.
With the formulas in App.~\ref{app:love rep}, we can transform the amplitude in the plane wave basis into the spherical wave basis with definite helicities as
\begin{equation}
    \begin{aligned}
    i \mathcal{A}(\omega, \ell=2,m,h\rightarrow \ell=2,m,h) & = - i \frac{\omega^5}{40M_{\rm pl}^2 \pi} M (G M)^4 \\ 
    & \quad \times \[\Lambda_{\hat{s}^0}^E + \frac{1}{2} i m H_{\hat{s}^1}^E + \frac{1}{6}(m^2-4)\( - \Lambda_{\hat{s}^2}^E - i m H_{\hat{s}^3}^E + \Lambda_{\hat{s}^4}^E (m^2 - 1)\)\] \\
    & \quad + {\rm magnetic} ~.
    \end{aligned}
    \label{eq : helicity basis 222}
\end{equation}
where we have used the operator form of the tidal response tensor $\lambda^{ij}_{kl} = \langle i,j |\bold{\Lambda} | k,l\rangle$
with
\begin{equation}
    \bold{\Lambda} = \Lambda_{\hat{s}^0} + i\frac{1}{2} H_{\hat{s}^1} \bold{J}_z + \frac{1}{6}(\bold{J}_z^2-4)[-\Lambda_{\hat{s}^2} -i \bold{J}_z H_{\hat{s}^3} + \Lambda_{\hat{s}^4} (\bold{J}_z^2-1)] ~,
\end{equation}
given in App.~\ref{app:love rep}. By matching this expression to the BHPT phase shift in Sec.~\ref{subsec:BHPT phase shift}
\begin{equation}
\label{eq:match to phase}
    i\mathcal{A}(\omega,\ell=2,m,h\rightarrow \ell=2,m,h) = 1 - {}_2\eta_{2 m}\exp(2 i {}_2\delta_{2m}) ~
\end{equation}
we can get the vanishing of Kerr Love numbers
\begin{eBox}
\begin{equation}
    \begin{aligned}
    \label{eq: Kerr Love numbers}
        \Lambda_{\hat{s}^0}^{E/B} = \Lambda_{\hat{s}^2}^{E/B} = \Lambda_{\hat{s}^4}^{E/B} = 0 ~,
    \end{aligned}
\end{equation}
\end{eBox}
while the dissipation numbers are given by
\begin{eBox}
\begin{equation}\label{eq:LO dissipation number}
    H_{\hat{s}^1}^{E/B} = - \frac{8}{45} (\chi + 3 \chi^3) ~, \quad H_{\hat{s}^3}^{E/B} = \frac{2}{3}\chi^3 ~.
\end{equation}
\end{eBox}
It is important here also to mention that one can also get the above results from the near zone approximation established by A.A.Starobinksy and D.N.Page in \cite{Starobinskil:1974nkd,Starobinsky:1973aij, Page:1976df,Page:1976ki}.

\subsection{NLO Dissipation Numbers: Large Spin Contribution and Shrinking Superradiance Parameter Space}
\label{subsec:match NLO dissipation}
In this section, we go beyond leading order Love and dissipation numbers and compute the NLO dissipation numbers that 
appear in the scattering phase at $\mathcal{O}\Big((r_s \omega)^6\Big)$.
We first reproduce the results of \cite{Saketh:2022xjb}. Then we discuss its relationship to the so-called superradiance factor recently
discussed in \cite{Chia:2020yla,Charalambous:2021mea}
in the context of tidal responses. 
Very similar to the calculation in the previous section, we immediately get the scattering amplitude from Eq.~\eqref{eq: Kerr tidal action} and~\eqref{eq:KerrAns}
\begin{equation}
\label{eq: NLO diagram}
  i \mathcal{M}(\boldsymbol{k}_{\rm in}\rightarrow \boldsymbol{k}_{\rm out},h \rightarrow h) = 
  \vcenter{\hbox{\begin{tikzpicture}[scale=0.7]
        \begin{feynman}
            \vertex (i) at (0,0);
            \vertex (e) at (0,3);
            \node[circle, draw=Orange, fill = Orange, scale=0.5] (w1) at (0, 1.0);
            \node[circle, draw=Orange, fill = Orange, scale=0.5] (w2) at (0, 2.0);
            \vertex (f1) at (1.5,2.8) {};
            \vertex (fs) at (1.5,0.2) {};

            \diagram*{
                (i) -- [double, double, thick] (w1),
                (w1) -- [dashed, Green, ultra thick,edge label = $\lambda_\omega^{E/B}$] (w2),
                (w2) -- [double, double, thick] (e),
                (f1) -- [boson, MyYellow, ultra thick] (w2),
                (fs) -- [boson, MyYellow, ultra thick] (w1)
            };
        \end{feynman}
    \end{tikzpicture}}}
    = \frac{\omega^5}{4 M_{\rm pl}^2}(\lambda^E_\omega)^{ij,kl}\epsilon_h^{kl}(\boldsymbol{k}_{\rm in}){\epsilon}_{h}^{*ij}(\boldsymbol{k}_{\rm out}) M (GM)^5 + {\rm magnetic} ~.
\end{equation}
Transforming this expression into the spherical basis states with definite helicity and using the tidal response tensor  $(\lambda_\omega)^{ij}_{kl} = \langle i,j|\bold{\Lambda}_{\omega}|k,l\rangle$ with
\begin{equation}
    \bold{\Lambda}_{\omega} = H_{\hat{s}^0,\omega} + i\frac{1}{2} \Lambda_{\hat{s}^1,\omega} \bold{J}_z + \frac{1}{6}(\bold{J}_z^2-4)[-H_{\hat{s}^2,\omega} -i \bold{J}_z \Lambda_{\hat{s}^3,\omega} + H_{\hat{s}^4,\omega} (\bold{J}_z^2-1)] ~,
\end{equation}
we get 
\begin{equation}
    \begin{aligned}
    i \mathcal{A}(\omega,\ell=2,m,h\rightarrow \ell=2,m,h) & = \frac{\omega^6}{40 M_{\rm pl}^2 \pi} M (GM)^5 \\ 
    & \quad \times \[H_{\hat{s}^0}^E + \frac{1}{2} i m \Lambda_{\hat{s}^1}^E + \frac{1}{6}(m^2-4)\( - H_{\hat{s}^2}^E - i m \Lambda_{\hat{s}^3}^E + H_{\hat{s}^4}^E (m^2 - 1)\)\] \\
    & \quad + {\rm magnetic} ~.
    \end{aligned}
    \label{eq : helicity basis NLO 222}
\end{equation}
As we expected, the conservative response coefficient and the dissipation coefficient switches their spin dependence due to the additional time derivatives which produces an additional factor of $\omega$. 
We will discuss the conservative part in detail in in Sec.~\ref{subsec: RG elastic}. Now let us focus on the dissipative part. After matching the EFT expression above to the BHPT phase shift, we get
\begin{eBox}
\begin{equation}
    \begin{aligned}
    \label{eq:NLO diss}
    H_{\hat{s}^0,\omega}^{E/B} & = \frac{8}{405}\left(9+9 \kappa+97 \chi^2+117 \kappa \chi^2-6 \chi^4+54 \kappa \chi^4+36 \chi B_2+108 \chi^3 B_2\right) ~,  \\
    H_{\hat{s}^2,\omega}^{E/B} & = -\frac{4}{135}\left(115 \chi^2+135 \kappa \chi^2+5 \chi^4+90 \kappa \chi^4-24 \chi B_1+18 \chi^3 B_1+48 \chi B_2+144 \chi^3 B_2\right) ~, \\
    H_{\hat{s}^4,\omega}^{E/B} & = \frac{4}{135}\left(20 \chi^4+45 \kappa \chi^4-24 \chi B_1+18 \chi^3 B_1+12 \chi B_2+36 \chi^3 B_2\right) ~,
    \end{aligned}
\end{equation}
\end{eBox}
where $\kappa = \sqrt{1-\chi^2}$ and $B_m = \operatorname{Im}[\psi^{(0)}(3+i m \chi / \kappa)]$ (for $m=1,2$), $\psi^{(0)}$ is the polygamma function. This expression agrees with \cite{Saketh:2022wap}. In the Schwarzschild case, we have
\begin{eBox}
\begin{equation}
    H_{\hat{s}^0, \omega}^{E/B}(\chi = 0) = \frac{16}{45} ~.
    \label{eq :NLO dissipation number small spin}
\end{equation}
\end{eBox}
Interestingly, the polygamma function $\psi^{(0)}(3+ i m \chi/\kappa)$ goes to infinity as spin $\chi$ goes to extremality ($\chi\rightarrow 1$), indicating a large spin contribution to the dissipation. 

The superradiance effect~\cite{zel1971generation} dictates 
the functional form of dissipation for Kerr BH in the static limit. 
Specifically, the non-vanishing of $H_{\hat{s}^1}^{E/B},H_{\hat{s}^3}^{E/B}$
directly follows from frame-dragging \cite{Chia:2020yla,Goldberger:2020fot,Charalambous:2021mea}. 
As shown in \cite{Starobinsky:1973aij,Starobinskil:1974nkd,Page:1976df,Page:1976ki,Maldacena:1997ih,Mano:1996gn,Chia:2020yla,Goldberger:2020fot,Charalambous:2021mea} using the near zone calculations, 
the low-frequency absorption probability can be written as
\begin{equation}
     {}_s \Gamma_{\ell m} = 2 (-P_+) \( \frac{(\ell + s)! (\ell - s)!}{(2\ell)! (2\ell + 1)!}\)^2 \Big(2 \omega ( r_+ -  r_-)\Big)^{2\ell + 1} \prod_{j=1}^{\ell} (j^2 + 4P_+^2) ~,
     \label{eq : near zone absorption}
\end{equation}
with the superradiance factor $P_+$
\begin{equation}
    P_+  = \frac{a m-2 M r_+ \omega}{r_+ - r_-} = \frac{m \Omega_{\rm H} - \omega}{4\pi T_{\rm H}} ~.
\end{equation}

\begin{figure}[ht!]
    \centering
    \includegraphics[scale = 1.2]{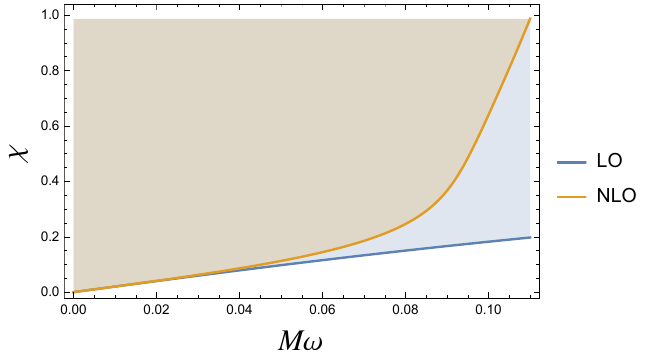}
    \caption{Comparison of LO and NLO supperadiance conditions, see Eq.~\eqref{eq:NLO diss}. In this figure, we show the example with $\ell =2,m=2$. The colored region corresponds to the superradiance parameter space, with solid lines showing the critical values, i.e. $m\Omega_H=\omega$ at LO.
    }
    \label{fig:modified superradiance}
\end{figure}

The appearance of the overall factor $m\Omega_H-\omega$ dictates a constraint on tidal response at leading and next-to-leading order in the small spin limit. This is seen by relating the above expression for absorption probability with the tidal response in the EFT to NLO as
\begin{alignat}{3}
{}_{2}\Gamma_{22} = 1-{}_{2}\eta_{22}^2 & = 2  \mr{Re}[i\mc{A}(\omega,\ell=m=2,h\rightarrow \ell=m=2,h)], 
\\ & =  M (GM)^4 \frac{\omega^5}{20 M_{pl}^2\pi} \Big [H_{\hat{s}^1}^E + (GM\omega) H_{\hat{s}^0}^E \Big] + {\rm magnetic} + \mc{O}(\omega^7),
\end{alignat}
where we have used Eqs.~(\ref{eq : helicity basis 222}), \eqref{eq:match to phase} and (\ref{eq : helicity basis NLO 222}). Now, taking the limit $\chi\rightarrow 0$ in Eq.~(\ref{eq : near zone absorption}), where we have $\Omega_H\rightarrow \chi/(4GM)$, and expand until $\mathcal{O}(\omega^6)$ order we have
\begin{alignat}{3}
- \frac{128}{225}(G M \omega)^5 \Big(\chi-2 (GM\omega)\Big) + \mc{O}(\omega^7) =  M (GM)^4 \frac{\omega^5}{20 M_{\rm pl}^2\pi} \Big[H_{\hat{s}^1}^E + (GM\omega) H_{\hat{s}^0}^E\Big] + {\rm magnetic} + \mc{O}(\omega^7) ~.
\end{alignat}
After using E/B duality, we arrive at the constraint
\begin{equation}
    \begin{aligned}
    \label{eq: low-spin diss consistency}
        H_{\hat{s}^0,\omega}^{E/B} = -2 \chi H_{\hat{s}^1}^{E/B}  ~,~\text{for} \quad \chi \rightarrow 0 ~,
    \end{aligned}
\end{equation} 
as desired. This is consistent with the small spin limit of expressions we obtained in Eqs.~(\ref{eq:LO dissipation number}),~(\ref{eq :NLO dissipation number small spin}).
When $m \Omega_{\rm H} > \omega$, BHs lose energy to external perturbations, i.e. superradiance takes place. 
The parameter space in the $\chi-(M\omega)$ plane
that corresponds to this regime of superradiance 
is presented by the blue shaded region in 
Fig.~\ref{fig:modified superradiance}. This superradiance condition is based on the near zone calculations. 
Consequently, it requires modification if one includes corrections from the far zone iteratively \cite{Chatziioannou:2012gq,Chatziioannou:2016kem}. 
The NLO Kerr dissipation number $H_{\hat{s}^0,\omega}$ in Eq.~\eqref{eq:NLO diss} includes these corrections and captures the modified superradiance condition by accounting for the large spin correction. 
The modified superradiance parameter space is plotted as the orange region in Fig.~\ref{fig:modified superradiance}. 
We see that the parameter space for superradiance shrinks due to the large spin corrections coming from the polygamma functions $\psi^{(0)}(3 + i m \chi / \kappa)$ in Eq.~\eqref{eq:NLO diss}. 
This is a particularly intriguing observation as the polygamma function indicates a non-analytic behavior in the spin magnitude $\chi$ and even exhibits a singular point for the 
extremal spin value ($\chi=1$). 
A similar pattern has been noted in \cite{Bautista:2022wjf}. 
As discussed in Sec.~\ref{sec:BHPT}, these non-analytic functions are derived from what we refer to as the near zone part. Its derivation was
formally carried out only for the sub-extremal cases ($0 \leq \chi <1$). 
In contrast, the far zone pieces that are comprised of power laws in $\chi$ and $\kappa = \sqrt{1 - \chi^2}$, exhibit only a branch point at $\chi=1$, except for the logarithmic tail effect. 
This structure is suitable for an analytic continuation in $\chi$ to the super-extremal region $\chi 
\gg 1$. 
To summarize, the  near zone terms only make sense in the physical sub-extremal region while the far zone terms can be formally 
extended into the extremal and super extremal region. 


\subsection{Tail Effects}
\label{subsec:tail}

In this section, we focus on the  \orange{orange} terms in Sec.~\ref{subsec:BHPT phase shift}. We show that these 
terms correspond to the leading tail effect due to the scattering off the long-range Newtonian potential. 
Let's consider the following diagram (the 1-loop integration can be found in \cite{Goldberger:2009qd} and \cite{Porto:2012as})
\begin{equation}
\begin{aligned}
       i\mathcal{M} = & \quad     \vcenter{\hbox{\begin{tikzpicture}[scale=0.7]
        \begin{feynman}
            \vertex (i) at (0,0);
            \vertex (e) at (0,3);
            \node[circle, draw=Orange, fill = Orange, scale=0.5] (w1) at (0, 1.0);
            \node[circle, draw=Orange, fill = Orange, scale=0.5] (w3) at (0, 2.0);
            \vertex(f1) at (2.0,2.8);
            \vertex (fs) at (2.0,0.2); 
            \vertex[label= left:$m$] (w2) at (0,0.25);
            \vertex (f3) at (1.0, 0.70);

            \diagram*{
                (i) -- [double, double, thick] (w1),
                (w1) -- [dashed, Green, ultra thick,edge label = $\lambda^{E/B}$] (w3),
                (w3) -- [double, double, thick] (e),
                (w2) -- [boson, MyYellow, ultra thick] (f3),
                (f1) -- [boson, MyYellow, ultra thick] (w3),
                (fs) -- [boson, MyYellow, ultra thick] (w1),
            };
        \end{feynman}
    \end{tikzpicture}}}
    + 
                \vcenter{\hbox{\begin{tikzpicture}[scale=0.7]
        \begin{feynman}
            \vertex (i) at (0,0);
            \vertex (e) at (0,3);
            \node[circle, draw=Orange, fill = Orange, scale=0.5] (w1) at (0, 1.0);
            \node[circle, draw=Orange, fill = Orange, scale=0.5] (w3) at (0, 2.0);
            \vertex(f1) at (2.0,2.8);
            \vertex (fs) at (2.0,0.2); 
            \vertex[label= left:$m$] (w2) at (0,2.75);
            \vertex (f3) at (1.0, 2.4);

            \diagram*{
                (i) -- [double, double, thick] (w1),
                (w1) -- [dashed, Green, ultra thick,edge label = $\lambda^{E/B}$] (w3),
                (w3) -- [double, double, thick] (e),
                (f1) -- [boson, MyYellow, ultra thick] (w3),
                (fs) -- [boson, MyYellow, ultra thick] (w1),
                (w2) -- [boson, MyYellow, ultra thick] (f3)
            };
        \end{feynman}
    \end{tikzpicture}}} \\
    & = \vcenter{\hbox{\begin{tikzpicture}[scale=0.7]
        \begin{feynman}
            \vertex (i) at (0,0);
            \vertex (e) at (0,3);
            \node[circle, draw=Orange, fill = Orange, scale=0.5] (w1) at (0, 1.0);
            \node[circle, draw=Orange, fill = Orange, scale=0.5] (w2) at (0, 2.0);
            \vertex (f1) at (1.5,2.8) {};
            \vertex (fs) at (1.5,0.2) {};

            \diagram*{
                (i) -- [double, double, thick] (w1),
                (w1) -- [dashed, Green, ultra thick,edge label = $\lambda^{E/B}$] (w2),
                (w2) -- [double, double, thick] (e),
                (f1) -- [boson, MyYellow, ultra thick] (w2),
                (fs) -- [boson, MyYellow, ultra thick] (w1)
            };
        \end{feynman}
    \end{tikzpicture}}} {\times 2 (16 \pi G m \omega^2) \int \frac{d^{d-1} \boldsymbol{q}}{(2\pi)^{d-1}} \frac{1}{\boldsymbol{q}^2} \frac{1}{(\omega + i 0)^2 - (\boldsymbol{q} + \boldsymbol{k})^2} } \\
    & =   \vcenter{\hbox{\begin{tikzpicture}[scale=0.7]
        \begin{feynman}
            \vertex (i) at (0,0);
            \vertex (e) at (0,3);
            \node[circle, draw=Orange, fill = Orange, scale=0.5] (w1) at (0, 1.0);
            \node[circle, draw=Orange, fill = Orange, scale=0.5] (w2) at (0, 2.0);
            \vertex (f1) at (1.5,2.8) {};
            \vertex (fs) at (1.5,0.2) {};

            \diagram*{
                (i) -- [double, double, thick] (w1),
                (w1) -- [dashed, Green, ultra thick,edge label = $\lambda^{E/B}$] (w2),
                (w2) -- [double, double, thick] (e),
                (f1) -- [boson, MyYellow, ultra thick] (w2),
                (fs) -- [boson, MyYellow, ultra thick] (w1)
            };
        \end{feynman}
    \end{tikzpicture}}} 
    \times 2 (i G_N m \omega) \[ - \frac{(\omega + i 0)^2}{\pi \mu^2} e^{\gamma_E}\]^{(d-4)/2} \times \[ \frac{2}{(d-4)_{\rm IR}} - \frac{11}{6} + \cdots \] \\
    & =  \vcenter{\hbox{\begin{tikzpicture}[scale=0.7]
        \begin{feynman}
            \vertex (i) at (0,0);
            \vertex (e) at (0,3);
            \node[circle, draw=Orange, fill = Orange, scale=0.5] (w1) at (0, 1.0);
            \node[circle, draw=Orange, fill = Orange, scale=0.5] (w2) at (0, 2.0);
            \vertex (f1) at (1.5,2.8) {};
            \vertex (fs) at (1.5,0.2) {};

            \diagram*{
                (i) -- [double, double, thick] (w1),
                (w1) -- [dashed, Green, ultra thick,edge label = $\lambda^{E/B}$] (w2),
                (w2) -- [double, double, thick] (e),
                (f1) -- [boson, MyYellow, ultra thick] (w2),
                (fs) -- [boson, MyYellow, ultra thick] (w1)
            };
        \end{feynman}
    \end{tikzpicture}}} 
    \times 2 G M \omega \[ {\rm sign}(\omega) \pi + i\( \frac{2}{(d-4)_{\rm IR}} + \log\frac{\omega^2}{\pi \mu^2} + \gamma_E - \frac{11}{6} \) \] ~,
    \end{aligned}
\end{equation}
where $\boldsymbol{k}^2 = \omega^2$ and in the last expression we have performed an expansion around\footnote{In this expression, we have used the retarded propagator to make causality manifest. The $i0$-prescription here tells the relative position between the physical sheet and the branch cut. When $\omega>0$, we choose the prescription $(-1)^{(d-4)/2} = e^{-i \pi (d-4)/2}$. When $\omega<0$, we choose another prescription $(-1)^{(d-4)/2} = e^{i\pi(d-4)/2}$. To compare with BHPT, we only need $ \omega >0$ as the BHPT results we presented in Section \ref{subsec:BHPT phase shift} always assumes $\omega>0$.} $d=4$. We now clearly see the IR divergences. 
As pointed out in \cite{Goldberger:2009qd,Porto:2012as}, the imaginary part of the above 1-loop integral will exponentiate into a universal pure phase out and can be dropped out in the physical observable such as the cross sections $\sigma \sim |\mathcal{M}|^2$. 
Indeed, we do not find the associated logarithms in the BHPT phase shift\footnote{It is important to not confuse the logarithms in the green terms with the tail effect logarithm discussed here. Although they seem to coincide at leading order in for spinning BHs, the logarithms in the green part do not exponentiate.} since they are unphysical, but we do find the finite terms obtained by multiplying the lower order amplitudes by the factor $2GM\omega\pi$, as the orange terms in the BHPT phase shift. The leading tail effect thus nicely explains all the \orange{orange} terms appear in the BHPT reuslts in Sec.~\ref{subsec:BHPT phase shift}. Most importantly, this effect explains the presence of certain odd-looking spin dependencies in the BHPT phase shift. For instance, the \orange{orange} piece appearing at NLO at $(r_s\omega)^6$ is odd in spin, as opposed to all the other terms which are even in spin. Thus, naively, it seems to transform differently from other terms at the same order under time-reversal ($\chi \rightarrow -\chi$, $\omega\rightarrow -\omega$). But recognizing the orange term as due to the leading order tail effect, we realize that that the actual dependence on frequency is $(r_s\omega)^5\times r_s|\omega|$ which makes the overall result even under time-reversal. 

\subsection{RG Running of Dissipation Numbers at NNLO}
\label{subsec:RGabs}

In the next two sections, we explain all the logarithmic dependencies observed in the BHPT results from Sec.~\ref{subsec:BHPT phase shift}. First, we will examine the logarithmic terms that appear in the absorption probability $1-({}_2\eta_{\ell m})^2$, highlighted in \purple{purple}. Subsequently, in Sec.~\ref{subsec: RG elastic}, we will discuss the logarithmic terms in the elastic phase shift ${}_2\delta_{\ell m}$, 
marked in \green{green}.

We will start with the RG running of dissipation numbers. 
For simplicity, 
we will be using the optical theorem and the fluctuation-dissipation theorem \cite{Goldberger:2005cd, Porto:2007qi, Ivanov:2022hlo} to calculate absorption probabilities. We have
\begin{equation}
    {}_2\Gamma_{\ell m} (\omega) = 1 - {}_2\eta_{\ell m}^2 = 2{\rm Re}\[   \vcenter{\hbox{\begin{tikzpicture}[scale=0.7]
        \begin{feynman}
            \vertex (i) at (0,0);
            \vertex (e) at (0,3);
            \node[circle, draw=Orange, fill = Orange, scale=0.5, label = left:$Q$] (w1) at (0, 1.0);
            \node[circle, draw=Orange, fill = Orange, scale=0.5, label = left:$Q$] (w2) at (0, 2.0);
            \vertex (f1) at (1.5,2.8) {};
            \vertex (fs) at (1.5,0.2) {};

            \diagram*{
                (i) -- [double, double, thick] (w1),
                (w1) -- [dashed, Green, ultra thick] (w2),
                (w2) -- [double, double, thick] (e),
                (f1) -- [boson, MyYellow, ultra thick] (w2),
                (fs) -- [boson, MyYellow, ultra thick] (w1)
            };
        \end{feynman}
    \end{tikzpicture}}} \] = \Bigg| \vcenter{\hbox{\begin{tikzpicture}[scale=0.8]
        \begin{feynman}
            \vertex (i) at (0,0);
            \vertex (e) at (0,2.0);
            \node[circle, draw=Orange, fill = Orange, scale=0.5, label=left:$Q$] (w1) at (0, 1.0);
            \vertex (fs) at (1.5,0.2) {}; 

            \diagram*{
                (i) -- [double, double, thick] (w1),
                (w1) -- [dashed, double, double, thick] (e),
                (fs) -- [boson,MyYellow, ultra thick] (w1)
            };
        \end{feynman}
    \end{tikzpicture}}} \Bigg|^2 ~, \quad \omega>0 ~.
\end{equation}
With this method, we can now compute the 2-loop corrections to the absorption as \footnote{Here, $\lambda_{E}$ is a short notation for the dissipation degrees of freedom $H_{\hat{s}^1}^E, H_{\hat{s}^3}^E$ in Eq.~\eqref{eq: def LO diss}.}
\begin{equation}
\begin{aligned}\label{eq:abs log div}
    & \quad \Bigg|
\vcenter{\hbox{\begin{tikzpicture}[scale=0.8]
        \begin{feynman}
            \vertex (i) at (0,0);
            \vertex (e) at (0,2.0);
            \node[circle, draw=Orange, fill = Orange, scale=0.5, label=left:$\lambda^E$] (w1) at (0, 1.0);
            \vertex (fs) at (1.5,0.2) {}; 

            \diagram*{
                (i) -- [double, double, thick] (w1),
                (w1) -- [dashed, double, double, thick] (e),
                (fs) -- [boson,MyYellow, ultra thick] (w1)
            };
        \end{feynman}
    \end{tikzpicture}}} 
    + 
         \vcenter{\hbox{\begin{tikzpicture}[scale=0.8]
        \begin{feynman}
            \vertex (i) at (0,0);
            \vertex (e) at (0,2.0);
            \node[circle, draw=Orange, fill = Orange, scale=0.5, label=left:$\lambda^E$] (w1) at (0, 1.0);
            \vertex (fs) at (1.5,0.2) {};
            \vertex[label= left:$m$] (w2) at (0,0.25);
            \vertex (f1) at (0.75, 0.8) {};

            \diagram*{
                (i) -- [double, double, thick] (w1),
                (w1) -- [dashed, double, double, thick] (e),
                (fs) -- [boson,MyYellow, ultra thick] (w1),
                (w2) -- [boson, MyYellow, ultra thick] (f1),
            };
        \end{feynman}
    \end{tikzpicture}}}
    +
    \vcenter{\hbox{\begin{tikzpicture}[scale=0.8]
        \begin{feynman}
            \vertex (i) at (0,0);
            \vertex (e) at (0,2.0);
            \node[circle, draw=Orange, fill = Orange, scale=0.5, label=left:$\lambda^E$] (w1) at (0, 1.0);
            \vertex (fs) at (1.5,0.2) {};
            \vertex[label= left:$m$] (w2) at (0,0.5);
            \vertex (f1) at (0.75, 0.8) {};
             \vertex[label= left:$m$] (w3) at (0,0.0);

            \diagram*{
                (i) -- [double, double, thick] (w1),
                (w1) -- [dashed, double, double, thick] (e),
                (fs) -- [boson,MyYellow, ultra thick] (w1),
                (w2) -- [boson, MyYellow, ultra thick] (f1),
                (w3) -- [boson, MyYellow, ultra thick] (f1),
            };
        \end{feynman}
    \end{tikzpicture}}}
    +
    \vcenter{\hbox{\begin{tikzpicture}[scale=0.8]
        \begin{feynman}
            \vertex (i) at (0,0);
            \vertex (e) at (0,2.0);
            \node[circle, draw=Orange, fill = Orange, scale=0.5, label=left:$\lambda^E$] (w1) at (0, 1.0);
            \vertex (fs) at (1.5,0.2) {};
            \vertex[label= left:$m$] (w2) at (0,0.5);
            \vertex (f1) at (0.75, 0.8) {};
             \vertex[label= left:$m$] (w3) at (0,0.0);
             \vertex (ib) at (0.4,0.45);

            \diagram*{
                (i) -- [double, double, thick] (w1),
                (w1) -- [dashed, double, double, thick] (e),
                (fs) -- [boson,MyYellow, ultra thick] (w1),
                (w2) -- [boson, MyYellow, ultra thick] (ib),
                (w3) -- [boson, MyYellow, ultra thick] (ib),
                (ib) -- [boson, MyYellow, ultra thick] (f1),
                };
        \end{feynman}
    \end{tikzpicture}}}
    +
    \vcenter{\hbox{\begin{tikzpicture}[scale=0.8]
        \begin{feynman}
            \vertex (i) at (0,0);
            \vertex (e) at (0,2.0);
            \node[circle, draw=Orange, fill = Orange, scale=0.5, label=left:$\lambda^E$] (w1) at (0, 1.0);
            \vertex (fs) at (1.7,0.25) {};
             \vertex (tfs) at (1.2,0.6);
            \vertex[label= left:$m$] (w2) at (0,0.5);
            \vertex (f1) at (0.75, 0.8);
             \vertex[label= left:$m$] (w3) at (0,0.0);

            \diagram*{
                (i) -- [double, double, thick] (w1),
                (w1) -- [dashed, double, double, thick] (e),
                (tfs) -- [boson,MyYellow, ultra thick] (w1),
                (w2) -- [boson, MyYellow, ultra thick] (f1),
                (w3) -- [boson, MyYellow, ultra thick] (tfs),
                (tfs) -- [boson,MyYellow, ultra thick] (fs),
                };
        \end{feynman}
    \end{tikzpicture}}}
    \Bigg|^2 + |{\rm magnetic}|^2
\end{aligned}
\end{equation}
As mentioned previously in \cite{Goldberger:2009qd}, we notice that the third and fourth diagram in the above equation
\begin{equation}
\begin{aligned}
        \vcenter{\hbox{\begin{tikzpicture}[scale=0.8]
        \begin{feynman}
            \vertex (i) at (0,0);
            \vertex (e) at (0,2.0);
            \node[circle, draw=Orange, fill = Orange, scale=0.5, label=left:$\lambda^E$] (w1) at (0, 1.0);
            \vertex (fs) at (1.5,0.2) {};
            \vertex[label= left:$m$] (w2) at (0,0.5);
            \vertex (f1) at (0.75, 0.8) {};
             \vertex[label= left:$m$] (w3) at (0,0.0);

            \diagram*{
                (i) -- [double, double, thick] (w1),
                (w1) -- [dashed, double, double, thick] (e),
                (fs) -- [boson,MyYellow, ultra thick] (w1),
                (w2) -- [boson, MyYellow, ultra thick] (f1),
                (w3) -- [boson, MyYellow, ultra thick] (f1),
            };
        \end{feynman}
    \end{tikzpicture}}}
    +
    \vcenter{\hbox{\begin{tikzpicture}[scale=0.8]
        \begin{feynman}
            \vertex (i) at (0,0);
            \vertex (e) at (0,2.0);
            \node[circle, draw=Orange, fill = Orange, scale=0.5, label=left:$\lambda^E$] (w1) at (0, 1.0);
            \vertex (fs) at (1.5,0.2) {};
            \vertex[label= left:$m$] (w2) at (0,0.5);
            \vertex (f1) at (0.75, 0.8) {};
             \vertex[label= left:$m$] (w3) at (0,0.0);
             \vertex (ib) at (0.4,0.45);

            \diagram*{
                (i) -- [double, double, thick] (w1),
                (w1) -- [dashed, double, double, thick] (e),
                (fs) -- [boson,MyYellow, ultra thick] (w1),
                (w2) -- [boson, MyYellow, ultra thick] (ib),
                (w3) -- [boson, MyYellow, ultra thick] (ib),
                (ib) -- [boson, MyYellow, ultra thick] (f1),
                };
        \end{feynman}
    \end{tikzpicture}}} 
    \sim \vcenter{\hbox{\begin{tikzpicture}[scale=0.8]
        \begin{feynman}
            \vertex (i) at (0,0);
            \vertex (e) at (0,2.0);
            \node[circle, draw=Orange, fill = Orange, scale=0.5, label=left:$\lambda^E$] (w1) at (0, 1.0);
            \vertex (fs) at (1.5,0.2) {}; 

            \diagram*{
                (i) -- [double, double, thick] (w1),
                (w1) -- [dashed, double, double, thick] (e),
                (fs) -- [boson,MyYellow, ultra thick] (w1)
            };
        \end{feynman}
    \end{tikzpicture}}} 
    & \times {\int \frac{d^{d-1}\boldsymbol{q}}{(2\pi)^{d-1}} \frac{1}{(\omega + i0)^2 - (\boldsymbol{k}+\boldsymbol{q})^2}} \\
    & \times {\int \frac{d^{d-1}\boldsymbol{p}}{(2\pi)^{d-1}} \frac{1}{\boldsymbol{p}^2} \frac{1}{(\boldsymbol{p}+\boldsymbol{q})^2}}
\end{aligned}
\end{equation}
only has a UV divergence, while the last diagram contains both UV and IR divergences,
\begin{equation}
\begin{aligned}
        \vcenter{\hbox{\begin{tikzpicture}[scale=0.8]
        \begin{feynman}
            \vertex (i) at (0,0);
            \vertex (e) at (0,2.0);
            \node[circle, draw=Orange, fill = Orange, scale=0.5, label=left:$\lambda^E$] (w1) at (0, 1.0);
            \vertex (fs) at (1.7,0.25) {};
             \vertex (tfs) at (1.2,0.6);
            \vertex[label= left:$m$] (w2) at (0,0.5);
            \vertex (f1) at (0.75, 0.8);
             \vertex[label= left:$m$] (w3) at (0,0.0);

            \diagram*{
                (i) -- [double, double, thick] (w1),
                (w1) -- [dashed, double, double, thick] (e),
                (tfs) -- [boson,MyYellow, ultra thick] (w1),
                (w2) -- [boson, MyYellow, ultra thick] (f1),
                (w3) -- [boson, MyYellow, ultra thick] (tfs),
                (tfs) -- [boson,MyYellow, ultra thick] (fs),
                };
        \end{feynman}
    \end{tikzpicture}}}
    \sim \vcenter{\hbox{\begin{tikzpicture}[scale=0.8]
        \begin{feynman}
            \vertex (i) at (0,0);
            \vertex (e) at (0,2.0);
            \node[circle, draw=Orange, fill = Orange, scale=0.5, label=left:$\lambda^E$] (w1) at (0, 1.0);
            \vertex (fs) at (1.5,0.2) {}; 

            \diagram*{
                (i) -- [double, double, thick] (w1),
                (w1) -- [dashed, double, double, thick] (e),
                (fs) -- [boson,MyYellow, ultra thick] (w1)
            };
        \end{feynman}
    \end{tikzpicture}}} 
    & \times {\int \frac{d^{d-1}\boldsymbol{q}}{(2\pi)^{d-1}} \frac{1}{\boldsymbol{q}^2}\frac{1}{(\omega+i0)^2 - (\boldsymbol{k} + \boldsymbol{q})^2}} \\
    & \times {\int \frac{d^{d-1}\boldsymbol{p}}{(2\pi)^{d-1}} \frac{1}{\boldsymbol{p}^2} \frac{1}{(\omega + i0)^2 - (\boldsymbol{k} + \boldsymbol{p} + \boldsymbol{q})^2}} ~.
\end{aligned}
\end{equation}
Adding all these terms together, we get the final expression (the detailed calculation can be found in \cite{Goldberger:2009qd})
\begin{equation}
    \begin{aligned}
        \text{Eq.~\eqref{eq:abs log div}} & =  \Bigg|\vcenter{\hbox{\begin{tikzpicture}[scale=0.7]
        \begin{feynman}
            \vertex (i) at (0,0);
            \vertex (e) at (0,2.0);
            \node[circle, draw=Orange, fill = Orange, scale=0.5, label=left:$\lambda^{E}$] (w1) at (0, 1.0);
            \vertex (fs) at (1.5,0.2) {}; 

            \diagram*{
                (i) -- [double, double, thick] (w1),
                (w1) -- [dashed, double, double, thick] (e),
                (fs) -- [boson,MyYellow, ultra thick] (w1)
            };
        \end{feynman}
    \end{tikzpicture}}}
    \Bigg|^2 \times \Bigg(1 + 2 \pi G M \omega {\rm sign} (\omega) -\frac{428( G M \omega)^2}{210}\Bigg[\frac{1}{(d-4)_{\rm UV}}+\gamma_E+\log\Bigg(\frac{\omega^2}{\pi\mu^2}\Bigg)\Bigg] \\
    & \quad +(G M \omega)^2\Bigg[\frac{4\pi^2}{3}+\frac{634913}{44100}\Bigg]\Bigg) + |{\rm magnetic}|^2 ~.
    \end{aligned}
\end{equation}

Interestingly, we observe that the IR divergences cancel in the above equation. However, we do encounter UV divergences, indicating that a renormalization is necessary. 
By the power counting in $(r_s \omega)$, we can 
estimate that the UV divergence appears $(r_s \omega)^2$ (2-loop) orders higher than the leading dissipation effect. 
This suggests that the divergence can be absorbed into $(\lambda_{\omega^2}^{E/B})_{ijkl}$, a term we introduced in Eq.~\eqref{eq:KerrAns}. As discussed in Sec.~\ref{sec:EFT}, $(\lambda^{E/B})_{ijkl}$ and $(\lambda_{\omega^2}^{E/B})_{ijkl}$ share the same symmetry properties, i.e. the spin-even part corresponds to conservative effects while the spin-odd part corresponds to dissipation. Thus, if we further consider the following process
\begin{equation}
        \begin{aligned}
    \Bigg|\vcenter{\hbox{\begin{tikzpicture}[scale=0.7]
        \begin{feynman}
            \vertex (i) at (0,0);
            \vertex (e) at (0,2.0);
            \node[circle, draw=Orange, fill = Orange, scale=0.5, label=left:$\lambda_{\omega^2}^{E}$] (w1) at (0, 1.0);
            \vertex (fs) at (1.5,0.2) {}; 

            \diagram*{
                (i) -- [double, double, thick] (w1),
                (w1) -- [dashed, double, double, thick] (e),
                (fs) -- [boson,MyYellow, ultra thick] (w1)
            };
        \end{feynman}
    \end{tikzpicture}}}
    \Bigg|^2
    +
       \Bigg|\vcenter{\hbox{\begin{tikzpicture}[scale=0.7]
        \begin{feynman}
            \vertex (i) at (0,0);
            \vertex (e) at (0,2.0);
            \node[circle, draw=Orange, fill = Orange, scale=0.5, label=left:$\lambda_{\omega^2}^{B}$] (w1) at (0, 1.0);
            \vertex (fs) at (1.5,0.2) {}; 

            \diagram*{
                (i) -- [double, double, thick] (w1),
                (w1) -- [dashed, double, double, thick] (e),
                (fs) -- [boson,MyYellow, ultra thick] (w1)
            };
        \end{feynman}
    \end{tikzpicture}}}
    \Bigg|^2
    & = - M (G M)^6 \frac{\omega^7}{20\pi M_{\rm pl}^2}\[\frac{1}{2} m H_{\hat{s}^1,\omega^2}^E-\frac{1}{6} m (m^2-4)H_{\hat{s}^3,\omega^2}^E\] \\
    & \quad + |{\rm magnetic}|^2  ~,
    \end{aligned}
\end{equation}
after renormalizing the diverging piece, we can extract the RG running of NNLO dissipation numbers $H_{\hat{s}^1,\omega^2}$ and $H_{\hat{s}^3, \omega^2}$ introduced in Eq.~\eqref{eq: def NNLO diss}
\begin{eBox}
\begin{equation}
    \label{eq:NNLO diss RG}
    \frac{\partial H_{\hat{s}^1,\omega^2}^{E/B}}{\partial \log \mu} = \frac{428}{105}H_{\hat{s}^1} = - \frac{3424}{4725} (\chi + 3\chi^3) ~, \quad \frac{\partial H_{\hat{s}^3,\omega^2}^{E/B}}{\partial \log \mu} = \frac{428}{105} H_{\hat{s}^3, \omega^2}^{E/B} = \frac{856}{315} \chi^3 ~. 
\end{equation}
\end{eBox}
Such RG running effects can also be extracted from BHPT (\purple{purple} term at $(r_s \omega)^7$) from Eq.~\eqref{eq: 22 abs} and \eqref{n21}
\begin{equation}
    \begin{aligned}
    & \quad p(1 + {\rm BH} \rightarrow 0 + {\rm BH}')|_{(r_s \omega)^7\log(r_s \omega)} \\
    & =  M (G M)^6 \frac{\omega^7}{20 \pi M_{\rm pl}^2} \[ \frac{1}{2} m \( \frac{3424}{4725}  (\chi + 3 \chi^3)\) + \frac{1}{6} \( \frac{856}{315} \chi^3\)m(m^2 - 4) \] \log(2\sqrt{1-\chi^2} r_s \omega) \times 2 ~.
   \end{aligned}
\end{equation}
Introducing the renormalization scale $\mu$, we can split the logarithm into 
two pieces, 
\begin{equation}
    \log(2\sqrt{1-\chi^2} r_s \omega) =  \underbrace{\log\( 2 \sqrt{1 - \chi^2} r_s \mu\)}_{\text{NNLO dissipation counter term}} + \underbrace{\log\( \frac{\omega}{\mu}\)}_{\text{loop divergence}} ~.
\end{equation}
The first term is the counter term that produces 
the RG running of the dissipation numbers at NNLO. 
The second term can be interpreted as a genuine 
logarithmic divergence from the loops. 
Including the constant piece from BHPT Eqs.~\eqref{eq: 22 abs} and \eqref{n21}, we get 
\begin{equation}
    \begin{aligned}
        H_{\hat{s}^1,\omega^2}(\mu) & = - \frac{3424}{4725} (\chi + 3 \chi^3) \log\(4 G M \sqrt{1-\chi^2} \mu \) + 80 \mathcal{A}_{22}(\chi) ~,  \\
        H_{\hat{s}^3,\omega^2}(\mu) & = \frac{856}{315} \chi^3 \log\(4 G M \sqrt{1-\chi^2} \mu\) + 80 (2 \mathcal{A}_{21}(\chi) - \mathcal{A}_{22}(\chi)) ~,
    \end{aligned}
\end{equation}
The constant terms $\mathcal{A}_{22}(\chi)$ and $\mathcal{A}_{21}(\chi)$ are discussed Sec.~\ref{sec:constants}. 
We show that their interpretation is somewhat 
intricate due to the spherical-spheroidal mixing discussed in Sec.~\ref{subsec: higher mixing}.

The RG equation is useful because it can sum large logarithms. 
In the worldline EFT context, similar ideas were discussed in the gravitational radiation context in \cite{Goldberger:2009qd}. 
Here, we apply them to the horizon absorption. 
Let us first choose a specific matching scale $\mu_0$. Intuitively, at the 2-loop level, all the logarithmic divergences will come from the following diagrams (we only estimate the diverging piece)
\begin{equation}
    \begin{aligned}
        & \quad \Bigg|
\vcenter{\hbox{\begin{tikzpicture}[scale=0.8]
        \begin{feynman}
            \vertex (i) at (0,0);
            \vertex (e) at (0,2.0);
            \node[circle, draw=Orange, fill = Orange, scale=0.5, label=left:$\lambda^E$] (w1) at (0, 1.0);
            \vertex (fs) at (1.5,0.2) {}; 

            \diagram*{
                (i) -- [double, double, thick] (w1),
                (w1) -- [dashed, double, double, thick] (e),
                (fs) -- [boson,MyYellow, ultra thick] (w1)
            };
        \end{feynman}
    \end{tikzpicture}}} 
    + 
         \vcenter{\hbox{\begin{tikzpicture}[scale=0.8]
        \begin{feynman}
            \vertex (i) at (0,0);
            \vertex (e) at (0,2.0);
            \node[circle, draw=Orange, fill = Orange, scale=0.5, label=left:$\lambda^E$] (w1) at (0, 1.0);
            \vertex (fs) at (1.5,0.2) {};
            \vertex[label= left:$m$] (w2) at (0,0.25);
            \vertex (f1) at (0.75, 0.8) {};

            \diagram*{
                (i) -- [double, double, thick] (w1),
                (w1) -- [dashed, double, double, thick] (e),
                (fs) -- [boson,MyYellow, ultra thick] (w1),
                (w2) -- [boson, MyYellow, ultra thick] (f1),
            };
        \end{feynman}
    \end{tikzpicture}}}
    +
    \vcenter{\hbox{\begin{tikzpicture}[scale=0.8]
        \begin{feynman}
            \vertex (i) at (0,0);
            \vertex (e) at (0,2.0);
            \node[circle, draw=Orange, fill = Orange, scale=0.5, label=left:$\lambda^E$] (w1) at (0, 1.0);
            \vertex (fs) at (1.5,0.2) {};
            \vertex[label= left:$m$] (w2) at (0,0.5);
            \vertex (f1) at (0.75, 0.8) {};
             \vertex[label= left:$m$] (w3) at (0,0.0);

            \diagram*{
                (i) -- [double, double, thick] (w1),
                (w1) -- [dashed, double, double, thick] (e),
                (fs) -- [boson,MyYellow, ultra thick] (w1),
                (w2) -- [boson, MyYellow, ultra thick] (f1),
                (w3) -- [boson, MyYellow, ultra thick] (f1),
            };
        \end{feynman}
    \end{tikzpicture}}}
    +
    \vcenter{\hbox{\begin{tikzpicture}[scale=0.8]
        \begin{feynman}
            \vertex (i) at (0,0);
            \vertex (e) at (0,2.0);
            \node[circle, draw=Orange, fill = Orange, scale=0.5, label=left:$\lambda^E$] (w1) at (0, 1.0);
            \vertex (fs) at (1.5,0.2) {};
            \vertex[label= left:$m$] (w2) at (0,0.5);
            \vertex (f1) at (0.75, 0.8) {};
             \vertex[label= left:$m$] (w3) at (0,0.0);
             \vertex (ib) at (0.4,0.45);

            \diagram*{
                (i) -- [double, double, thick] (w1),
                (w1) -- [dashed, double, double, thick] (e),
                (fs) -- [boson,MyYellow, ultra thick] (w1),
                (w2) -- [boson, MyYellow, ultra thick] (ib),
                (w3) -- [boson, MyYellow, ultra thick] (ib),
                (ib) -- [boson, MyYellow, ultra thick] (f1),
                };
        \end{feynman}
    \end{tikzpicture}}}
    +
    \vcenter{\hbox{\begin{tikzpicture}[scale=0.8]
        \begin{feynman}
            \vertex (i) at (0,0);
            \vertex (e) at (0,2.0);
            \node[circle, draw=Orange, fill = Orange, scale=0.5, label=left:$\lambda^E$] (w1) at (0, 1.0);
            \vertex (fs) at (1.7,0.25) {};
             \vertex (tfs) at (1.2,0.6);
            \vertex[label= left:$m$] (w2) at (0,0.5);
            \vertex (f1) at (0.75, 0.8);
             \vertex[label= left:$m$] (w3) at (0,0.0);

            \diagram*{
                (i) -- [double, double, thick] (w1),
                (w1) -- [dashed, double, double, thick] (e),
                (tfs) -- [boson,MyYellow, ultra thick] (w1),
                (w2) -- [boson, MyYellow, ultra thick] (f1),
                (w3) -- [boson, MyYellow, ultra thick] (tfs),
                (tfs) -- [boson,MyYellow, ultra thick] (fs),
                };
        \end{feynman}
    \end{tikzpicture}}}
    \Bigg|^2 \\
    & \sim \Bigg| \vcenter{\hbox{\begin{tikzpicture}[scale=0.8]
        \begin{feynman}
            \vertex (i) at (0,0);
            \vertex (e) at (0,2.0);
            \node[circle, draw=Orange, fill = Orange, scale=0.5, label=left:$\lambda^E$] (w1) at (0, 1.0);
            \vertex (fs) at (1.5,0.2) {}; 

            \diagram*{
                (i) -- [double, double, thick] (w1),
                (w1) -- [dashed, double, double, thick] (e),
                (fs) -- [boson,MyYellow, ultra thick] (w1)
            };
        \end{feynman}
    \end{tikzpicture}}} 
    \Bigg|^2 \times \[ - \frac{428}{105} (G M \omega)^2 \log\(\frac{\omega}{\mu_0}\) \]
    \end{aligned}
\end{equation}

\begin{equation}
       \begin{aligned}
        & \quad \frac{1}{2!} \Bigg|
\vcenter{\hbox{\begin{tikzpicture}[scale=0.8]
        \begin{feynman}
            \vertex (i) at (0,0);
            \vertex (e) at (0,2.0);
            \node[circle, draw=Orange, fill = Orange, scale=0.5, label=left:$\lambda_{\omega^2}^E$] (w1) at (0, 1.0);
            \vertex (fs) at (1.5,0.2) {}; 

            \diagram*{
                (i) -- [double, double, thick] (w1),
                (w1) -- [dashed, double, double, thick] (e),
                (fs) -- [boson,MyYellow, ultra thick] (w1)
            };
        \end{feynman}
    \end{tikzpicture}}} 
    + 
         \vcenter{\hbox{\begin{tikzpicture}[scale=0.8]
        \begin{feynman}
            \vertex (i) at (0,0);
            \vertex (e) at (0,2.0);
            \node[circle, draw=Orange, fill = Orange, scale=0.5, label=left:$\lambda_{\omega^2}^E$] (w1) at (0, 1.0);
            \vertex (fs) at (1.5,0.2) {};
            \vertex[label= left:$m$] (w2) at (0,0.25);
            \vertex (f1) at (0.75, 0.8) {};

            \diagram*{
                (i) -- [double, double, thick] (w1),
                (w1) -- [dashed, double, double, thick] (e),
                (fs) -- [boson,MyYellow, ultra thick] (w1),
                (w2) -- [boson, MyYellow, ultra thick] (f1),
            };
        \end{feynman}
    \end{tikzpicture}}}
    +
    \vcenter{\hbox{\begin{tikzpicture}[scale=0.8]
        \begin{feynman}
            \vertex (i) at (0,0);
            \vertex (e) at (0,2.0);
            \node[circle, draw=Orange, fill = Orange, scale=0.5, label=left:$\lambda_{\omega^2}^E$] (w1) at (0, 1.0);
            \vertex (fs) at (1.5,0.2) {};
            \vertex[label= left:$m$] (w2) at (0,0.5);
            \vertex (f1) at (0.75, 0.8) {};
             \vertex[label= left:$m$] (w3) at (0,0.0);

            \diagram*{
                (i) -- [double, double, thick] (w1),
                (w1) -- [dashed, double, double, thick] (e),
                (fs) -- [boson,MyYellow, ultra thick] (w1),
                (w2) -- [boson, MyYellow, ultra thick] (f1),
                (w3) -- [boson, MyYellow, ultra thick] (f1),
            };
        \end{feynman}
    \end{tikzpicture}}}
    +
    \vcenter{\hbox{\begin{tikzpicture}[scale=0.8]
        \begin{feynman}
            \vertex (i) at (0,0);
            \vertex (e) at (0,2.0);
            \node[circle, draw=Orange, fill = Orange, scale=0.5, label=left:$\lambda_{\omega^2}^E$] (w1) at (0, 1.0);
            \vertex (fs) at (1.5,0.2) {};
            \vertex[label= left:$m$] (w2) at (0,0.5);
            \vertex (f1) at (0.75, 0.8) {};
             \vertex[label= left:$m$] (w3) at (0,0.0);
             \vertex (ib) at (0.4,0.45);

            \diagram*{
                (i) -- [double, double, thick] (w1),
                (w1) -- [dashed, double, double, thick] (e),
                (fs) -- [boson,MyYellow, ultra thick] (w1),
                (w2) -- [boson, MyYellow, ultra thick] (ib),
                (w3) -- [boson, MyYellow, ultra thick] (ib),
                (ib) -- [boson, MyYellow, ultra thick] (f1),
                };
        \end{feynman}
    \end{tikzpicture}}}
    +
    \vcenter{\hbox{\begin{tikzpicture}[scale=0.8]
        \begin{feynman}
            \vertex (i) at (0,0);
            \vertex (e) at (0,2.0);
            \node[circle, draw=Orange, fill = Orange, scale=0.5, label=left:$\lambda_{\omega^2}^E$] (w1) at (0, 1.0);
            \vertex (fs) at (1.7,0.25) {};
             \vertex (tfs) at (1.2,0.6);
            \vertex[label= left:$m$] (w2) at (0,0.5);
            \vertex (f1) at (0.75, 0.8);
             \vertex[label= left:$m$] (w3) at (0,0.0);

            \diagram*{
                (i) -- [double, double, thick] (w1),
                (w1) -- [dashed, double, double, thick] (e),
                (tfs) -- [boson,MyYellow, ultra thick] (w1),
                (w2) -- [boson, MyYellow, ultra thick] (f1),
                (w3) -- [boson, MyYellow, ultra thick] (tfs),
                (tfs) -- [boson,MyYellow, ultra thick] (fs),
                };
        \end{feynman}
    \end{tikzpicture}}}
    \Bigg|^2 \\
    & \sim \frac{1}{2!} \Bigg| \vcenter{\hbox{\begin{tikzpicture}[scale=0.8]
        \begin{feynman}
            \vertex (i) at (0,0);
            \vertex (e) at (0,2.0);
            \node[circle, draw=Orange, fill = Orange, scale=0.5, label=left:$\lambda_{\omega^2}^E$] (w1) at (0, 1.0);
            \vertex (fs) at (1.5,0.2) {}; 

            \diagram*{
                (i) -- [double, double, thick] (w1),
                (w1) -- [dashed, double, double, thick] (e),
                (fs) -- [boson,MyYellow, ultra thick] (w1)
            };
        \end{feynman}
    \end{tikzpicture}}} 
    \Bigg|^2 \times \[ - \frac{428}{105} (G M \omega)^2 \log\(\frac{\omega}{\mu_0}\) \] \\
    & \sim \frac{1}{2!} \Bigg| \vcenter{\hbox{\begin{tikzpicture}[scale=0.8]
        \begin{feynman}
            \vertex (i) at (0,0);
            \vertex (e) at (0,2.0);
            \node[circle, draw=Orange, fill = Orange, scale=0.5, label=left:$\lambda^E$] (w1) at (0, 1.0);
            \vertex (fs) at (1.5,0.2) {}; 

            \diagram*{
                (i) -- [double, double, thick] (w1),
                (w1) -- [dashed, double, double, thick] (e),
                (fs) -- [boson,MyYellow, ultra thick] (w1)
            };
        \end{feynman}
    \end{tikzpicture}}} 
    \Bigg|^2 \times \[ - \frac{428}{105} (G M \omega)^2 \log\(\frac{\omega}{\mu_0}\) \]^2 \\
    & \cdots \cdots 
    \end{aligned}
\end{equation}
In the above equation, $1/(2!)$ is the symmetry factor. 
One can iterate the above corrections further to get
\begin{equation}
    \begin{aligned}
        & \quad \frac{1}{n!}\Bigg|
\vcenter{\hbox{\begin{tikzpicture}[scale=0.8]
        \begin{feynman}
            \vertex (i) at (0,0);
            \vertex (e) at (0,2.0);
            \node[circle, draw=Orange, fill = Orange, scale=0.5, label=left:$\lambda_{\omega^{2n}}^E$] (w1) at (0, 1.0);
            \vertex (fs) at (1.5,0.2) {}; 

            \diagram*{
                (i) -- [double, double, thick] (w1),
                (w1) -- [dashed, double, double, thick] (e),
                (fs) -- [boson,MyYellow, ultra thick] (w1)
            };
        \end{feynman}
    \end{tikzpicture}}} 
    + 
         \vcenter{\hbox{\begin{tikzpicture}[scale=0.8]
        \begin{feynman}
            \vertex (i) at (0,0);
            \vertex (e) at (0,2.0);
            \node[circle, draw=Orange, fill = Orange, scale=0.5, label=left:$\lambda_{\omega^{2n}}^E$] (w1) at (0, 1.0);
            \vertex (fs) at (1.5,0.2) {};
            \vertex[label= left:$m$] (w2) at (0,0.25);
            \vertex (f1) at (0.75, 0.8) {};

            \diagram*{
                (i) -- [double, double, thick] (w1),
                (w1) -- [dashed, double, double, thick] (e),
                (fs) -- [boson,MyYellow, ultra thick] (w1),
                (w2) -- [boson, MyYellow, ultra thick] (f1),
            };
        \end{feynman}
    \end{tikzpicture}}}
    +
    \vcenter{\hbox{\begin{tikzpicture}[scale=0.8]
        \begin{feynman}
            \vertex (i) at (0,0);
            \vertex (e) at (0,2.0);
            \node[circle, draw=Orange, fill = Orange, scale=0.5, label=left:$\lambda_{\omega^{2n}}^E$] (w1) at (0, 1.0);
            \vertex (fs) at (1.5,0.2) {};
            \vertex[label= left:$m$] (w2) at (0,0.5);
            \vertex (f1) at (0.75, 0.8) {};
             \vertex[label= left:$m$] (w3) at (0,0.0);

            \diagram*{
                (i) -- [double, double, thick] (w1),
                (w1) -- [dashed, double, double, thick] (e),
                (fs) -- [boson,MyYellow, ultra thick] (w1),
                (w2) -- [boson, MyYellow, ultra thick] (f1),
                (w3) -- [boson, MyYellow, ultra thick] (f1),
            };
        \end{feynman}
    \end{tikzpicture}}}
    +
    \vcenter{\hbox{\begin{tikzpicture}[scale=0.8]
        \begin{feynman}
            \vertex (i) at (0,0);
            \vertex (e) at (0,2.0);
            \node[circle, draw=Orange, fill = Orange, scale=0.5, label=left:$\lambda_{\omega^{2n}}^E$] (w1) at (0, 1.0);
            \vertex (fs) at (1.5,0.2) {};
            \vertex[label= left:$m$] (w2) at (0,0.5);
            \vertex (f1) at (0.75, 0.8) {};
             \vertex[label= left:$m$] (w3) at (0,0.0);
             \vertex (ib) at (0.4,0.45);

            \diagram*{
                (i) -- [double, double, thick] (w1),
                (w1) -- [dashed, double, double, thick] (e),
                (fs) -- [boson,MyYellow, ultra thick] (w1),
                (w2) -- [boson, MyYellow, ultra thick] (ib),
                (w3) -- [boson, MyYellow, ultra thick] (ib),
                (ib) -- [boson, MyYellow, ultra thick] (f1),
                };
        \end{feynman}
    \end{tikzpicture}}}
    +
    \vcenter{\hbox{\begin{tikzpicture}[scale=0.8]
        \begin{feynman}
            \vertex (i) at (0,0);
            \vertex (e) at (0,2.0);
            \node[circle, draw=Orange, fill = Orange, scale=0.5, label=left:$\lambda_{\omega^{2n}}^E$] (w1) at (0, 1.0);
            \vertex (fs) at (1.7,0.25) {};
             \vertex (tfs) at (1.2,0.6);
            \vertex[label= left:$m$] (w2) at (0,0.5);
            \vertex (f1) at (0.75, 0.8);
             \vertex[label= left:$m$] (w3) at (0,0.0);

            \diagram*{
                (i) -- [double, double, thick] (w1),
                (w1) -- [dashed, double, double, thick] (e),
                (tfs) -- [boson,MyYellow, ultra thick] (w1),
                (w2) -- [boson, MyYellow, ultra thick] (f1),
                (w3) -- [boson, MyYellow, ultra thick] (tfs),
                (tfs) -- [boson,MyYellow, ultra thick] (fs),
                };
        \end{feynman}
    \end{tikzpicture}}}
    \Bigg|^2 \\
    & \sim \frac{1}{n!} \Bigg| \vcenter{\hbox{\begin{tikzpicture}[scale=0.8]
        \begin{feynman}
            \vertex (i) at (0,0);
            \vertex (e) at (0,2.0);
            \node[circle, draw=Orange, fill = Orange, scale=0.5, label=left:$\lambda^E$] (w1) at (0, 1.0);
            \vertex (fs) at (1.5,0.2) {}; 

            \diagram*{
                (i) -- [double, double, thick] (w1),
                (w1) -- [dashed, double, double, thick] (e),
                (fs) -- [boson,MyYellow, ultra thick] (w1)
            };
        \end{feynman}
    \end{tikzpicture}}} 
    \Bigg|^2 \times \[ - \frac{428}{105} (G M \omega)^2 \log\(\frac{\omega}{\mu_0}\) \]^{2n}
    \end{aligned}
\end{equation}
Summing all these diagrams together, we get the leading logarithmic RG running behavior (log from 2-loops) in the absorption probability
\begin{equation}\label{eq: leading log resum}
    \sim  \Bigg| \vcenter{\hbox{\begin{tikzpicture}[scale=0.8]
        \begin{feynman}
            \vertex (i) at (0,0);
            \vertex (e) at (0,2.0);
            \node[circle, draw=Orange, fill = Orange, scale=0.5, label=left:$\lambda^E$] (w1) at (0, 1.0);
            \vertex (fs) at (1.5,0.2) {}; 

            \diagram*{
                (i) -- [double, double, thick] (w1),
                (w1) -- [dashed, double, double, thick] (e),
                (fs) -- [boson,MyYellow, ultra thick] (w1)
            };
        \end{feynman}
    \end{tikzpicture}}} 
    \Bigg|^2 \times \( \frac{\omega}{\mu_0}\)^{- \frac{428}{105}(G M \omega)^2} ~.
\end{equation}
The above explicit resummation can also be written in a more general form of renormalized non-local (NL) retarded Green's function \cite{Goldberger:2009qd,Goldberger:2020geb} that accounts for the dissipations. We parametrize the bare retarded Green function in terms of the renormalized (physical) Green function as
\begin{equation}
   \langle Q_{ij} Q_{kl} \rangle_{\rm ret,Bare}^{\rm NL}(\omega) = Z (\omega,\mu)^2 \langle Q_{ij} Q_{kl} \rangle_{\rm ret}^{\rm NL}(\omega,\mu) ~.
\end{equation}
where we have introduced the matching scale $\mu$. By calculating the same diagram in Eq.~\eqref{eq:abs log div} with $\lambda^E$ replaced by $Q^E$, we can get the renormalization constant in $\overline{\rm MS}$ scheme as
\begin{equation}
    Z^{\overline{\rm MS}} = 1 + \frac{107}{105} (G M \omega)^2 \times \[ \frac{1}{(d-4)_{\rm UV}} + \gamma_E - \log (4\pi)\] ~,
\end{equation}
which absorbs the UV divergence in dimensional regularization. 
By requiring the physical absorption probability to be independent on the sliding scale $\mu$, we get the $\beta$ function
\begin{eBox}
\begin{equation}
    \frac{d {}_2\Gamma_{\ell m}}{d\log \mu } =0 \quad \Rightarrow \frac{d}{d \log \mu} \langle Q_{ij} Q_{kl} \rangle_{\rm ret}^{\rm NL}(\omega,\mu) = - \frac{428}{105} (G M \omega)^2\langle Q_{ij} Q_{kl} \rangle_{\rm ret}^{\rm NL}(\omega,\mu) ~.
\end{equation}
\end{eBox}
The solution to this RG equation perfectly agrees the leading log (2-loop) resummation we see in Eq.~\eqref{eq: leading log resum}. 
For general compact objects, 
the same set of diagrams should generate RG running in the conservative sector, which will be proportional to the tree-level Love number Wilson coefficients, see Ref.~\cite{Mandal:2023hqa} for an explicit calculation.
However, for black holes they vanish due to the vanishing of static Love numbers, 
and hence the RG running of their Love numbers
will appear only at the 6-loop
order. 

To summarize, the \purple{purple} logarithmic terms in the BHPT phase shifts Eq.~\eqref{eq: 22 abs} and Eq.~\eqref{n21} correspond to the RG running of 
the non-local retarded Green function at the 2-loop order.

\subsection{RG Running of Dynamical Love Numbers}
\label{subsec: RG elastic}

The BHPT results in Sec.~\ref{subsec:BHPT phase shift} contain another type of logarithmic dependence in the elastic phase shifts given in Eqs.~\eqref{eq: 22 elastic},~\eqref{eq: 21 elastic} and~\eqref{eq: 20 elastic}. In 
what follows, we will first match the RG running of the appropriate 
beta function coefficients by comparing the logarithms and then discuss the origin of the associated divergences.

As we have discussed in Sec.~\ref{subsec:MST}, the near-far factorization guarantees that the logarithmic terms will only appear in the near zone part, except for those arising from the loop corrections to the scattering off long-range Newtonian potential (tails). 
Thus, the matching of the logarithmic piece for the conservative effect is similar to what we have done for the static tidal Love numbers in Sec.~\ref{subsec:match love and dissipation}. Consider now the same diagram as in Eq.~\eqref{eq: NLO diagram}, but focus on the conservative sector. 
Matching to the BHPT elastic phase shift in Eq.~\eqref{eq: 22 elastic},~\eqref{eq: 21 elastic} and Eq.~\eqref{eq: 20 elastic}. We get
the following $\beta$ functions
\begin{eBox}
\begin{alignat}{3}
\label{eq: RG of DTLNs 1}
\frac{\partial \Lambda_{\hat{s}^1, \omega}^{E/B}}{\partial \log(\mu)} = -\frac{32}{45} (\chi + 3 \chi^3), \quad \frac{\partial \Lambda_{\hat{s}^3, \omega}^{E/B}}{\partial \log(\mu)} = \frac{8}{3} \chi^3  ~.
\end{alignat}
\end{eBox}
Similarly, we can also consider the scattering phase shift
contribution at $(r_s \omega)^7$ order from the diagram
\begin{equation}
      \vcenter{\hbox{\begin{tikzpicture}[scale=0.7]
        \begin{feynman}
            \vertex (i) at (0,0);
            \vertex (e) at (0,3);
            \node[circle, draw=Orange, fill = Orange, scale=0.5] (w1) at (0, 1.0);
            \node[circle, draw=Orange, fill = Orange, scale=0.5] (w2) at (0, 2.0);
            \vertex (f1) at (1.5,2.8) {};
            \vertex (fs) at (1.5,0.2) {};

            \diagram*{
                (i) -- [double, double, thick] (w1),
                (w1) -- [dashed, Green, ultra thick,edge label = $\lambda_{\omega^2}^{E/B}$] (w2),
                (w2) -- [double, double, thick] (e),
                (f1) -- [boson, MyYellow, ultra thick] (w2),
                (fs) -- [boson, MyYellow, ultra thick] (w1)
            };
        \end{feynman}
    \end{tikzpicture}}} ~.
\end{equation}
With similar methods we developed in Sec.~\ref{subsec:match love and dissipation} and \ref{subsec:match NLO dissipation}, we can get the RG running for $\Lambda_{\hat{s}^0, \omega^2}, \Lambda_{\hat{s}^2, \omega^2}$ and $\Lambda_{\hat{s}^4, \omega^2}$
\begin{eBox}
\begin{equation}
\label{eq : NNLO elastic RG}
    \begin{aligned}
        \frac{\partial \Lambda_{\hat{s}^0, \omega^2}^{E/B}}{\partial \log \mu} & = \frac{32}{405}   \left(9 +97 \chi ^2 -6 \chi ^4\right) ~, \\
        \frac{\partial \Lambda_{\hat{s}^2, \omega^2}^{E/B} }{\partial \log \mu} & = - \frac{16}{27}  \left(\chi ^2+23\right) \chi ^2 ~,  \\
        \frac{\Lambda_{\hat{s}^4}^{E/B}}{\partial \log \mu} & = \frac{64}{27}  \chi ^4 ~.
    \end{aligned}
\end{equation}
\end{eBox}
As an important check of our results, we can now take the Schwarzschild limit $\chi=0$, and get
\begin{equation}
\label{eq:Schw DLN RG}
    \frac{\partial \Lambda_{\hat{s}^0, \omega^2}^{E/B}}{\partial \log \mu} = \frac{32}{45} ~,
\end{equation}
which perfectly agrees with the off-shell graviton one-point function
calculations of \cite{Chakrabarti:2013lua} upon
taking into account the difference in our conventions. 

It is worth noting that the numerical value of the
Wilson coefficient of the $\int d\tau \dot{E}^2$ operator in Eq.~\eqref{eq : NNLO elastic RG} depends on the black hole spin. It is important to distinguish such implicit dependencies of magnitudes of Wilson coefficients on spin
from spin-dependent expressions generated 
directly by relevant tensors, c.f.~\eqref{eq:fivebasis}. 
One should keep this
subtlety in mind when interpreting the MST solution as a function of spin
in the context of effective field theory.

The logarithmic running of dynamical Love numbers
is generated by UV divergences of the GR loops. 
For the sake of simplicity, let us focus on the Schwarzschild case. As explained in Sec.~\ref{sec:EFT}, Eq.~\eqref{eq : countercons}, the conservative effect can be represented as a local contact term on the worldline. When expressed in the language of diagrams, this suggests that the RG running of $\Lambda_{\omega^2}^{E/B}$ can be interpreted as the logarithmic dependence of the following diagram
\begin{equation}
    \vcenter{\hbox{\begin{tikzpicture}[scale=0.7]
        \begin{feynman}
            \vertex (i) at (0,0);
            \vertex (e) at (0,3);
            \node[circle, draw=repGreen, fill = repGreen, scale=1, label=left:$\Lambda_{\omega^2}^{E/B}$] (w1) at (0, 1.5);
            \vertex(f1) at (2.0,2.8);
            \vertex (fs) at (2.0,0.2); 

            \diagram*{
                (i) -- [double, double, thick] (w1),
                (w1) -- [double, double, thick] (e),
                (f1) -- [boson, MyYellow, ultra thick] (w1),
                (fs) -- [boson, MyYellow, ultra thick] (w1)
            };
        \end{feynman}
    \end{tikzpicture}}} ~.
\end{equation}
Unlike the logarithms discussed in the RG flow of the dissipation numbers, this logarithms cannot be understood as loop corrections to the lower order tidal scatterings, since they vanish identically. In the context of BHPT, we have rigorously verified that varying the boundary conditions at the event horizon does not alter the RG running described by Eq.~\eqref{eq: RG of DTLNs 1}. Therefore, this logarithms is insensitive to the lower order tidal coefficients.
But we find that this can act as a counter term for UV divergences in the scattering amplitude due to the gravitational non-linearities, e.g.
\begin{equation}
               i \mathcal{M}(\boldsymbol{k}_{\rm in}\rightarrow \boldsymbol{k}_{\rm out},h \rightarrow h) =  \begin{tikzpicture}[baseline={([yshift=-0.4 ex]current bounding box.center)}]
        \begin{feynman}
            \vertex (i) at (0,0);
            \vertex (e) at (0,3);
            \vertex[dot, MyYellow,label=left:$m$] (f1) at (0,0.3) {};
            \vertex[dot, MyYellow,label=left:$m$] (f2) at (0,0.7) {};
            \vertex[dot, MyYellow,label=left:$m$] (f3) at (0,1.1) {};
            \vertex[dot, MyYellow,label=left:$m$] (f4) at (0,1.5) {};
            \vertex[dot, MyYellow,label=left:$m$] (f5) at (0,1.9) {};
            \vertex[dot, MyYellow,label=left:$m$] (f6) at (0,2.3) {};
            \vertex[dot, MyYellow,label=left:$m$] (f7) at (0,2.7) {};
            \vertex (c1) at (1.0,1.5);
            \vertex (c3) at (3.0,3.0);
            \vertex (c4) at (3.0,0.0);
            \vertex (c2) at (2.0,1.5);

            \diagram*{
                (i) -- [double, double, thick] (e),
                (f1) -- [boson, MyYellow, ultra thick] (c1),
                (f2) -- [boson, MyYellow, ultra thick] (c1),
                (f3) -- [boson, MyYellow, ultra thick] (c1),
                (f4) -- [boson, MyYellow, ultra thick] (c1), 
                (f5) -- [boson, MyYellow, ultra thick] (c1), 
                (f6) -- [boson, MyYellow, ultra thick] (c1), 
                (f7) -- [boson, MyYellow, ultra thick] (c1), 
                (c2) -- [boson, MyYellow, ultra thick] (c3),
                (c2) -- [boson, MyYellow, ultra thick] (c4),
                (c1) -- [boson, MyYellow, ultra thick] (c2)
            };
        \end{feynman}
    \end{tikzpicture}
    + \text{other diagrams at 6-loop order} ~,
\end{equation}
According to the RG running computed in Eq.~\eqref{eq:Schw DLN RG}, the coefficient in from the log from the combination of all 
of such diagrams should be 
\begin{equation}
    i \mathcal{M}^{\rm EFT}_{\rm 6-loop}|_{\rm log}(\boldsymbol{k}_{\rm in}\rightarrow \boldsymbol{k}_{\rm out},h \rightarrow h) = i \pi \frac{512}{45} (G M )^7 \omega^6 \log\(\frac{\omega}{\mu}\) \cos^4\(\frac{\theta}{2}\) ~,
\end{equation}
where $\mu$ is the matching scale. 
We leave the explicit calculation of these 6-loop diagrams 
for future work.

We would like to finish with an interesting observation that 
the RG running observed here is universal in the sense that it comes from the non-linearity of GR. 
Thus, it should produce the same RG running of dynamical 
Love numbers for all spherically symmetric compact object at the $G^7$ order. 
In~\cite{Chakrabarti:2013lua}, the authors also tried to calculate the dynamical response of a non-rotating Neutron star, see their Eq.~(15), but miss this universal relation. It would be interesting to explore this generality further.

\section{Quadrupole-Octupole Mixing}
\label{sec:mix}

In this section, we shall study the Quadrupole-Octupole mixing effects in the tidal reponse function that originate from the breaking of the spherical symmetry by Kerr BHs. This is a particular examples of the general
phenomena of the spherical-spheroidal mixing. 

\subsection{Imposing Spheroidal Separability and Matching}

As mentioned at the end of Sec.~\ref{subsec:BHPT phase shift}, the scattering phase shifts in BHPT are given in spheroidal harmonic basis because the presence of spin breaks spherical symmetry. 
Thus, there will be mixing terms in the scattering amplitude that transform the $\ell=2$ states to $\ell = 3$ states and vice-versa as we have discussed in Sec.~\ref{subsec: tidal coefficients}. However, since the Teukolsky equation is separable in the spheroidal basis, we need to impose the same spheroidal separability in the EFT. To do that, we can construct a new basis using the relation between the spherical and spheroidal harmonics at order $a\omega=  G M \omega \chi =r_s\omega \chi/2$ given by
\begin{alignat}{3}
& {}_{- 2}S_\ell^m(\cos\theta, a\omega) = {}_{- 2}Y_{\ell}^m(\cos\theta) + (a\omega)\Bigg[\frac{2\sqrt{(\ell+1)^2-4}\sqrt{(\ell+1)^2-m^2}}{(\ell+1)^2\sqrt{2(\ell+1)-1}\sqrt{2(\ell+1)+1}}{}_{- 2}Y_{\ell+1}^m(\cos\theta)\nnm \\&-\frac{2\sqrt{\ell^2-4}\sqrt{\ell^2-m^2}}{\ell^2\sqrt{2\ell-1}\sqrt{2\ell+1}}{}_{-2}Y_{\ell-1}^m(\cos\theta)\Bigg] +\mathcal{O}\Big((a\omega)^2\Big) ~.
\end{alignat}
Recall that the state $|\omega,\ell,m,h\rangle$ has the angular wavefunction $\sim{}_{-h}Y_{\ell}^{m}(\cos\theta)$, therefore we can define a new orthogonal spheroidal basis as 
\begin{alignat}{3}
\nnm |\omega,\ell,m,h \rangle_{\mr{sp}} =  |\omega,\ell,m,h \rangle +  \frac{h}{2} (a\omega) \Bigg[& \frac{2\sqrt{(\ell+1)^2-4}\sqrt{(\ell+1)^2-m^2}}{(\ell+1)^2\sqrt{2(\ell+1)-1}\sqrt{2(\ell+1)+1}}|\omega,\ell+1,m,h\rangle \\& -\frac{2\sqrt{\ell^2-4}\sqrt{\ell^2-m^2}}{\ell^2\sqrt{2\ell-1}\sqrt{2\ell+1}}|\omega,\ell-1,m,h\rangle\Bigg]+\mathcal{O}\Big((a\omega)^2\Big),
\label{eq: somethingelse}
\end{alignat}
where we are using the subscript `$\text{sp}$' to denote the new spheroidal basis states with definite helicity. It turns out more convenient to work in the basis with definite parities. We can transform the above equation into the parity basis using Eq.~\eqref{eq: h2P} in App.~\ref{app : single particle} and get
\begin{alignat}{3}
\nnm|\omega,\ell,m,P\rangle_{\text{sp}} = |\omega,\ell,m,P\rangle + (a\omega) \Bigg[& \frac{2\sqrt{(\ell+1)^2-4}\sqrt{(\ell+1)^2-m^2}}{(\ell+1)^2\sqrt{2(\ell+1)-1}\sqrt{2(\ell+1)+1}}|\omega,\ell+1,m,P\rangle \\& -\frac{2\sqrt{\ell^2-4}\sqrt{\ell^2-m^2}}{\ell^2\sqrt{2\ell-1}\sqrt{2\ell+1}}|\omega,\ell-1,m,P\rangle\Bigg]+\mathcal{O}\Big((a\omega)^2\Big).
\end{alignat}
In the worldline EFT, spheroidal separability can be imposed simply be demanding that the S-matrix be diagonal in the spheroidal basis. Restricting to leading order mixing between quadrupolar ($\ell=2$) and octupolar ($\ell=3$) sectors, this means that
\begin{equation}
{}_{\rm sp}\langle \omega,2,m,P|T|\omega, 3,m',P'\rangle_{\text{sp}} = {}_{\rm sp}\langle \omega,3,m,P| T |\omega,2,m',P'\rangle_{\text{sp}} = 0,
\end{equation}
which implies
\begin{equation}
\langle \omega,2,m,P| T |\omega,3,m',P' \rangle =  \langle \omega,3,m,P| T |\omega,2,m',P' \rangle = (a\omega) \frac{2\sqrt{9-m^2}}{9\sqrt{7}}\langle \omega,2,m,P| T |\omega,2,m',P'\rangle ~.
\label{eq: sph cond}
\end{equation}

In the following, we will show that these relations can be used to fix the tidal mixing coefficients in $\nu^{E/B}$ and $\xi^{E/B}$ we introduced in Sec.~\ref{subsec: tidal coefficients}. The tidal tensor $(\nu^{E})^{ij}{}_{kl}$ relates the scattering process from octupolar sector ($\ell=3$) to the quadrupolar sector ($\ell=2$). and thus responsible for $\mathcal{A}(\omega,\ell=3,m,P\rightarrow \ell=2,m',P')$. Similar to the discussions in Sec.~\ref{subsec:match love and dissipation}, we introduce the retarded Green function 
\begin{equation}
     \vcenter{\hbox{\begin{tikzpicture}[scale=0.7]
        \begin{feynman}
            \node[circle, draw=Orange, fill = Orange, scale=0.5, label=left:$Q^{E}_{ij}$] (w1) at (0.0, 0.0);
            \node[circle, draw=Orange, fill = Orange, scale=0.5, label=right:$Q^{B}_{klm}$] (w2) at (2.0, 0.0);

            \diagram*{
                (w1) -- [dashed, blue, ultra thick] (w2),
            };
        \end{feynman}
    \end{tikzpicture}}}
    \equiv i \langle [ Q^{E}_{ij}(\tau) Q^{B}_{klm}(0)] \rangle \theta(\tau) ~.
\end{equation}
According to Eq.~\eqref{eq:KerrAns}, in the frequency domain, the above retarded Green function is just $-M (GM)^5 (\nu^E)_{ij\langle kl} \hat{s}_{m \rangle}$. Then we can evaluate the following Feynman diagram
\begin{equation}
  i \mathcal{M}(\boldsymbol{k}_{\rm in}\rightarrow \boldsymbol{k}_{\rm out},h \rightarrow h') = 
  \vcenter{\hbox{\begin{tikzpicture}[scale=0.7]
        \begin{feynman}
            \vertex (i) at (0,0);
            \vertex (e) at (0,3);
            \node[circle, draw=Orange, fill = Orange, scale=0.5, label = left:$Q^B_{klm}$] (w1) at (0, 1.0);
            \node[circle, draw=Orange, fill = Orange, scale=0.5, label = left:$Q^{E}_{ij}$] (w2) at (0, 2.0);
            \vertex (f1) at (1.5,2.8) {};
            \vertex (fs) at (1.5,0.2) {};

            \diagram*{
                (i) -- [double, double, thick] (w1),
                (w1) -- [dashed, Blue, ultra thick] (w2),
                (w2) -- [double, double, thick] (e),
                (f1) -- [boson, MyYellow, ultra thick] (w2),
                (fs) -- [boson, MyYellow, ultra thick] (w1)
            };
        \end{feynman}
    \end{tikzpicture}}}
    = -i M(GM)^5\frac{\omega^4}{4 M_{\rm pl}^2}\Bigg[\frac{h}{2} (\nu^E)^{ijkl}\hat{s}^m\epsilon^h_{(kl}(\boldsymbol{k}_{\rm in})k_{m)}{\epsilon}^{* h'}_{ij}(\boldsymbol{k}_{\rm out}) \Bigg] ~.
\end{equation}
We now switch to spherical basis states with definite parities
\begin{alignat}{3}
\label{eq : nu amp 322}
\mc{A}(\omega,\ell =3,m,P=+1 \rightarrow \ell=2,m,P=+1) &= -\frac{\sqrt{9-m^2}(GM\omega)^6}{4 G \sqrt{7}\pi M_{\rm pl}^2}  \Bigg\{ \tilde{\Lambda}_{s^0}^E + i\frac{1}{2}\tilde{H}_{s^1}^Em  \\& + \frac{1}{6}(m^2-4)\Big[-\tilde{\Lambda}_{s^2}^E-i  \tilde{H}_{s^3}^E + \tilde{\Lambda}_{s^4}^E (m^2-1)\Big]\Bigg\} ~.
\end{alignat}
Imposing the constraints in Eq.~\eqref{eq: sph cond} and comparing with $ \mathcal{A}(\omega, \ell=2,m,P=+\rightarrow \ell=2,m,P
 )$ derived from Eq.~\eqref{eq : helicity basis 222}
\begin{equation}
\begin{aligned}
\label{eq : lambda amp 222}
  \mathcal{A}(\omega, \ell=2,m,P=+1 \rightarrow \ell=2,m,P= + 1
 )  & = - \frac{\omega^5}{40M_{\rm pl}^2 \pi} M (G M)^4  \times \Bigg[\Lambda_{\hat{s}^0}^E + \frac{1}{2} i m H_{\hat{s}^1}^E  \\
 & \quad + \frac{1}{6}(m^2-4)\( - \Lambda_{\hat{s}^2}^E - i m H_{\hat{s}^3}^E + \Lambda_{\hat{s}^4}^E (m^2 - 1)\)\Bigg],
\end{aligned}
\end{equation}
we immediately get the following relation
\begin{eBox}
 \begin{alignat}{3}
\tilde{H}_{s^{1/3}}^E=\frac{2}{3}\chi H_{s^{1/3}}^E,~ \tilde{\Lambda}_{s^{0/2/4}}^E=\frac{2}{3}\chi\Lambda_{s^{0/2/4}}^E \implies (\nu^E)^{ij}{}_{kl}=\frac{2}{3}\chi(\lambda^E)^{ij}{}_{kl}.
 \label{eq: nuEfix}
\end{alignat}   
\end{eBox}
By comparing the above with the $P=-1$ scatterings, we can straightforwardly get
\begin{eBox}
\begin{alignat}{3}
 \tilde{H}_{s^{1/3}}^B= -\frac{2}{3}\chi H_{s^{1/3}}^B,~ \tilde{\Lambda}_{s^{0/2/4}}^B= - \frac{2}{3}\chi\Lambda_{s^{0/2/4}}^B \implies (\nu^B)^{ij}{}_{kl}=-\frac{2}{3}\chi(\lambda^B)^{ij}{}_{kl}.
  \label{eq: nuBfix}
\end{alignat}  
\end{eBox}
Similarly, we can also get the relation for $(\xi^{E/B})_{ijkl}$ tensors
\begin{eBox}
 \begin{alignat}{3}
{\Lambda'}^{B/E}_{s^i}=\tilde {\Lambda}^{E/B}_{s^i},~{H'}^{B/E}_{s^i}=\tilde{H}^{E/B}_{s^i} \implies (\xi^{B/E})^{ij}{}_{kl}=(\nu^{E/B})^{ij}{}_{kl}.
 \label{eq: xiB fix}
\end{alignat}   
\end{eBox}
The same results have been previously obtained in a more classical manner in \cite{Saketh:2022xjb}. 

A comment is in order. We have found that the 
LO quadrupole-octupole mixing terms 
are proportional to the static Love numbers.
This can be thought of as a consistency condition 
in the EFT. Since the static Love numbers of Kerr BHs 
vanish, the leading order conservative quadrupole-octupole coupling 
vanishes as well. 

\subsection{Comments on Higher Order Mixings}
\label{subsec: higher mixing}

At the end of Sec.~\ref{sec:BHPT}, we have briefly mentioned that all the results in Sec.~\ref{sec:matching} can be derived within spherical harmonics. This is because 
\begin{alignat}{3}
{}_{\rm sp}\langle \omega,2,m,P | T | \omega,2,m',P'\rangle_{\text{sp}} &=\langle \omega,2,m,P | T | \omega,2,m',P'\rangle\Bigg[1+\frac{4(9-m^2)}{567}\Bigg(\frac{\chi}{2}r_s\omega\Bigg)^2\Bigg] \\& +\mathcal{O}[(r_s\omega)^8], \nnm
\end{alignat}
where we have used the expansion for ${}_{-2}S_{2}^{m}(\cos\theta, a\omega)$ to $(a \omega)^2$ along with the relations in Eq.~\eqref{eq: sph cond}. Thus, we see that the $\ell=2\rightarrow 2$ amplitude remains unchanged to NLO at $(r_s \omega)^6$  regardless of whether we use 
the spherical or spheroidal basis. 
However, at $(r_s \omega)^7$ order, the constant pieces will be affected by the above corrections from spheroidal harmonics, but not the logarithms. Thus, all the RG running discussions in Sec.~\ref{sec:matching} will not be affected.

\section{Constant Parts of RG Running Couplings}
\label{sec:constants}

In this section, we discuss the constant pieces $\mathcal{A}_{\ell m}(\chi), \mathcal{B}_{\ell m}(\chi)$ and $\mathcal{C}_{\ell m}(\chi)$ for $\ell = 2, m = 0,1,2$ mentioned in Sec.~\ref{sec:BHPT} Eqs.~\eqref{eq: 22 abs}-\eqref{eq: 20 elastic}. Unlike the logarithmic terms which can be nicely interpreted as the RG runnings of dissipation numbers and dynamical Love numbers in Sec.~\ref{subsec:RGabs} and Sec.~\ref{subsec: RG elastic}, the physical meaning of these constant, scheme-dependent pieces is 
more uncertain. 
To match these terms, we need to carry out the appropriate  
EFT multi-loop calculations and match to the BHPT results 
in a given scheme. 
This goes beyond the scope  of this paper. 
An alternative would be to absorb the entire 
constant 
scattering contributions at a given order into the 
constant part of wordline couplings. 
This would simply be a particular 
scheme choice. In this section we try to follow this path, 
but find certain obstacles, mainly due to the mixing effects. 
With these caveats in mind, we proceed to the discussion of our results. 
We stress that main goal of this section is to estimate the constant, 
scheme-dependent parts of dynamical Love numbers and NNLO dissipation
numbers, and outline challenges that one should encounter 
in a rigorous matching of these quantities. 

We first list the analytic expression for $\mathcal{B}_{\ell m}(x)$
extracted from the analytic BHPT results, 
\begin{equation}
    \begin{aligned}
    \mathcal{B}_{22}(\chi) & = \frac{2}{225} \left(3 \chi ^2+1\right) \chi  \Re\left(\psi ^{(0)}\left(\frac{2 i \chi }{\sqrt{1-\chi ^2}}-2\right)\right)+\left(\frac{4 \gamma_E }{75}-\frac{313543499}{2939328000}\right) \chi ^3 \\
    & \quad +\left(\frac{4 \gamma_E }{225}-\frac{70741499}{653184000}\right) \chi -\frac{107 \chi  \zeta (3)}{630} ~,
   \end{aligned}
\end{equation}
\begin{equation}
    \begin{aligned}
    \mathcal{B}_{21}(\chi) & = \frac{7 \chi \left(4-3 \chi ^2\right) \Re\left(\psi ^{(0)}\left(\frac{i \chi }{\sqrt{1-\chi ^2}}-2\right)\right)}{6300}+\left(\frac{28979207}{734832000}-\frac{\gamma_E }{150}\right) \chi ^3 \\
    & \quad +\left(\frac{2 \gamma_E }{225}-\frac{70741499}{1306368000}\right) \chi - \frac{107}{1260} \chi \zeta(3) ~,
   \end{aligned}
\end{equation}
where the tail contribution has been removed by considering the time-reversal symmetry properties and the Sommerfeld enhancement factor \cite{Goldberger:2009qd}. We are yet to interpret the $\zeta(3)$ factor in the above expressions. 
Moreover, we stress that the RG running of dynamical Love numbers
implies the breakdown of the near-far factorization 
for the constant pieces, which means the 
near zone and far zone pieces for $\ell = 2$ shall mix, and therefore 
the far zone pieces at high PM 
order should be carefully taken into account.  
Thus, in general one  needs to consider the full expansion of the spheroidal harmonics in $(a\omega)$ to $(a\omega)^6$ order in the far zone part. We defer this task to future work.

\begin{figure}[ht!]
    \centering
    \subfloat[Fitting $\mathcal{A}_{22}(\chi)$]{\includegraphics[width=0.5\textwidth]{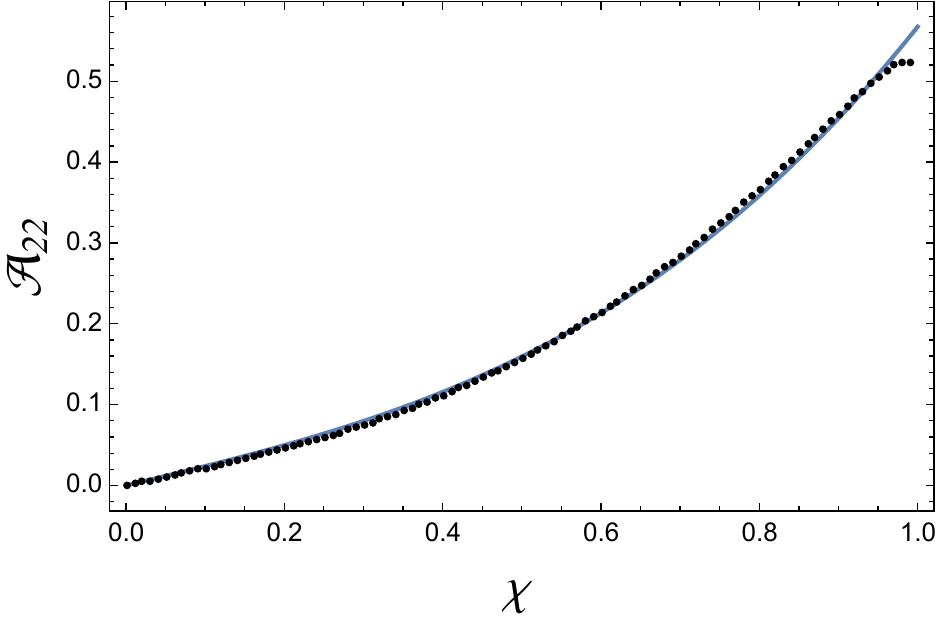}\label{fig:fitA22}}
    \hfill
    \subfloat[Fitting $\mathcal{A}_{21}(\chi)$]{\includegraphics[width=0.5\textwidth]{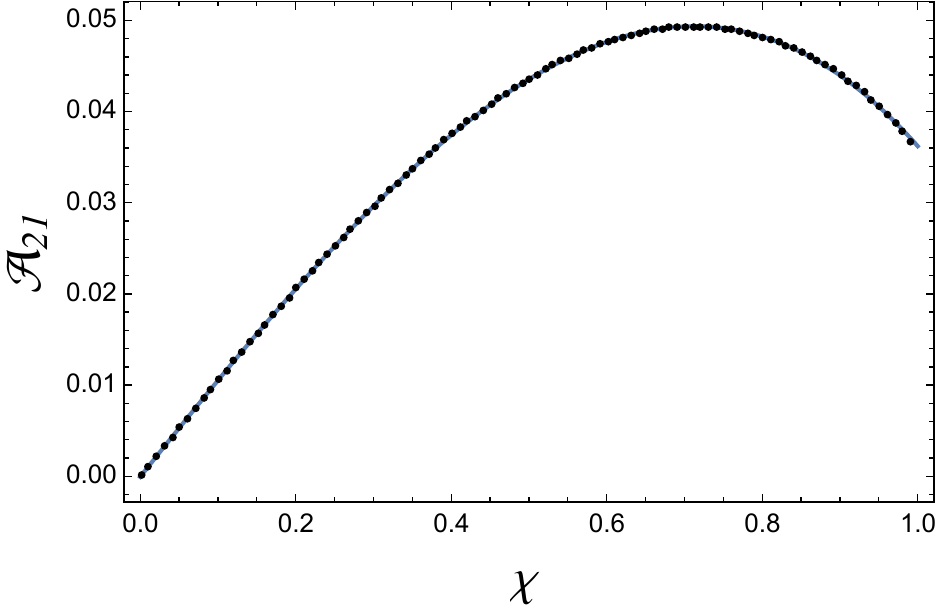}\label{fig:fitA21}}
    \caption{Numerical estimation of $\mathcal{A}_{22}$ and $\mathcal{A}_{21}$ as functions of $\chi$. Dotted lines are the numerical value with blue curves showing the fitting result.}
    \label{fig:fitA}
\end{figure}

\begin{figure}[ht!]
    \centering
    \subfloat[Fitting $\mathcal{C}_{22}(\chi)$]{\includegraphics[width=0.5\textwidth]{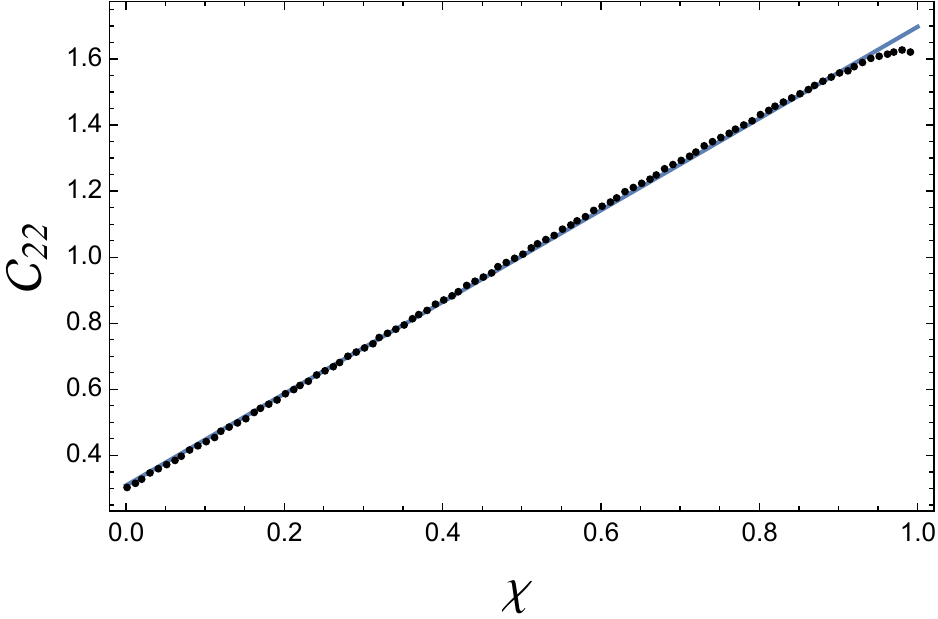}\label{fig:fitC22}}
    \hfill
    \subfloat[Fitting $\mathcal{C}_{21}(\chi)$]{\includegraphics[width=0.5\textwidth]{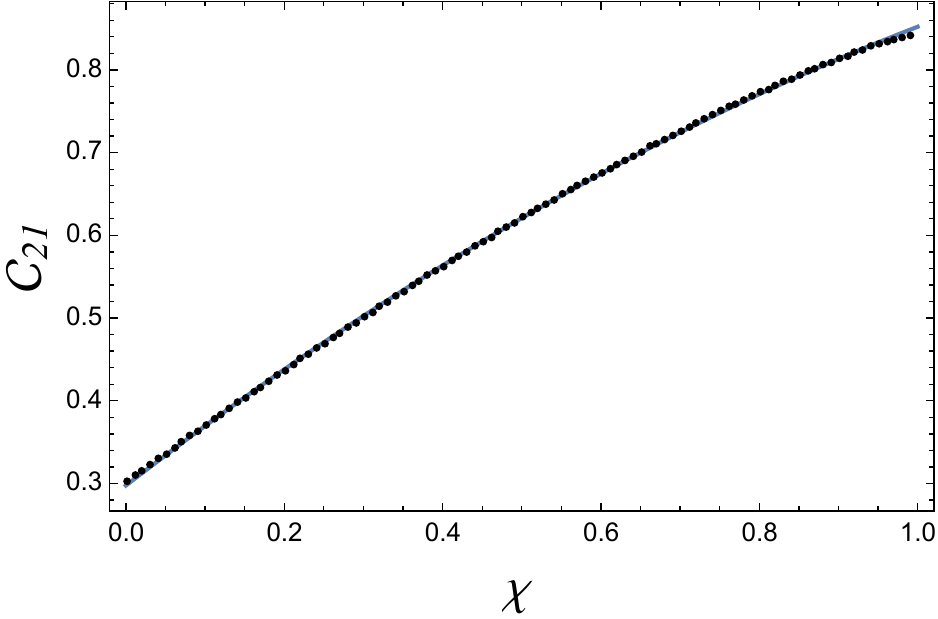}\label{fig:fitC21}}
    \hfill
    \subfloat[Fitting $\mathcal{C}_{20}(\chi)$]{\includegraphics[width=0.5\textwidth]{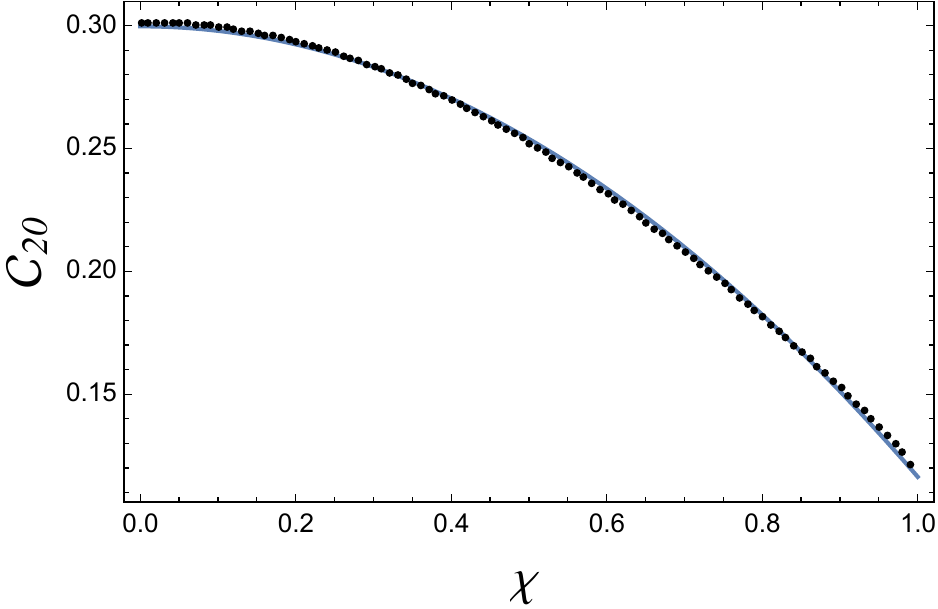}\label{fig:fitC20}}    
    \caption{Numerical estimation of $\mathcal{C}_{22}$, $\mathcal{C}_{21}$ and $\mathcal{C}_{20}$ as functions of $\chi$. Dotted lines are the numerical value with blue curves showing the fitting result.}
    \label{fig:fitC}
\end{figure}

For the constant piece of $\mathcal{A}_{\ell m}(\chi)$ the analytic 
expression can be easily obtained \cite{constants}, but it appears cumbersome and 
will not be presented here. Instead, we will present a numerical 
estimation in Fig.~\ref{fig:fitA}, which can be fit with a simple function
\begin{equation}
   \mathcal{A}_{22}(\chi) = 0.236\chi + 0.331 \chi^3  ~, \quad \mathcal{A}_{21}(\chi) =  0.104 \chi - 0.068 \chi^3  ~.
\end{equation}
This expression involves linear and cubic in spin terms which are consistent with the time-reversal symmetry discussed in Sec.~\ref{sec:EFT}. Indeed, at $(r_s \omega)^7$ order, the linear and cubic in 
spin terms correspond to dissipation.
Finally, we also obtain the numerical results for $\mathcal{C}_{\ell m}(\chi)$ in Fig.~\ref{fig:fitC},
which can be fit as
\begin{equation}
    \begin{aligned}
        \mathcal{C}_{22}(\chi) & = 1.389 \chi + 0.309 ~, \quad \mathcal{C}_{21}(\chi) =  0.739 \chi + 0.184 (1-\chi^2) + 0.114 ~, \\
        \mathcal{C}_{20}(\chi) & =  0.183 (1- \chi^2) + 0.116 ~.
    \end{aligned}
\end{equation}
Interestingly, we observe that for in the Schwarzschild limit $\chi=0$, $\mathcal{C}_{22},\mathcal{C}_{21}$ and $\mathcal{C}_{20}$ basically give the same number around $0.30$, which is consistent with the 
emergence of the spherical symmetry so that results for different $m$ modes become degenerate. 
However, we find it particular confusing to have the linear in spin term appear in $\mathcal{C}_{22}(\chi)$ and $\mathcal{C}_{21}(\chi)$. This term does not satisfy the time-reversal symmetry expected 
from the conservative tidal effects at $(r_s \omega)^7$ as we discussed in Sec.~\ref{sec:EFT}. One possible origin of such term will be the higher order tail effects contaminating the constant terms at this order. We leave a detailed study of this behavior for future work.

\section{Conclusions and Outlook}
\label{sec:con}

In this study, for the first time, we have taken a deep look at how Kerr BHs respond to changes in their gravitational environment at NLO and NNLO in frequency. 
We examined the dynamical tidal response 
in the framework of the worldline EFT. 
By comparing the scattering amplitudes in the EFT with phase shifts in BHPT, we have determined some key features of the black holes tidal response at higher orders in frequency, and to all order in the 
value of the BH spin. 
First of all, we reproduced the known results for the  
static tidal Love numbers, as well as LO and NLO tidal dissipation numbers. 
We also identified the leading order tail effects in the tidal response, using the explicit 1-loop calculation in the EFT. 
More interestingly, we scrutinized the RG running pattern of NNLO tidal dissipation numbers and dynamical tidal Love numbers. 
For dissipation numbers, we showed that the RG running can be understood as the UV divergence in the 2-loop corrections to the leading order absorptive tidal scattering diagrams in the EFT. 
For dynamical tidal Love numbers, we argue that the running should come from the UV divergence of higher order loop diagrams in the EFT due to non-linearities of GR. We explicitly 
identified the scheme-invariant beta functions 
of the NLO and NNLO local EFT couplings. 
Finally, we discussed the quadrupole-octupole mixing effects in the context of the Kerr tidal response.

The worldline EFT is generic, and the tidal response coefficients, once fixed, may be used for computing the dynamics of arbitrary 
compact objects. 
From a practical perspective, the most relevant system is an inspiraling binary, whose dynamics are typically studied in the post-Newtonian (PN) expansion.  In the following, we summarize all the relevant  coefficients for Kerr BHs and show the related PN order relevant for GW waveform modeling. Then we discuss possible future directions.

\subsection{Summary of All Tidal Coefficients}
In table~\ref{table: summary tidal number 1}, we show all the coefficients that are responsible for conservative tidal interactions and the related PN order in the equation of motion (EoM). The RG running in the conservative sectors start from the 6.5PN order.
\begin{table}[h]
  \centering
  \begin{threeparttable}
    \label{tab1:threshold}
    \begin{tabular*}{\textwidth}{l@{\extracolsep{\fill}}*{3}{c}}
    \toprule
      \textbf{Names} & \textbf{Notation} & \textbf{Explicit expression} & \textbf{PN order in EoM} \\
      \midrule
      \multirow{2}[0]{*}{TLNs} & $\Lambda_{\hat{s}^0}^{E} , \Lambda_{\hat{s}^2}^{E}, \Lambda_{\hat{s}^4}^{E}$ & \multirow{2}[0]{*}{Eq.~\eqref{eq: Kerr Love numbers}} & 5 \\
      & $\Lambda_{\hat{s}^0}^B, \Lambda_{\hat{s}^2}^B, \Lambda_{\hat{s}^4}^B$ & & 6 \\
      \midrule
      \multirow{4}[0]{*}{DTLNs} & $\Lambda_{\hat{s}^1,\omega}^{E},\Lambda_{\hat{s}^3,\omega}^{E}$ & \multirow{2}[0]{*}{Eq.~\eqref{eq: RG of DTLNs 1}} & 6.5 \\
      & $\Lambda_{\hat{s}^1,\omega}^B, \Lambda_{\hat{s}^3,\omega}^B$ & & 7.5 \\ 
      & $\Lambda_{\hat{s}^0,\omega^2}^{E} , \Lambda_{\hat{s}^2,\omega^2}^{E}, \Lambda_{\hat{s}^4,\omega^2}^{E}$ & \multirow{2}[0]{*}{Eq.~\eqref{eq : NNLO elastic RG}}& 8 \\
      & $\Lambda_{\hat{s}^0,\omega^2}^{B} , \Lambda_{\hat{s}^2,\omega^2}^{B}, \Lambda_{\hat{s}^4,\omega^2}^{B}$ & & 9 \\
      \midrule
      \multirow{2}[0]{*}{CTMNs} & $\tilde{\Lambda}^{E/B}_{\hat{s}^0}, \tilde{\Lambda}^{E/B}_{\hat{s}^2}, \tilde{\Lambda}^{E/B}_{\hat{s}^4}$ & Eq.~\eqref{eq: nuEfix}, Eq.~\eqref{eq: nuBfix} & \multirow{2}[0]{*}{6.5} \\
      &  ${\Lambda'}^{E/B}_{\hat{s}^0}, {\Lambda'}^{E/B}_{\hat{s}^2}, {\Lambda'}^{E/B}_{\hat{s}^4}$  & Eq.~\eqref{eq: xiB fix} & \\
      \bottomrule
    \end{tabular*}
  \end{threeparttable}
  \caption{Summary of all conservative tidal response coefficients including static tidal Love numbers (TLNs), ``dynamical" tidal Love numbers (DTLNs) and conservative tidal mixing numbers (CTMNs). We also estimated the relative PN order in which these coefficients will first appear in the (EoM) with general spin orientation.}
  \label{table: summary tidal number 1}
\end{table}

In table \ref{table: summary tidal number 2}, we list all the coefficients that contribute to the dissipative tidal interactions and the related PN order in the energy flux as well as in the EoM through radiation reaction. The RG running in the dissipative sectors starts to contribute at 5.5 PN order in the energy flux. 

\begin{table}[h]
  \centering
  \begin{threeparttable}
    \label{tab1:threshold}
    \begin{tabular*}{\textwidth}{l@{\extracolsep{\fill}}*{3}{c}}
    \toprule
      \textbf{Names} & \textbf{Notation} & \textbf{Explicit expression} & \textbf{PN order in Flux~(EoM)}\\
      \midrule
      \multirow{2}[0]{*}{LO TDNs} & $H_{\hat{s}^1}^{E} , H_{\hat{s}^3}^{E}$ & \multirow{2}[0]{*}{Eq.~\eqref{eq:LO dissipation number}} & 2.5~(5) \\
      & $H_{\hat{s}^1}^{B} , H_{\hat{s}^3}^{B}$ & & 3.5~(6) \\
      \midrule
      \multirow{2}[0]{*}{NLO TDNs} & $H_{\hat{s}^0,\omega}^{E},H_{\hat{s}^2,\omega}^{E}, H_{\hat{s}^4,\omega}^E$ & \multirow{2}[0]{*}{Eq.~\eqref{eq:NLO diss}} & 4~(6.5) \\
      & $H_{\hat{s}^0,\omega}^B, H_{\hat{s}^2,\omega}^B, H_{\hat{s}^4,\omega}^B$ & & 5~(7.5) \\ 
      \midrule
      \multirow{2}[0]{*}{NNLO TDNs} & $H_{\hat{s}^1,\omega^2}^{E} , H_{\hat{s}^3,\omega^2}^{E}$ & \multirow{2}[0]{*}{Eq.~\eqref{eq:NNLO diss RG}}& 5.5~(8) \\
      & $H_{\hat{s}^1,\omega^2}^{B} , H_{\hat{s}^3,\omega^2}^{B}$ & & 6.5~(9) \\
      \midrule
      \multirow{2}[0]{*}{DTMNs} & $\tilde{H}^{E/B}_{\hat{s}^0}, \tilde{H}^{E/B}_{\hat{s}^2}, \tilde{H}^{E/B}_{\hat{s}^4}$ & Eq.~\eqref{eq: nuEfix}, Eq.~\eqref{eq: nuBfix} & \multirow{2}[0]{*}{4~(6.5)} \\
      &  ${H'}^{E/B}_{\hat{s}^0}, {H'}^{E/B}_{\hat{s}^2}, {H'}^{E/B}_{\hat{s}^4}$  & Eq.~\eqref{eq: xiB fix} & \\
      \bottomrule
    \end{tabular*}
  \end{threeparttable}
  \caption{Summary of all dissipative tidal response coefficients including LO/NLO/NNLO tidal dissipation numbers (TDNs) and dissipative tidal mixing numbers (DTMNs). We estimated the relative PN order where these coefficients will first appear in the energy flux with general spin orientation. We also show the PN order in which they enter the EoM through radiation-reaction in brackets.}
  \label{table: summary tidal number 2}
\end{table}

\subsection{Future Directions}
The dynamical tidal response is of particular importance in studying the structure of scatterings on Kerr BHs and the GW astrophysics. Our work can be extended in the following directions:
\begin{itemize}
    \item \textbf{Constraining tidal dissipation numbers from data}: To better understand the nature of compact objects, one needs to consistently include all finite size effects such as spin-induced multipole moments, conservative tidal deformations and tidal dissipations \cite{LIGOScientific:2021djp}. Even though the literature studying the extraction of  spin-induced multipoles and the tidal Love numbers from the current Gravitational-wave Transient Catalog (GWTC) catalog \cite{LIGOScientific:2018mvr,LIGOScientific:2020ibl,LIGOScientific:2021usb,LIGOScientific:2021djp} is vast, studies for constraining dissipation numbers are missing. Moreover, recently pointed out in Refs.~\cite{Ripley:2023qxo,Most:2021zvc,Most:2022yhe}, within the scenario of large bulk viscosity of the Neutron stars, the tidal dissipations could give non-negligible contributions.

    \item \textbf{Loop calculations and the analytic structure of 
    scattering amplitudes in GR}: In Sec.~\ref{subsec: RG elastic}, we pointed out that the RG running of dynamical Love numbers for Schwarzschild black holes should originate from the 6-loop UV divergences in the bulk diagrams. It would be interesting to carry out an explicit calculation of these diagrams. 
    These calculations can also help us better understand the near-far factorization and its apparent breakdown discussed in Sec.~\ref{sec:BHPT}, as well as the analytic structure of 
    scattering amplitudes from 
    the MST solution to the Teukolsky equation.
    \item \textbf{Extensions to the non-linear perturbations}: With the recently growing interest of the non-linearities in the Kerr BH perturbations \cite{Lagos:2022otp,Mitman:2022qdl,Cheung:2022rbm,DeLuca:2023mio}, it would be interesting to explore the non-linear tidal response function of Kerr BHs.
    \item \textbf{Extensions to arbitrary compact objects} : While the worldline EFT framework presented here applies to most of compact objects of interest, the analytic GR results are 
    currently available only for black holes. 
    Motivated by the current interest in searches for exotic compact objects, it would be very interesting to extend the methods in the paper to generic compact objects, along the lines in \cite{Maggio:2020jml}. 
\end{itemize}

\section*{Acknowledgements}
We thank  Rafael Aoude, Clifford Cheung, Horng Sheng Chia, Andreas Helset, Lam Hui, Austin Joyce, Jung-Wook Kim, Yue-Zhou Li, Julio-Parra Martinez,  Raj Patil, Eric Poisson, Rafael Porto, Alexander Ochirov, Canxin Shi, Jan Steinoff, Justin Vines, Huan Yang, Heng-Rui Zhu for useful discussions. We thank Chris Kavanagh in particular for sharing with us the code for generating solutions to the Teukolsky equation. We also thank Huan Yang for organizing the Strong Gravity Seminar in Perimeter Institute where some of the results were presented.

\appendix

\section{Worldline Action For Spinning Particles With Tidal Moments}
\label{app:tetrads}
In the main text, we briefly explained the inclusion of (tidally induced) multipole moments and their respective couplings with the tidal fields. Here, we provide some additional details regarding the worldline action for a spinning particle including tidal effects and our choice of tetrad for describing the tidal response. We follow the presentation given in \cite{Saketh:2022xjb} for the inclusion of tidally induced multipole moments.


A general worldline action for a spinning particle with tidally induced\footnote{In principle, this action may also be used for including spin-induced multipole moments.} quadrupole and octupole degrees of freedom in worldline theory may be written in the implicit form
\begin{alignat}{3}
S &= \int d\tau \mathcal{L}(u^{\mu}=\dot{z}^{\mu},\Omega^{\mu\nu}=\epsilon_A{}^{\mu}\frac{D\epsilon^{A\nu}}{D\tau}, Q_{E,B}^{\mu_L}, \frac{D Q_{E,B}^{\mu_L}}{D\tau},R_{\mu\nu\rho\sigma},\nabla_{\lambda}R_{\mu\nu\rho\sigma}),~(\ell=2,3) \\ &= \int d\tau L(u^{\mu},\Omega^{\mu\nu}, Q_{E,B}^{\mu_L}, \frac{D Q_{E,B}^{\mu_L}}{D\tau})  + \int d\tau Q_E^{\mu\nu}E_{\mu\nu} + \int d\tau Q_B^{\mu\nu\rho}B_{\mu\nu\rho} +(E\leftrightarrow B), \nnm 
\end{alignat}
where $\epsilon_A{}^{\mu}$ is a co-rotating orthonormal tetrad satisfying $\epsilon_A{}^{\mu}\epsilon^{B\nu}=g^{\mu\nu},~\epsilon_A{}^{\mu}\epsilon_{B\mu}= \eta_{AB}$, which move and spins along with the particle. Varying the action with respect to the tetrad $\epsilon_A{}^{\mu}(\tau)$, and the worldline $z^{\mu}(\tau)$, while respecting all constraints yields the well known pole-dipole MPD equations of motion for momentum and spin, along with the contribution of tidal forces and torques given by
\begin{alignat}{3}
\label{seqn}
 \frac{D S^{\mu\nu}}{D\tau} &=  2 p^{[\mu}u^{\nu]}+\frac{4}{3}R^{[\mu}{}_{\lambda\rho\sigma}J_{\lambda}{}^{\nu]\tau\rho\sigma}  +\frac{2}{3}\nabla^{\lambda}R^{[\mu}{}_{\tau\rho\sigma}J_{\lambda}{}^{\nu]\tau\rho\sigma}+\frac{1}{6}\nabla^{[\mu}R_{\lambda\tau\rho\sigma}J^{\nu]\lambda\tau\rho\sigma}. \\
\frac{D p_{\mu}}{D\tau} &= -\frac{1}{2}R_{\mu\nu\rho\sigma}u^{\nu}S^{\rho\sigma}-\frac{1}{6}J^{\lambda\nu\rho\sigma}\nabla_{\mu}R_{\lambda\nu\rho\sigma}-\frac{1}{12}J^{\tau\lambda\nu\rho\sigma}\nabla_{\mu}\nabla_{\tau}R_{\lambda\nu\rho\sigma}.
\end{alignat}
where $p_{\mu} =(\partial \mathcal{L}/\partial u^{\mu})|_{\Omega^{\mu\nu}}, ~S_{\mu\nu} = \frac{1}{2}(\partial \mathcal{L}/\partial \Omega^{\mu\nu})-\sum_{\ell=2}^{3}\ell [P_{E}^{[\mu}{}_{\mu_{L-1}}Q^{\nu]\mu_{L-1}}_{E}+(E \leftrightarrow B)]$, with $P_{E/B}^{\mu_L}= (\partial \mathcal{L}/ \partial \dot{Q}_{\mu_L}^{E/B})$, and\footnote{We are using $\langle abcd \rangle_R$ ($\langle abcd \rangle_{\nabla R}$)  to represent the symmetrization of indices according to the symmetries of the Riemann tensor (covariant derivative of the Riemann tensor).} 
\begin{alignat}{3}
& J^{\mu\nu\rho\sigma} =  -6\frac{\partial \mathcal{L}}{\partial R_{\mu\nu\rho\sigma}} =- 3 u^{[\mu}Q_{E}^{\nu][\rho}u^{\sigma]}+\frac{3}{2}Q_{B}^{\alpha\langle\rho}\epsilon_{\beta\alpha}{}^{\mu\nu}u^{|\beta|}u^{\sigma\rangle_R},\label{4quad}\\& J^{\lambda\mu\nu\rho\sigma} = -12\frac{\partial \mathcal{L}}{\partial \nabla_{\lambda}R_{\mu\nu\rho\sigma}} = - 6 u^{\langle\mu}Q_{E}^{\nu\rho\lambda}u^{\sigma\rangle_{\nabla R}}+3 Q_{B}^{\alpha\langle\rho\lambda}\epsilon_{\beta\alpha}{}^{\mu\nu}u^{|\beta|}u^{\sigma\rangle_{\nabla R}}.\label{4oct}
\end{alignat}
To solve the equations of motion, we also need to impose a ``spin supplementary condition'' (SSC) to ensure that $S_{\mu\nu}$ it is a spatial tensor with the right number of degrees of freedom. Here, we impose the ``covariant'' or Tulczyjew-Dixon SSC at the level of the equations of motion upon the total physical spin angular momentum as $S^{\mu\nu}p_{\mu}=0$. This enforces that the spin tensor is a spatial antisymmetric tensor in the frame defined by $p_{\mu}$, with 3 degrees of freedom, as expected. Similarly, we also need to place constraints on the quadrupole (octupole) moments, to render them orthogonal to $u^{\mu}$ , symmetric and trace-free as expected from the tidal fields with which they couple. It is convenient to impose this by appropriately choosing the tidal response such that $Q^{\mu\nu}_Eu_{\nu}$ or $Q^{\mu\nu}_Ep_{\nu}=0$\footnote{Note that the difference between $p_{\mu}$ and $u_{\mu}$ becomes relevant only in the presence of external curvature. However, in this work, since we restrict ourselves to linearly induced tides and comparisons with linear perturbation theory, we can neglect the difference in the tidal response. For the same reason, we can treat the black hole as a stationary object with fixed spin following a geodesic in the tidal response.}.

For instance, consider a generic tidal response for the electric quadrupole moment $Q_E^{\mu\nu}$ to next-to-leading order ignoring quadrupole-octupole mixing. Consistent with parity, we can write 
\begin{equation}
Q^{\mu\nu}_E = -M (GM)^5\Bigg[(\lambda_E)^{\mu\nu,\rho\sigma}E_{\rho\sigma}- (GM)(\lambda_{E,\omega})^{\mu\nu,\rho\sigma}\frac{DE_{\rho\sigma}}{D\tau}\Bigg].
\label{eq: app demo}
\end{equation}
Here, we require $\lambda_E^{\mu\nu,\rho\sigma}$ to be symmetric and trace-free in $\mu\nu$ and $\rho\sigma$ respectively, and also orthogonal to $u^{\mu}$, thus curtailing undesired degrees of freedom. Given the space-like nature of the multipole moments, it is convenient to rewrite the above response equation in a suitably chosen orthonormal tetrad $e_{I}{}^{\mu}$, with $e_{I=0}{}^{\mu}= u^{\mu}$. Then, we can express both the multipole moments and tidal fields as spatial STF tensors in the tetrad as $Q_E^{ij} = Q_E^{\mu\nu}e^i{}_{\mu}e^j{}_{\mu}$,~$E_{ij} =E_{\mu\nu}e^i{}_{\mu}e^j{}_{\mu}$ where $i,j=1,2,3$. We can then rewrite Eq.~\eqref{eq: app demo} by contracting with $e^i{}_{\mu}e^j{}_{\nu}$ as 
\begin{alignat}{3}
 \nnm Q_E^{ij} &= -M(GM)^5 \Bigg[(\lambda_E)^{ij,kl}E_{kl}- (GM)(\lambda_{E,\omega})^{ij,mn}e_{m}{}^{\rho}e_{n}{}^{\sigma}\frac{D(E_{kl}e^{k}{}_{\rho}e^{l}{}_{\sigma})}{D\tau}\Bigg], 
 \\& = \nnm-M(GM)^5\Bigg[(\lambda_E)^{ij,kl}E_{kl}- (GM)(\lambda_{E,\omega})^{ij,kl}\frac{d(E_{kl})}{d\tau}-\\& - (GM)(\lambda_{E,\omega})^{ij,ml}\hat{\Omega}_{m}{}^{k}E_{kl} -  (GM)(\lambda_{E,\omega})^{ij,kn}\hat{\Omega}_{n}{}^{l}E_{kl}\Bigg], 
\end{alignat}
where $\hat{\Omega}^{ij}\equiv e^{i\mu}(De^{j}{}_{\mu}/D\tau)$ is the angular velocity of the space-like vectors in the tetrad, measuring their evolution along the worldline. The simplest choice is to let them be parallel transported along the worldline so that $\hat{\Omega}^{ij}=0$, which is consistent with the orthonormality conditions on the tetrad provided the black hole follows a geodesic. We can regard this choice of tetrad as a comoving tetrad, corresponding to an inertial observer co-translating (but not corotating) along the worldline. Then, we simply have 
\begin{alignat}{3}
 Q^{ij}_E = -M(GM)^5\Bigg[(\lambda_E)^{ij,kl}E_{kl}- (GM)(\lambda_{E,\omega})^{ij,kl}\frac{\partial(E_{kl})}{\partial t}\Bigg], 
\end{alignat}
where the ordinary proper-time derivative reduces to a simple partial time-derivative w.r.t coordinate time measured by the observer at infinity. In BHPT, this coordinate time corresponds to time measured at asymptotic infinity by an inertial stationary (w.r.t black hole) observer (e.g., the $t$ coordinate in the Kerr metric in Boyer-Lindquist coordinates). In this work, we consistently describe the tidal response and associated couplings in this comoving tetrad. In \cite{Goldberger:2020fot}, the spatial tetrad vectors are not parallel transported along the worldline, but instead identified with the corotating tetrad. Thus, the covariant-time derivative does not reduce to a simple partial derivative in the response. The two choices are related by a spatial rotation about the rotation axis $\epsilon_i{}^{\mu}=R_i{}^{j}(\tau)e_{j}^{\mu}$, $\epsilon_0{}^{\mu}\approx e_0{}^{\mu}=u^{\mu}$. This does not affect any physical results or conclusions.

\section{Plane Wave, Sperical Wave and Operator Form of Response Tensors}
\label{app:love rep}

\subsection{Basis for single particle states}
\label{app : single particle}
\textbf{Plane wave basis} - The physical graviton plane wave states $|\boldsymbol{k},h\rangle$, with fixed momentum $k=(\omega,\boldsymbol{k})$ and helicity $h=\pm 2$ are chosen to have the normalization and wavefunction
\begin{alignat}{3}
\label{eq:plane norm}
\langle \boldsymbol{k}',h'|\boldsymbol{k},h\rangle = 2|\boldsymbol{k}|(2\pi)^3 \delta^{(3)}(\boldsymbol{k}'-\boldsymbol{k}) \delta_{hh'},\quad \langle 0|h_{ij}(\boldsymbol{x})|\boldsymbol{k},h\rangle = \frac{1}{M_{\rm pl}} \exp(-ik\cdot x) \epsilon^h_{ij}(\boldsymbol{k}),
\end{alignat}
where $\epsilon^h_{ij}(\boldsymbol{k}) $ is the purely spatial symmetric polarization tensor satisfying $\epsilon^{i}{}_{i}=0$ and $\epsilon^{ij}k_{j}=0$. Finally, the completeness relation is given by
\begin{alignat}{3}
 \sum_h\int \frac{d^3 \boldsymbol{k}}{(2\pi)^3}\frac{|\boldsymbol{k},h\rangle \langle \boldsymbol{k},h|}{2|\boldsymbol{k}|} = \bold{I} 
\end{alignat}

\textbf{Spherical basis} -
We also define a different basis for single particle states, with fixed energy $\omega$, total angular momentum $\ell$, and angular momentum along z-axis $m$, i.e., labeled as $|\omega,\ell,m,h\rangle$ with normalization
\begin{alignat}{3}
\langle \omega,\ell,m,h |\omega',\ell',m',h'\rangle = 2\pi \delta(\omega-\omega') \delta_{\ell \ell'}\delta_{mm'}\delta_{hh'}.
\end{alignat}
Note that although it is meaningless to define the helicity in the spherical basis, we still use the label $h=\pm 2$ to indicate that they are a superposition of plane waves with definite helicity. The completeness relation is then given by
\begin{alignat}{3}
\label{comps}
\int \frac{d\omega}{2\pi}\sum_{\ell,m,h}|\omega,\ell,m,h\rangle\langle \omega,\ell,m,h| = \bold{I} ~.
\end{alignat}
\textbf{Transformation between plane wave states and spherical states} -
For simplicity, we first consider a plane wave state with momentum along the $\hat{z}$-direction. i.e., $|\omega \hat{\boldsymbol{z}},h\rangle$. Its wave function is given by $\exp(-i \omega t + k r \cos(\theta))\epsilon^h_{ij}(\boldsymbol{k})$. First of all, it is important to notice the plane wave state $|\omega \hat{\boldsymbol{z}}, h \rangle$ only has overlap with spherical state $|\omega, \ell, m,h \rangle$ when $m = h$. This is because they are both the eigenstate of $\hat{z}$-direction angular momentum $\bold{J}_z$. Thus, we can expand it in spherical basis simply as
\begin{alignat}{3}
|\omega \hat{\boldsymbol{z}}, h = \pm 2 \rangle = \sum_{\ell=2}^{\infty} C_h^{\ell}(\omega)|\omega,\ell,m=h,h=\pm 2 \rangle, 
\label{p2s}
\end{alignat}
where the coefficients $C_h^{\ell}(\omega)$ will be fixed later by the normalization of the plane wave states. Now, to obtain the expansion for momentum along arbitrary direction, we can simply multiply with the following rotation operator with definition of Euler angle follow \cite{enwiki:1158957702}
\begin{equation}
    \mathbf{R}(\phi,\theta,\psi) =  \exp(-i\phi \bold{J}_z)\exp(-i\theta \bold{J}_y)\exp(-i \psi \bold{J}_z) ~.
\end{equation}
which rotates the vector $\hat{\boldsymbol{z}}$ into the direction $\hat{\boldsymbol{k}}=(\theta,\phi)$ in spherical-polar coordinates. Note that the parameter $\psi$ is of no consequence when acting on a vector that is already along $\hat{z}$-axis, or more generally an eigenvector of $\bold{J}_z$. Also, from now on, we use the short hand notation 
\begin{equation}
    \mathbf{R}(\hat{\boldsymbol{k}},0) \equiv \mathbf{R}(\phi,\theta,0) ~
\end{equation}
for rotational matrix.
Acting this operator upon the expansion in Eq.~(\ref{p2s}), we get
\begin{alignat}{3}
|\boldsymbol{k},h = \pm 2\rangle = \bold{R}(\hat{\boldsymbol{k}},0) |\omega \hat{\boldsymbol{z}}, h = \pm 2\rangle = \sum_{\ell=2}^{\infty} C^{\ell}_{h}(\omega)\bold{R}(\hat{\boldsymbol{k}},0)|\omega,\ell,m=h,h=\pm 2 \rangle.
\end{alignat}
We find it useful to introduce the Wigner D-matrix as
\begin{equation}
\langle \ell,m'| \bold{R}(\hat{\boldsymbol{k}},0) |\ell,m \rangle \equiv D^{\ell}_{m'm}(\hat{\boldsymbol{k}},0) ~.
\end{equation}
with
\begin{equation}
D^{\ell}_{mh}(\hat{\boldsymbol{k}},0)=(-1)^m \sqrt{\frac{4\pi}{2\ell+1}}{}_{h}Y_{\ell,-m}(\hat{\boldsymbol{k}}),
\label{wig2sp}
\end{equation}
where ${}_{h}Y_{\ell m}$ are the spin-weighted spherical harmonics.
After using the completeness relation for the spherical basis in Eq.~(\ref{comps}), we can rewrite this as 
\begin{alignat}{3}
\label{eq: sph 2 plane}
|\boldsymbol{k},h\rangle = \sum_{\ell=2}^{\infty}\sum_{m=-\ell}^\ell C^\ell_h(\omega) D^{\ell}_{mh}(\hat{\boldsymbol{k}},0)|\omega,\ell,m,h\rangle,
\end{alignat}
where we have used $\langle \omega',\ell',m',h'|\bold{R}(\hat{k},0)|\omega=|\vec{\boldsymbol{k}}|,\ell,m,h \rangle = 2\pi \delta(\omega-\omega') \delta_{hh'} \delta_{\ell \ell'} D^{\ell}_{m'm}(\hat{\boldsymbol{k}},0) $ which follows from the invariance of the helicity label $h$ and energy (or frequency $\omega$) under rotations.

We can now fix the coefficients $C^{\ell}_h(\omega)$ by considering the normalization of plane waves in Eq.~\eqref{eq:plane norm} and get 
\begin{alignat}{3}
C_{h}^\ell (|\boldsymbol{k}|) = 2\pi\sqrt{\frac{(2 \ell +1)}{2\pi|\boldsymbol{k}|}},
\end{alignat}
which is consistent with \cite{Goldberger:2020fot}. Now, we can write down the final expression that relates plane wave states and spherical wave states  
\begin{alignat}{3}
|\boldsymbol{k},h\rangle = \sum_{\ell=2}^\infty \sum_{m=-\ell}^\ell 2\pi\sqrt{\frac{(2\ell+1)}{2\pi|\boldsymbol{k}|}} D^{\ell}_{mh}(\hat{\boldsymbol{k}},0)||\boldsymbol{k}|,l,m,h\rangle.
\label{sp2p}
\end{alignat}

In principle, we are now set for writing amplitudes calculated for plane waves in spherical basis. However, to simplify them further, we want to write the polarization tensors $\epsilon^h_{ij}(\boldsymbol{k})$ in terms of the Wigner D-matrix as well. So, let us establish the conventions used for the polarization tensors, and their relation to Wigner D-matrix.

\textbf{Relating polarization tensors to Wigner-D matrix} -
Let's now discussing the relation between polarization tensors and the Wigner-D matrix. Before going to details, let's first make clear the following concept:
\begin{itemize}
    \item  \textbf{Rank-$\ell$ STF tensors}: symmetric trace-free (STF) tensor with L indices. For example, the electric tidal field $E_{ij}$ is a rank-2 STF tensor. Furthermore, all the rank-L STF tensors form a vector space with dimention $2\ell + 1$, and thus can be used as a vector space for $SO(3)$ representation. The transformation law between the spherical harmonics and tensorial spherical harmoncis is given by
    \begin{equation}
        Y_{\ell m} = \mathcal{Y}_{i_1 \cdots i_\ell}^{\ell m} n^{i_1} \cdots n^{i_\ell} ~,
    \end{equation}
    in which $Y_{i_1 \cdots i_\ell}^{\ell m}$ can be viewed as the base vector in rank-$\ell$ STF space.
\end{itemize}

Going back to the graviton polarization tensor, since it encodes the spin of gravitons, we can define it as the eigenstate of $\hat{z}$-direction angular momentum operator as
\begin{equation}
{J_z}|_{\rm rank-2}\epsilon_{h=\pm 2}(\hat{\boldsymbol{z}}) = h \epsilon_h(\hat{\boldsymbol{z}}) ~,
\end{equation}
where
\begin{alignat}{3}
J_z|_{\rm rank-2} = I \otimes J_z + J_z \otimes I ~.
\label{r1tor2}
\end{alignat}
The right hand needs to be evaluated in the fundamental representation.
In component form, we can write it as 
\begin{equation}
    {J_z|_{\rm rank-2}}^{ij}{}_{kl} = \delta^{i}_{k} {J_z}^{j}{}_{l} + {J_z}^i{}_k \delta^{j}_\ell ~.
\end{equation}
We can get a nicer formula by further symmetrizing $(i,j)$ and $(k,l)$ to get
\begin{alignat}{3}
{J_z|_{\rm rank-2}}^{ij}{}_{kl} = \frac{1}{2}(\delta^{i}_{k}{{J}_z}^{j}{}_{l}+\delta^{i}_{l}{J_z}^{j}{}_{k}+\delta^{j}_{k}{J_z}^{i}{}_{l}+\delta^{j}_{l}{J_z}^{i}{}_{k}) ~.
\end{alignat}
Since we know that in the spherical basis $\bold{J}_z |\ell = 2, m =h=\pm 2 \rangle = \pm 2 |\ell = 2, m =h= \pm 2\rangle$, then we naturally have
\begin{alignat}{3}
\epsilon^h_{ij} (\boldsymbol{z}) = \langle i,j |\ell=2,m=h=\pm 2\rangle,
\end{alignat}
where we have chosen the normalization such that
\begin{equation}
    \epsilon_{ij}^{+2} (\boldsymbol{z}) = 
    \begin{pmatrix}
        \frac{1}{2} & \frac{i}{2} & 0 \\
        \frac{i}{2} & - \frac{1}{2} & 0 \\
        0 & 0 & 0 
    \end{pmatrix}
    \propto \mathcal{Y}_{ij}^{22} ~,
    \quad 
    \epsilon_{ij}^{-2} (\boldsymbol{z}) = 
    \begin{pmatrix}
        \frac{1}{2} & -\frac{i}{2} & 0 \\
        -\frac{i}{2} & - \frac{1}{2} & 0 \\
        0 & 0 & 0 
    \end{pmatrix}
    \propto \mathcal{Y}_{ij}^{2-2}
\end{equation}
and ${\epsilon}^{*ij}(\hat{\boldsymbol{k}})\epsilon_{ij}(\hat{\boldsymbol{k}})=1$.

Now, given the polarization tensor for when the momentum is along $\hat{\boldsymbol{z}}$, we can derive the polarization tensor for arbitrary direction $\hat{\boldsymbol{k}}$ 
\begin{equation}
    \begin{aligned}
            \epsilon_{ij}^h(\boldsymbol{k}) & =  \langle i,j| \bold{R}(\phi,\theta,0)|\ell=2,m=h=\pm 2\rangle \\
      & = \sum_{m=-2}^2 D_{mh}^{\ell=2}(\hat{\boldsymbol{k}},0) \langle i,j |\ell=2,m\rangle.
\label{wig2p}
    \end{aligned}
\end{equation}

\textbf{States with definite Parity} - Sometimes, it is convenient instead to switch to spherical basis states with definite parity. To that end, we define a parity operator $\bold{P}$, which flips momentum and helicity. It acts on plane wave states as 
\begin{alignat}{3}
\bold{P}|\vec{k},h \rangle = |-\vec{k},-h\rangle,~\bold{P}^2=\bold{I}.
\end{alignat}
We can use Eq.~\eqref{eq: sph 2 plane} to identify its action on spherical basis states, which is 
\begin{alignat}{3}
\bold{P}|\omega,\ell,m,h\rangle = (-1)^\ell |\omega,\ell,m,-h\rangle.
\end{alignat}
Using this, we can easily define the parity even and odd states in the spherical basis as 
\begin{alignat}{3}
|\omega,\ell,m,P=\pm\rangle = \frac{1}{\sqrt{2}} \Big(|\omega,\ell,m,h=+2\rangle\pm(-1)^\ell |\omega,\ell,m,-h=-2\rangle \Big).
\label{eq: h2P}
\end{alignat}
\subsection{Operator Form of Tidal Response Tensors}
In Section \ref{sec:EFT}, we have shown that the tidal response tensor can be nicely written as the operator form. In this appendix, we are going to show part of the derivation.

\begin{itemize}
    \item \textbf{zero spin}: The simplest case would be the zeroth order in spin where
    \begin{equation}
        \langle i,j|\mathbf{\Lambda}_{\hat{s}^0} | k,l \rangle= \Lambda_{\hat{s}^0} \delta^{\langle i}_{\langle k}\delta^{j \rangle }_{l\rangle} ~.
    \end{equation}
    In the abstract notation, this means that 
    \begin{equation}
        \mathbf{\Lambda}_{\hat{s}^0} = \Lambda_{\hat{s}^0} \mathbf{I}~.
    \end{equation}

    \item \textbf{linear in spin}: Now, we are moving to the linear in spin case
    \begin{alignat}{3}
        \langle i,j | \mathbf{\Lambda}_{\hat{s}^1} | k,l \rangle = H_{\hat{s}^1} \hat{S}^{\langle i}{}_{\langle k}\delta^{j\rangle}_{l\rangle}.
    \end{alignat}
Recall that $\hat{S}_{ij}\equiv \epsilon^{ijk}\hat{s}_k$, where $\hat{s}$ is normal vector pointing along the spin of the particle in the rest frame. Since we have oriented the spin to be along the $\hat{z}$-axis, we have $\hat{s}^i=\delta^{i3}$. By noting the following relation
\begin{equation}
    \label{eq: def Jz}
    {J_z}^i_j = -i \epsilon^{i}{}_{jk}\hat{s}^k = - i \hat{S}^{i}{}_j ~,
\end{equation}
we immediately see that 
\begin{equation}
    \langle i,j | \mathbf{\Lambda}_{\hat{s}^1}|k,l \rangle  =  \frac{1}{2} i H_{\hat{s}^1} J_z|_{\rm rank-2}
\end{equation}
In the abstract notation, we can write this as
\begin{equation}
    \bold{\Lambda}_{\hat{s}^1} = \frac{1}{2} i H_{\hat{s}^1} \mathbf{J}_z
\end{equation}

\item \textbf{quadratic in spin}: The quadratic in spin part of the response tensor is given as
\begin{alignat}{3}
\langle i,j | \bold{\Lambda}_{\hat{s}^2} |k,l\rangle  = \Lambda_{\hat{s}^2} \hat{s}^{\langle i}\hat{s}_{\langle k}\delta^{j\rangle}_{l\rangle} = \frac{\Lambda_{\hat{s}^2}}{2} \(({\hat{s}^2})^{\langle i}{}_{\langle k}\otimes I^{j\rangle }{}_{l \rangle} +  I^{\langle i}{}_{\langle k} \otimes(\hat{s}^{2})^{j\rangle}{}_{l\rangle} \),
\end{alignat}
where we have defining $\hat{s}^2 \equiv \hat{s}\otimes\hat{s}$. First of all, we notice the following relation
\begin{alignat}{3}
    \hat{s}_i \hat{s}_k = \delta_{ik} - {J_z}_{ij}{J_{z}}^{j}_{k} ~,
\end{alignat}
which implies
\begin{alignat}{3}
{J}^2_z|_{\rm rank-2}= 2 (I\otimes I  + J_z \otimes J_z) - \hat{s}^2 \otimes I - I\otimes \hat{s}^2.
\end{alignat}
To proceed further, we symmetrize and remove the traces over $i,j$ and $k,l$ to get
\begin{alignat}{3}
\langle i,j|J_z^2|_{\rm rank-2}|k,l\rangle = 2 \langle i,j |I|_{\rm rank-2}|k,l\rangle + 2 J^{\langle i}_{z \langle k } J^{j \rangle}_{z  l\rangle } - 2\hat{s}^{\langle i}\hat{s}_{\langle k}\delta^{j\rangle}_{l\rangle},
\end{alignat}
and use the following relation
\begin{alignat}{3}
2 J^{\langle i}_{z \langle k } J^{j \rangle}_{z  l\rangle} = 2\langle i,j |I|_{\rm rank-2}|k,l\rangle  - 4\hat{s}^{\langle i}\hat{s}_{\langle k}\delta^{j\rangle}_{l\rangle}  
\label{conv:corr},
\end{alignat}
which can be derived using the Eq.~\eqref{eq: def Jz}. Eventually, we arrive at
\begin{alignat}{3}
\bold{\Lambda}_{\hat{s}^2}  =-\frac{1}{6} \Lambda_{\hat{s}^2} (\bold{J}_z^2-4~ \bold{I}).
\label{qisp:corr}
\end{alignat}

\item \textbf{cubic in spin}: Consider now the cubic in spin interactions
\begin{equation}
    \langle i,j | \mathbf{\Lambda}_{\hat{s}^3} | k,l \rangle = H_{\hat{s}^3} \hat{S}^{\langle i}{}_{\langle k} \hat{s}^{j \rangle} \hat{s}_{l \rangle} ~.
\end{equation}
To simplify this expression, we first notice that
\begin{equation}
    \begin{aligned}
         & \quad \hat{S} \otimes \hat{s} \otimes \hat{s} +  \hat{s} \otimes \hat{s} \otimes \hat{S}  \\
        & = i J_z|_{\rm rank-2} - i (J_z \otimes (J_z)^2 + (J_z)^2 \otimes J_z)
    \end{aligned}
\end{equation}
The last term can be simplified using the following relation:
\begin{alignat}{3}
J^3_z|_{\rm rank-2} = (J_z\otimes I + I\otimes J_z)^3 = {J}_z^3\otimes I + 3 J_z^2\otimes J_z + 3 J_z\otimes J_z^2 + I\otimes J_z^3.
\end{alignat}
We further notice that $J_z^3=J_z$ following from ${J}\cdot(\hat{s}\otimes\hat{s})=0$, we get  
\begin{alignat}{3}
 \hat{S} \otimes \hat{s} \otimes \hat{s} +  \hat{s} \otimes \hat{s} \otimes \hat{S} = -\frac{i}{3}(J_z|_{\rm rank-2}^3- 4 J_z|_{\rm rank-2}).
\end{alignat}
Since $\Lambda_{\hat{s}^3}$ is STF in $i,j$ and $k,l$, and thus we can simply symmetrize the indices and remove the trace to get
\begin{equation}
    \bold{\Lambda}_{\hat{s}^3} = -\frac{i}{6} H_{\hat{s}^3} (\bold{J}_z^3- 4 \bold{J}_z).
\end{equation}

\item
\textbf{quartic in spin} : 
Finally, the quartic-in-spin part of the response tensor is given by
\begin{alignat}{3}
\langle i,j |\bold{\Lambda}_{\hat{s}^4} |k,l \rangle = \Lambda_{\hat{s}^4} \hat{s}^{\langle i}\hat{s}_{\langle k}\hat{s}^{j\rangle}\hat{s}_{l\rangle}.
\end{alignat}
To rewrite this, we once again start with the expression for the appropriate power of $\bold{J}_z|_{\rm rank-2}$, as
\begin{alignat}{3}
     J_z^4|_{\rm rank-2} = 8 I \otimes I - 7~(\hat{s}^2\otimes I + I \otimes \hat{s}^2) + 6~\hat{s}^2\otimes \hat{s}^2 + 8 J_z\otimes J_z.
\end{alignat}
where we have used $J_z^3=J_z$ and $\bold{J}_z^2=\bold{I}-\hat{\bold{s}}^2$. Now, using Eq.~\eqref{conv:corr} and Eq.~\eqref{qisp:corr}, we finally arrive at
\begin{alignat}{3}
 \bold{\Lambda}_{s^4} = \frac{1}{6} \Lambda_{\hat{s}^4} (\bold{J}_z^2-4\bold{I})(\bold{J}_z^2-\bold{I}) ~.
\end{alignat}
\end{itemize}

\bibliographystyle{JHEP}
\bibliography{references}
\end{document}